\documentclass[prd,reprint,aps,amsmath,amssymb,tightenlines,showpacs,nofootinbib]{revtex4-1}

\usepackage{graphicx}% Include figure files
\usepackage{dcolumn}% Align table columns on decimal point
\usepackage{mathtools}
\usepackage{bm}% bold math
\usepackage[dvipsnames]{xcolor}
\usepackage{ulem}
\usepackage[colorlinks=true,citecolor=blue,linkcolor=blue]{hyperref}
\usepackage{comment}

\newcommand\myshade{85}
\colorlet{mylinkcolor}{PineGreen}
\colorlet{mycitecolor}{BrickRed}
\colorlet{myurlcolor}{violet}
\hypersetup{
  linkcolor  = mylinkcolor!\myshade!black,
  citecolor  = mycitecolor!\myshade!black,
  urlcolor   = myurlcolor!\myshade!black,
  colorlinks = true
}

\def\beq{\begin{equation}\begin{aligned}}
\def\eeq{\end{aligned}\end{equation}}

\newcommand{\be}{\begin{equation}} 
\newcommand{\ee}{\end{equation}}

\newcommand{\bk}{\mathbf{k}}
\newcommand{\bkp}{\mathbf{k'}}
\newcommand{\bs}[1]{\mathbf{#1}}

\newcommand{\bhat}[2]{\mathbf{\hat{#1}_{#2}}}

\newcommand{\VH}{\mathcal{V}_{\mathcal{H}}}
\newcommand{\VO}{\mathcal{V}_{\mathcal{O}}}

\newcommand{\mycomment}[1]{}

\DeclareMathOperator{\sech}{sech}

\def\ev{\, {\rm eV}}

\begin{document}

\title{Dark Radiation Isocurvature from Cosmological Phase Transitions} 

\author{Matthew~R.~Buckley}
\affiliation{NHETC, Department of Physics and Astronomy, Rutgers University, Piscataway, NJ 08854, USA}
\author{Peizhi~Du}
\affiliation{NHETC, Department of Physics and Astronomy, Rutgers University, Piscataway, NJ 08854, USA}
\author{Nicolas~Fernandez}
\affiliation{NHETC, Department of Physics and Astronomy, Rutgers University, Piscataway, NJ 08854, USA}
\author{Mitchell~J.~Weikert}
\affiliation{NHETC, Department of Physics and Astronomy, Rutgers University, Piscataway, NJ 08854, USA}

\begin{abstract}
Cosmological first order phase transitions are typically associated with physics beyond the Standard Model, and thus of great theoretical and observational interest. Models of phase transitions where the energy is mostly converted to dark radiation can be constrained through limits on the dark radiation energy density (parameterized by $\Delta N_{\rm eff}$). However, the current constraint ($\Delta N_{\rm eff} < 0.3$) assumes the perturbations are adiabatic. We point out that a broad class of non-thermal first order phase transitions that start during inflation but do not complete until after reheating leave a distinct imprint in the scalar field from bubble nucleation. Dark radiation inherits the perturbation from the scalar field  when the phase transition completes, leading to large-scale isocurvature that would be observable in the CMB. We perform a detailed calculation of the isocurvature power spectrum and derive constraints on $\Delta N_{\rm eff}$ based on CMB+BAO data. For a reheating temperature of $T_{\rm rh}$ and a nucleation temperature $T_*$, the constraint is approximately $\Delta N_{\rm eff}\lesssim 10^{-5} (T_*/T_{\rm rh})^{-4}$, which can be much stronger than the adiabatic result. We also point out that since perturbations of dark radiation have a non-Gaussian origin, searches for non-Gaussianity in the CMB could place a stringent bound on $\Delta N_{\rm eff}$ as well.
\end{abstract}

\maketitle

%%%%%%%%%%%%%%%%%%%%%%%%%%%%%%%%%%%%%%%%%%%%%%%%%%%%%%%%%%%%%%
%%%%%%%%%%%%%%%%%%%%%%%%%%%%%%%%%%%%%%%%%%%%%%%%%%%%%%%%%%%%%%
\section{Introduction}
\label{sec:intro}

Extensions of the Standard Model of particle physics often result in first order phase transitions (FOPT) in the early Universe.
For example, the phase transition (PT) in models of electroweak baryogenesis is predicted to be first order~\cite{KUZMIN198536,Shaposhnikov:1986jp,Shaposhnikov:1987tw,Cohen:1993nk,Trodden:1998ym,Riotto:1999yt,Dine:2003ax,Morrissey:2012db}. 
New physics that postulates new confining gauge groups -- such as in composite Higgs/Randall Sundrum models which address the Planck-Weak Hierarchy problem~\cite{Creminelli:2001th, Randall:2006py, Nardini:2007me, Konstandin:2010cd, Konstandin:2011dr, Bunk:2017fic, Baratella:2018pxi,Bruggisser:2018mrt, Megias:2018sxv, Agashe:2019lhy,Fujikura:2019oyi, Agashe:2020lfz, Ares:2020lbt,Agrawal:2021alq, Levi:2022bzt, Csaki:2023pwy,Eroncel:2023uqf,Mishra:2023kiu,Mishra:2024ehr}, or in dark matter models with dark QCD or dark $SU(N)$ gauge theories~\cite{Bai:2013xga,Kribs:2016cew,Cline:2021itd,Shelton:2010ta,DUTTA2011364,Holthausen:2013ota,Schwaller:2015tja,Hall:2019rld,Bai:2021ibt,Bottaro:2021aal,Pasechnik:2023hwv,Pasechnik:2023hwv} -- can also result in FOPT. A class of Early Dark Energy models~\cite{Karwal:2016vyq,Mortsell:2018mfj,Poulin:2018cxd}, introduced to relieve the tension between the early- and late-time measurements of $H_0$ \cite{Planck:2018vyg,Riess:2021jrx}, generally require PTs in their dark sector~\cite{Niedermann:2019olb,Niedermann:2020dwg,Niedermann:2021vgd}. Probing cosmological FOPTs is therefore a problem of wide applicability to physics beyond the Standard Model.

Cosmological FOPTs create gravitational waves through bubble collisions, shock waves, and turbulence produced during the transition~\cite{Kosowsky:1992rz,Kosowsky:1991ua,Kosowsky:1992vn,Kamionkowski:1993fg,Caprini:2015zlo,Caprini:2019egz,Caldwell:2022qsj}. For phase transitions at ${\cal O}(10-100)~{\rm MeV}$ temperatures, these gravitational waves have frequencies accessible to the Pulsar Timing Array (PTA)~\cite{NANOGrav:2023gor,NANOGrav:2023hvm,EPTA:2023fyk,Reardon:2023gzh,Xu:2023wog}. 
Future experiments, including LISA \cite{LISA:2017pwj,LISACosmologyWorkingGroup:2022jok}, DECIGO \cite{Kawamura:2006up,Kawamura:2020pcg}, and BBO \cite{Harry:2006fi}, will be able to probe transition temperature around the TeV scale or above.

In addition to these direct probes of the stochastic gravitational wave background generated by the FOPT, precision measurements of the Cosmic Microwave Background (CMB) can constrain phase transitions that generate massless or low-mass dark radiation (DR)~\cite{Bai:2021ibt}.\footnote{In this context, gravitational waves are also a form of dark radiation.} The energy density of DR is typically parameterized in terms of the extra number of effective neutrino species $\Delta N_{\rm eff}$ above the Standard Model expectation $N_{\rm eff}=3.044$ \cite{Froustey:2020mcq, Bennett:2020zkv,Akita:2020szl}. 
For free-streaming DR, the CMB sets the upper bound $\Delta N_{\rm eff}<0.3$ at 95\% confidence level (CL)~\cite{Planck:2018vyg}. 

Importantly, this constraint on the number of light degrees of freedom assumes the initial density perturbations of the dark radiation are {\it adiabatic}: with over- and under-densities proportional to those of the other radiation species.
If the DR is sourced by a FOPT that undergoes violent bubble nucleation, the initial spatial distribution of the perturbations will contain additional {\it isocurvature} modes.\footnote{For general discussions of isocurvature, see for example~\cite{Bucher:1999re,Wands:2000dp,Gordon:2000hv,Lyth:2002my,Malik:2004tf,Wands:2007bd}.} However, for FOPTs with energy scales far above the $\sim {\rm eV}$ scales of recombination and decoupling, the isocurvature perturbations are typically highly suppressed for wavenumbers large enough to leave observable imprints within the CMB \cite{Freese:2023fcr,Elor:2023xbz}. As a result -- even though isocurvature perturbations are tightly constrained by CMB measurements \cite{Planck:2018jri,Ghosh:2021axu}\footnote{There are also constraints on dark radiation isocurvature from Big Bang nucleosynthesis~\cite{Adshead:2020htj}.} -- these limits typically cannot be applied to the high-scale FOPTs, and the dominant constraint comes from the standard measurements on the number of relativistic degrees of freedom.

In this paper, we consider a broad class of high-scale FOPT models which generate large-scale isocurvature perturbations whose imprint {\it would} be visible within the CMB. These models have non-thermal (vacuum) phase transitions, where the scalar potential that triggers the transition is independent of temperature and couples to the inflaton field.
As a result of this coupling, the FOPT can begin during inflation, with a bubble nucleation rate per volume $\Gamma_{\rm PT}$ which we assume to satisfy
\begin{equation}\label{eq:incomplete condition}
\Gamma_{\rm PT} \ll H_{\rm inf}^4,
\end{equation}
during the inflationary epoch (when the Hubble parameter is $H_{\rm inf}$). 
This hierarchy of scales can be chosen so that the PT does {\it not} complete by the end of inflation and instead completes only after reheating.  In the scenario we consider in this paper, the energy density in the scalar field undergoing the FOPT is mostly converted into dark radiation when the PT completes, with perturbations inherited from the scalar field.

\begin{figure}
    \centering    \includegraphics[width=\columnwidth]{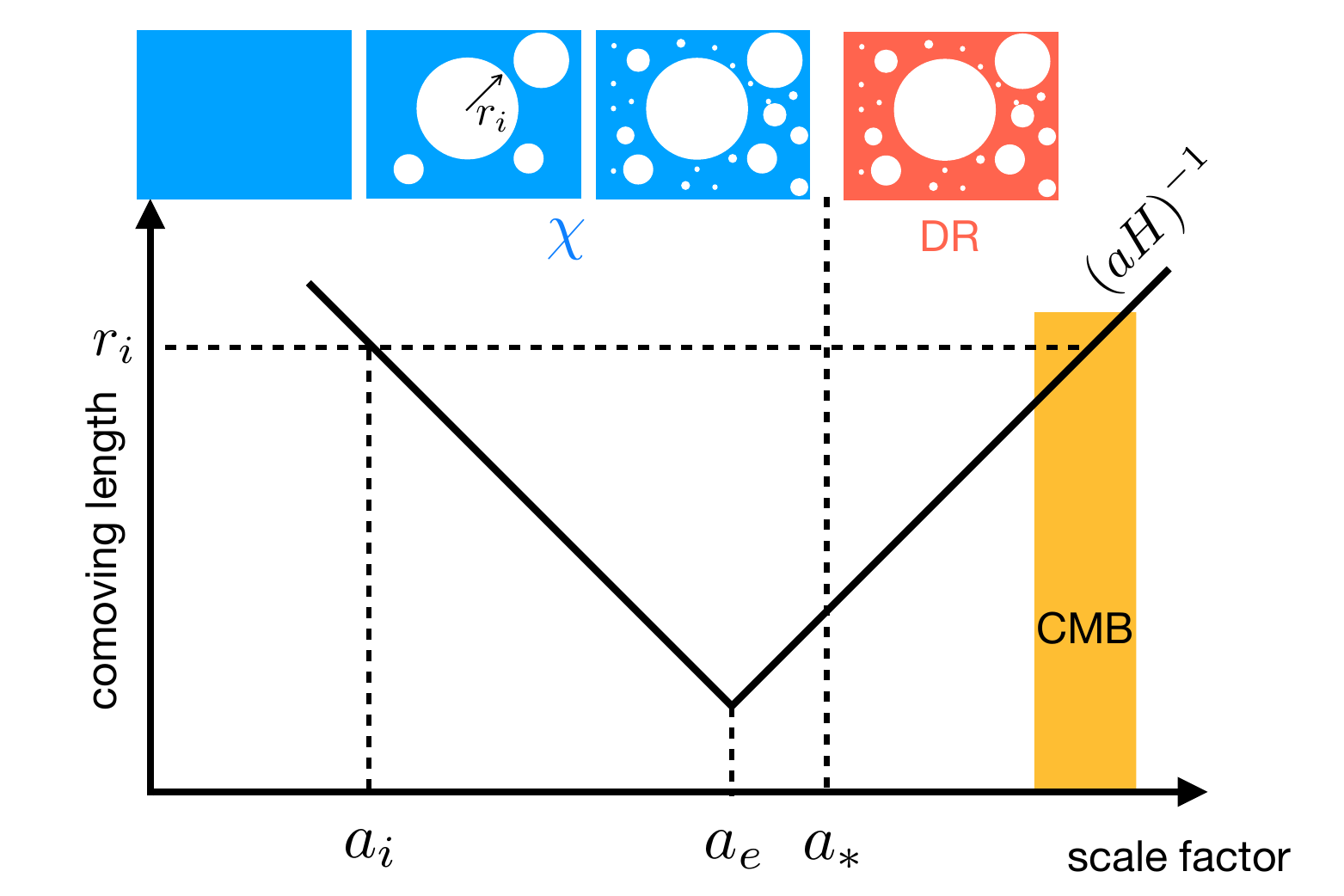}
    \caption{A schematic plot of comoving horizon size $(aH)^{-1}$ (black line) as a function of scale factor $a$. Three scale factors have been listed: $a_i$ denotes the start of the FOPT, $a_e$ shows the end of inflation, and $a_*$ corresponds to the nucleation temperature $T_*$ when phase transition completes. The shaded orange region shows the time window when modes that CMB can probe enter the horizon. The upper part of the plot shows the evolution of perturbations in the $\chi$ field due to nucleation of bubbles. $r_i\equiv (a_i H_{\rm inf})^{-1}$ is the size of the earliest and largest bubble, nucleated at $a_i$. After the completion of the PT around $a_*$, $\chi$ converts into DR with the same large-scale perturbations, which will imprint  DR isocurvature signals on the CMB after re-entering the horizon. (We note that the distribution of bubbles shown in this plot are schematic. We will present more realistic bubble distributions from simulations in Section~\ref{sec:bubbles}.)\label{fig:schematic}}
\end{figure}

Bubbles nucleated in the scalar field during inflation will expand to the horizon size in a Hubble time and remain frozen in comoving coordinates afterwards. As the comoving horizon is rapidly changing during inflation, this results in a wide range of bubble sizes, with the largest bubbles corresponding to the earliest nucleation events. The bubble distribution will later be imprinted in DR when the PT completes.
A schematic of cosmological history during and after inflation is shown in Figure~\ref{fig:schematic}. For phase transitions that start when the inflationary horizon is of order the comoving scale at recombination or larger, stochastic perturbations in DR can leave an observable imprint on the CMB.\footnote{For earlier works on effects of large bubbles from phase transitions during inflation, see Refs.~\cite{GUTH1983321,10.1093/mnras/253.4.637,Turner:1992tz,Copeland:1994vg,Baccigalupi:1999rz,Barir:2022kzo}.} As the bubbles are randomly distributed, the DR perturbations are uncorrelated with the fluctuations of the inflaton and behave as isocurvature. Since the CMB is consistent with adiabatic initial conditions with zero isocurvature \cite{Planck:2018jri}, we can place constraints on $\Delta N_{\rm eff}$ from this class of FOPTs that are much stronger than in the adiabatic scenario.
%%%%%%%%%%%%%%%%%%%%%%%%%%%%%%%%%%%%%%%%%%%%%%%%%%%%%%%%%%%%%%
%%%%%%%%%%%%%%%%%%%%%%%%%%%%%%%%%%%%%%%%%%%%%%%%%%%%%%%%%%%%%%
In this work, we present a detailed calculation of the DR isocurvature power spectrum from non-thermal PTs that start during inflation. We implement this model in CLASS \cite{Blas:2011rf,Lesgourgues:2011re,Lesgourgues:2011rh}, appropriately modified to include DR isocurvature. Using numerical simulations, we show how the CMB power spectra would change in the presence of DR isocurvature. We then perform a Markov Chain Monte Carlo (MCMC) scan using CMB and Baryon Acoustic Oscillation (BAO)  data to constrain the DR energy density (written in terms of $\Delta N_{\rm eff}$) and other parameters related to the phase transition. We find that when the temperature $T_*$ at which the phase transition completes (estimated from $\Gamma_{\rm PT}=H(T_{*})^4$), is close to the reheating temperature $T_{\rm rh}$, the isocurvature perturbations set a limit of  $\Delta N_{\rm eff} \lesssim 10^{-5}(T_*/T_{\rm rh})^{-4}$. This limit weakens as $T_*/T_{\rm rh}$ decreases and approaches the adiabatic constraint of $\Delta N_{\rm eff}<0.3$ when $T_* \ll T_{\rm rh}$. In addition to the isocurvature, a FOPT which begins during inflation will also result in non-Gaussianity in the CMB. While we defer a full analysis of such constraints to a future work, here we estimate the expected limits by analogy with neutrino density isocurvature non-Gaussianity \cite{Planck:2019kim,Montandon:2020kuk}.

This paper is organized as follows. In Section~\ref{sec:models_PT}, we introduce a model of non-thermal phase transitions and calculate the transition rate. In Section~\ref{sec:bubbles} we describe the dynamics of the phase transition during inflation and the resulting distribution of nucleated bubbles. In Section~\ref{sec:cosmological perturbation theory}, we calculate the two quantities that will lead to observational constraints on the model: the DR isocurvature power spectrum and bispectrum. 
Next, we implement the model in CLASS and show the effects of DR isocurvature on the CMB angular power spectra in Section~\ref{sec:observables}. In Section~\ref{sec:MCMC} we describe the datasets we use to set our limits and the details of our MCMC parameter scans. The resulting constraints on the DR isocurvature parameters from the CMB angular power spectra and bispectra are shown in Section~\ref{sec:results}. We conclude in Section~\ref{sec:conclusions}.

\section{Models of Non-thermal Phase Transitions}\label{sec:models_PT}

In this work, we consider models containing a scalar field $\chi$ that undergoes a FOPT that imprints significant isocurvature perturbations into the CMB. The minimum requirements for such models are:
\begin{itemize}
    \item[{\it i.}] The FOPT starts during inflation, at a moment when the comoving horizon ($r_i$) is above the minimum comoving length scale that the CMB can probe.
   \item[{\it ii.}] The PT rate remains small during inflation ($\gamma_{\rm PT}\equiv \Gamma_{\rm PT}/H_{\rm inf}^4\ll 1$). 
   \item[{\it iii.}] After reheating, the PT completes (at a temperature $T_*$, defined by $\Gamma_{\rm PT}/H(T_{*})^4=1$) and the energy density in the $\chi$ field converts into DR.
\end{itemize}
The interesting isocurvature effects we consider in this paper are present as long as these three requirements are met. 

As a specific implementation, we realize these features with a scalar potential that is non-thermal -- that is, independent of the temperature $T$ of the Universe. This non-thermality can be simply achieved if there are no sizeable couplings between the scalar field $\chi$ and the Standard Model fields. Without temperature dependence in the potential, the parameters of the model can be chosen such that the phase transition rate $\Gamma_{\rm PT}$ is constant and satisfies $\Gamma_{\rm PT}\ll H_{\rm inf}^4$. The phase transition in our study is triggered by the slow roll of the inflaton field $\phi$, through couplings between $\phi$ and $\chi$.
We further assume that the energy density of the $\chi$ field is subdominant to that of $\phi$ during inflation, and couplings between $\chi$ and other light fields in the dark sector allow the eventual conversion of vacuum energy into dark radiation after the phase transition completes. The example potential we consider is described in Section~\ref{sec:bare_potential} along with the resulting phase transition rate. The phase transition trigger during inflation is discussed in Section~\ref{sec:inflaton_trigger}.
%%%%%%%%%%%%%%%%%%%%%%%%%%%%%%%%%%%%%%%%%%%%%%%%%%%%%%%%%%%%%%
\begin{figure*}
    \centering
    \includegraphics[width=\columnwidth]{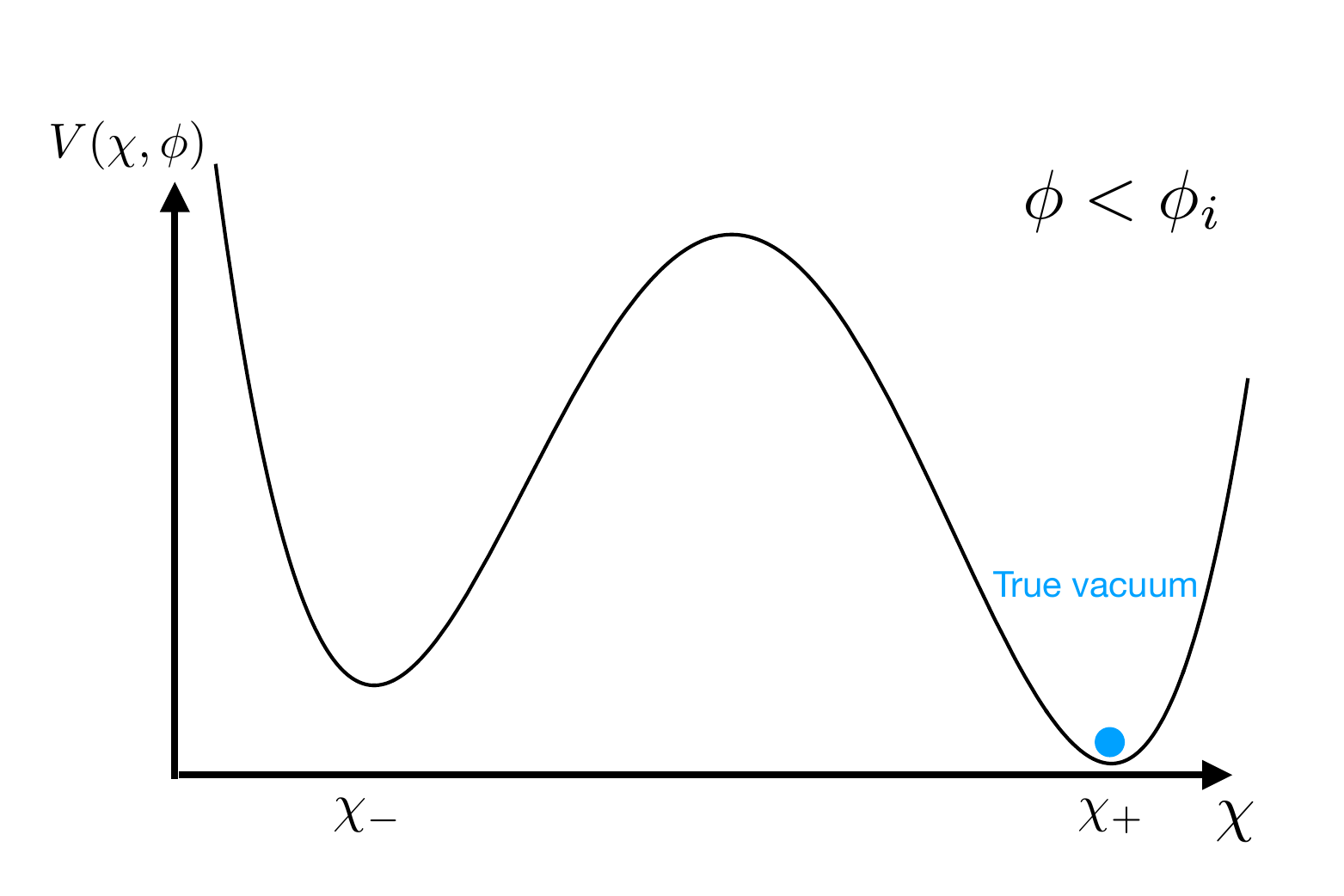}
    \includegraphics[width=\columnwidth]{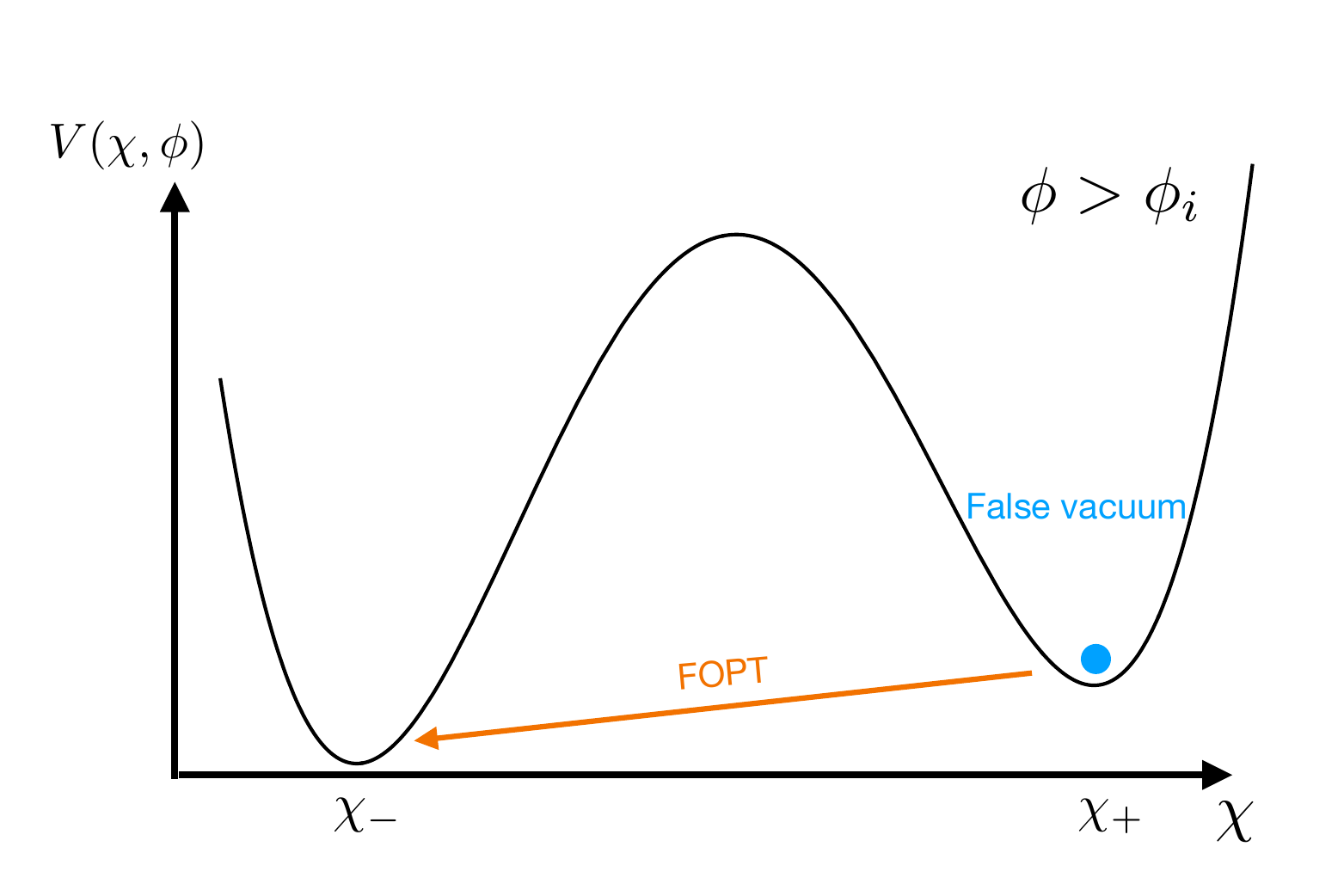}
    \caption{Schematic plot of the scalar potential $V(\chi,\phi)$ in Eq.~\eqref{eq:V_chi_phi}. When $\phi<\phi_i$, $\chi$ stays in the true vacuum near $\chi_+$. Soon after $\phi>\phi_i$, it turns into the false vacuum because $\mu(\phi)$ changes sign. Therefore, the FOPT starts and $\chi$ tunnels to true vacuum near $\chi_-$. \label{fig:schematic_potential}}
\end{figure*}
%%%%%%%%%%%%%%%%%%%%%%%%%%%%%%%%%%%%%%%%%%%%%%%%%%%%%%%%%%%%%%
\subsection{Temperature-Independent Phase Transition}\label{sec:bare_potential}

The temperature-independence of the scalar potential $V(\chi)$ can be achieved by requiring that $V$ is dominated by a bare tree-level potential. For this paper, we consider the simple example:
\begin{eqnarray}\label{eq:bare_potential_chi}
    V(\chi)=-\frac{1}{2}m^2 \chi^2+ \frac{\mu}{3} \chi^3+\frac{\lambda}{4} \chi^4.
\end{eqnarray}
The $Z_2$ symmetry of the potential is broken by $\mu$, which we assume to be small ($\mu \ll m$) and positive. This condition implies that the two vacuum states at $\chi_\pm \approx \pm m/\sqrt{\lambda}$ are nearly degenerate. That is, the vacuum energy difference $\Delta V \sim \mu m^3/\lambda^{3/2}$ is much smaller than the potential barrier $V_{\rm max}\sim m^4/\lambda$.

We assume the field is in the false vacuum state $\chi_+$ early in the inflationary phase. When the comoving horizon is $r_i$, the FOPT starts as $\chi$ tunnels to the true vacuum $\chi_-$. In Section~\ref{sec:inflaton_trigger}, we present one example of a dynamical mechanism for this transition, achieved by coupling $\chi$ and the inflaton $\phi$ through the $Z_2$-breaking $\mu$ term.

For $\Delta V \ll V_{\rm max}$, the thin-wall approximation is applicable and the quantum tunneling rate can be calculated in terms of the $O(4)$ symmetric bounce action~\cite{Linde:1981zj,Coleman:1977py,Callan:1977pt}
\begin{eqnarray}
S_4= \frac{27\pi^2 S_1^4}{2\Delta V^3} \sim  \lambda^{1/2} \left( \frac{m}{\mu} \right)^3 \gg 1,
\end{eqnarray}
where $S_1\sim m^3/\lambda$ is the bounce action along the radial direction. The quantum phase transition rate per volume is 
\begin{eqnarray}\label{eq:Gamma_PT_O4}
\Gamma_{\rm PT}\sim r_c^{-4} S_4^2  e^{-S_4},
\end{eqnarray}
where $r_c=3S_1/\Delta V$ is the size of the critical bubble for the $O(4)$ symmetric bubble under the thin-wall approximation. 
Since $\Gamma_{\rm PT}$ is exponentially sensitive to $S_4$, one can easily achieve the requisite small $\Gamma_{\rm PT}/H_{\rm inf}^4$ for a reasonable choice of parameters. Under these assumptions, and with the additional requirement that the number of $e$-folds after the tunneling begins is small enough (see Section~\ref{sec:bubbles} for details), the phase transition does not complete by the end of inflation. 
That is, at the end of inflation there are regions where the scalar field is in the false vacuum, separating bubbles of true vacuum nucleated during inflation. 

After inflation, we assume the Universe instantaneously reheats to a radiation-dominated state with a temperature $T_{\rm rh}$. We assume $\Gamma_{\rm PT}$ remains constant after inflation, and that the PT completes at a temperature $T_*$. As the horizon changes after inflation, a constant $\Gamma_{\rm PT}$ requires that the critical bubble size $r_c$ is less than the physical horizon size at any time. Otherwise, the bounce action will acquire horizon dependent corrections from the curved spacetime~\cite{PhysRevD.21.3305}. That is, we require
\begin{eqnarray}\label{eq:condition_Hinf}
    r_c^{-1} \sim \frac{\mu}{\lambda^{1/2}} \gg H_{\rm inf}.
\end{eqnarray}
This and all other requirements stated in this section can be satisfied by choosing appropriate values of $m$, $\lambda$, $\mu$, and $H_{\rm inf}$.

\subsection{Inflaton-Triggered Phase Transition}\label{sec:inflaton_trigger}

Our scenario requires a phase transition that is triggered during inflation, rather than by a coupling to a thermalized Standard Model bath. While more complex examples might be motivated by other theoretical considerations, a simple realization of this phenomenon is to couple the $\chi$ field to the $\phi$ through the small $Z_2$-breaking $\mu$ term in Eq.~\eqref{eq:bare_potential_chi}:
\begin{eqnarray}\label{eq:V_chi_phi}
    V(\chi,\phi)=-\frac{1}{2}m^2 \chi^2+ \frac{\mu(\phi)}{3} \chi^3+\frac{\lambda}{4} \chi^4 .
\end{eqnarray}
To trigger the FOPT, $\mu(\phi)$ must be some function of $\phi$ that switches sign as $\phi$ slow-rolls past some critical value of $\phi_i$. To simplify the later discussion, we focus on a specific form of $\mu(\phi)$ that behaves like a step function when $\phi$ passes $\phi_i$. One simple realization of such a $\mu(\phi)$ function is
\begin{eqnarray}
    \mu(\phi)=\mu\,\tanh\left(\frac{\phi-\phi_i}{\Delta\phi}\right)\,,
\end{eqnarray}
where the mass scale $\Delta\phi$ is chosen to be much smaller than change of $\phi$ in a Hubble time. We also require $\mu$ to be small such that the $\tfrac{1}{3}\mu(\phi)\chi^3$ term has a negligible contribution to the inflaton potential. We note that the isocurvature signals we consider in this work are insensitive to the specific choice of $\mu(\phi)$, provided that it behaves like a step function with a quick transition. 

For choices of $\mu(\phi)$ that have the necessary temperature-independent trigger behavior, this model has the novel feature that the phase transition begins during inflation when $\phi$ rolls past a critical value $\phi_i$. When this happens, $\chi_+$ is no longer the global minimum of the potential and a FOPT can occur as $\chi$ tunnels to the true minimum at $\chi_-$ (see Figure~\ref{fig:schematic_potential}). This tunneling occurs stochastically; as bubbles of true vacuum form at random times and locations, they expand to the horizon size and inflate along with it (as we will discuss in detail in the next section). Therefore, the largest bubbles after inflation are those that nucleated earliest, when $\phi \approx \phi_i$ and the comoving horizon size was $r_i$. 
Therefore, the distribution of bubbles has a characteristic feature around $r_i$. 
In our simple scenario, the PT rate remains constant and the phase transition can be fully specified by $\gamma_{\rm PT} \equiv \Gamma_{\rm PT}/H_{\rm inf}^4$ and $r_i$.

 We emphasize that all the interesting large-scale isocurvature effects considered in this paper are present as long as the three requirements stated at the beginning of this section are satisfied. To demonstrate the resulting constraints from the CMB data, we restrict ourselves to the simple toy model parameterized by $\gamma_{\rm PT}$ and $r_i$.

\section{Stochastic Bubble Distribution from Incomplete Phase Transition during inflation}\label{sec:bubbles}

In this section, we discuss the dynamics of bubbles nucleated during inflation and describe their stochastic distribution. From the distribution of bubbles, we will calculate density perturbations in the $\chi$ field, which will lead to DR
isocurvature with non-Gaussian statistics discussed in Section~\ref{sec:cosmological perturbation theory}.

For this calculation, we work in the homogeneous inflationary background, neglecting the effect of metric perturbations on the dynamics of the $\chi$ field during inflation.
This approximation is valid if $\bar{\rho}_\chi/(\bar\rho_\phi+\bar\rho_\chi)\ll 1$ and $\delta \rho_\chi/\bar{\rho}_\chi\gg \delta \rho_\phi/\bar{\rho}_\phi$. Here $\bar\rho$ denotes the averaged energy density of the corresponding field and $\delta\rho$ its perturbation. These conditions will be satisfied for all model parameters considered in this work that lead to constraints driven by DR isocurvature. Under this approximation, each bubble will expand with spherical symmetry. 

The bubble wall dynamics are calculated in detail in Appendix~\ref{app:bubble wall}: for the model parameters and approximations under which we are working, within a Hubble time the wall accelerates to a terminal velocity which is close to the speed of light.
Given the short acceleration time, we can take the bubble wall velocity to be $v_w\approx 1$ immediately after nucleation. Under this approximation, a bubble that nucleates at time $t'$ has a  comoving radius at time $t$: 
\begin{equation} \label{eq:comoving radius}
\begin{split}
    r(t, t') = & \frac{1}{H_{\rm inf}a(t')}\left(1 - \frac{a(t')}{a(t)}\right) + \frac{r_{c}}{a(t')}\\
    \approx & \frac{1}{H_{\rm inf}a(t')}.
\end{split}
\end{equation}
Here we drop the second term (as $a(t')/a(t)$ decays exponentially with time) and the third term (as $r_c \ll H_{\rm inf}^{-1}$ from Eq.~\eqref{eq:condition_Hinf}). After neglecting these terms, the radius only depends on the nucleation time $t'$ and so we will omit the $t$ argument: writing $r(t,t')\to r(t')$. This means that the bubble will quickly expand to the size of the comoving horizon at the time of nucleation. As the comoving horizon decreases during inflation, the bubble becomes super-horizon soon after nucleation and is fixed in size during the later evolution of the Universe.
%%%%%%%%%%%%%%%%%%%%%%%%%%%%%%%%%%%%%%%%%%%%%%%%%%%%%%%%%%%%%%
\begin{figure*}[th]
    \centering
    \includegraphics[width=.67\columnwidth]{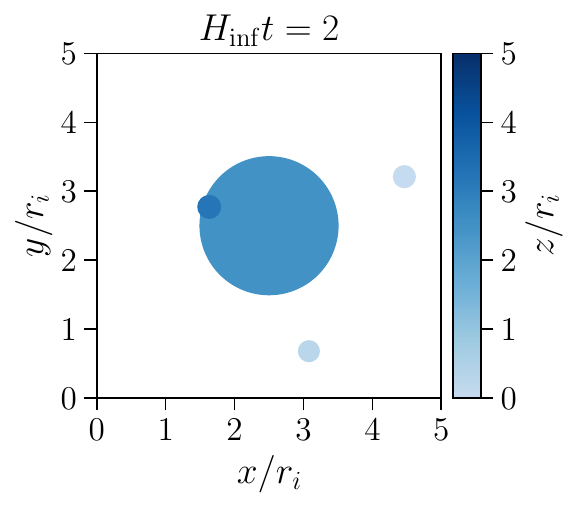}
    \includegraphics[width=.67\columnwidth]{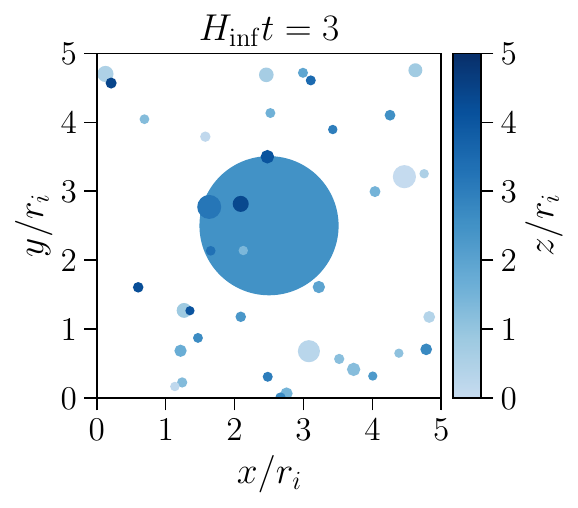}
    \includegraphics[width=.67\columnwidth]{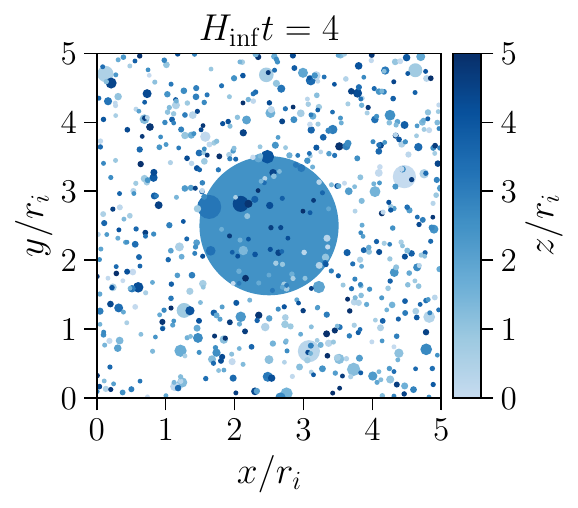}
    \caption{Simulated bubble distribution at different times within a cube with side-length $5r_i$ for $\gamma_{\rm PT} = 1\times 10^{-4}$. The simulation starts at $t=0$ and $r_i$ is the hypothetical radius of a bubble nucleated at $t=0$. The bubble distribution is projected to the $x$-$y$ plane with the colors representing the $z$ coordinate of the center of each bubble.}
    \label{fig:bubbles_sim}
\end{figure*}
%%%%%%%%%%%%%%%%%%%%%%%%%%%%%%%%%%%%%%%%%%%%%%%%%%%%%%%%%%%%%%
As the nucleation of bubbles is a stochastic process, the location and time of nucleation events are randomly distributed. 
Analytically, we can calculate the probability that a point in space remains in the false vacuum at a given time. Without loss of generality, we set $t=0$ to be the moment when the inflaton rolls past the trigger value of $\phi_i$ (see Section~\ref{sec:inflaton_trigger}) and bubbles can start nucleating. 
For a point $(\bs{x}, t)$ to be in false vacuum, a bubble cannot have nucleated at any time $t'<t$ in any location within a sphere of radius $r(t')$ centered at $\bs{x}$.
In terms of the Hubble-normalized phase transition rate $\gamma_{\rm PT} \equiv \Gamma_{\rm PT}/H_{\rm inf}^4$, the expected number of nucleation events within this region of space-time is 
\begin{equation}\label{eq:I(t)}
\begin{split}
    I(t) = & \frac{4\pi}{3}\Gamma_{\rm PT}\int\displaylimits_{0}^t dt' a(t')^3r(t')^3 \\ 
    =& \frac{4\pi}{3}\gamma_{\rm PT} H_{\rm inf} t \,.
\end{split}
\end{equation}
As a result, the probability of false vacuum at $(\bs{x}, t)$ is
\begin{equation}
        p_{\rm false}(t) = e^{-I(t)} = e^{-t/\tau_{\rm PT}} ,
\end{equation}
where
\begin{equation}
    \tau_{\rm PT}^{-1} \equiv \frac{4\pi}{3} \gamma_{\rm PT} H_{\rm inf} \,.
\end{equation}
Outside of the bubbles the $\chi$ field has an energy density of $\Delta V$, while inside the energy density is zero. Thus, the spatially averaged energy density of the $\chi$ field at time $t$ is
\begin{equation}
    \bar{\rho}_\chi(t) = \Delta V p_{\rm false}(t) = \Delta V e^{-t/\tau_{\rm PT}}.
\end{equation}
We are interested in the scenario where the phase transition remains incomplete when inflation ends at time $t_e$. This occurs if $t_e/\tau_{\rm PT}< 1$. In that case, the $\chi$ field will have a distribution of mostly non-overlapping bubbles with radii between $ r_e\equiv r(t_e)$ and $ r_i \equiv r(0)$. 
In this scenario, we can approximate the energy density of the $\chi$ field during inflation as
\begin{equation}\label{eq:rho chi}
    \rho_\chi(\bs{x}, t) = \Delta V \left[1-\sum_{I: t_I<t} \Theta(r_I-|\bs{x}-\bs{x_I}|)\right],
\end{equation}
where $r_I\equiv r(t_I)$ and $\bs{x}_I$ are the comoving radius and center location of bubble $I$, respectively, and the sum runs over all bubbles nucleated before $t$. 

To illustrate the time evolution of the distribution of bubbles, we perform a simulation of bubble nucleation for an incomplete FOPT during inflation. We simulate a cubic region of space with comoving volume  $(50\,r_i)^3$ and $\gamma_{\rm PT}=10^{-4}$. We start our simulation at $t=0$, nucleating bubbles at a rate given by
\begin{equation}\label{eq:rate of bubble nucleation}
    \frac{dN}{dt} = \mathcal{V}_{\rm false}(t) a(t)^3 \Gamma_{\rm PT},
\end{equation}
where $\mathcal{V}_{\rm false}(t)$ is the volume of space in the false vacuum at $t$. When a bubble is nucleated, the location of its center is sampled uniformly from the region of space that is in the false vacuum.
We show the resulting spatial distribution of bubbles at three different times in Figure~\ref{fig:bubbles_sim}. For illustrative purposes, we show only the $(5r_i)^3$ volume that contains the first (and largest) bubble nucleated in the original volume.
 
Next, we calculate the density contrast parameter at the end of inflation, which is defined as
\begin{eqnarray}\label{eq:delta chi}
    \delta_\chi(\bs{x}, t_e) & \equiv & \frac{\rho_\chi(\bs{x}, t_e)-\bar{\rho}_\chi(t_e)}{\bar{\rho}_\chi(t_e)}.
\end{eqnarray}
In Fourier space this is
\begin{equation}\label{eq:density parameter chi}
    \delta_\chi(\bs{k}, t_{e}) = -e^{t_e/\tau_{\rm PT}} \frac{4\pi}{k^3} \sum_{I: t_I<t_e} e^{-i\bk\cdot\bs{x_I}}\mathcal{A}(kr_I) + {\cal C}\times \delta^3(\bk),
\end{equation}
where $\mathcal{A}(y) \equiv \sin{y} - y \cos{y}$.
We ignore the term proportional to the delta function in Eq.~\eqref{eq:density parameter chi} in our analysis, as the $\bk = 0$ Fourier mode will not contribute to CMB observables with multipoles $\ell>0$. 

The two point function of $\delta_\chi$ is then given by
\begin{eqnarray} \label{eq:chi correlation sum}
     \langle \delta_{\chi}(\bk)\delta_{\chi}(\bkp)\rangle &=& e^{2t_e/\tau_{\rm PT}} \frac{(4\pi)^2}{{k'}^3k^3}\\
   & & \times \sum_{IJ} \langle e^{-i(\bk\cdot\bs{x_I} + \bkp \cdot \bs{x_J})}\mathcal{A}(kr_I)\mathcal{A}(k'r_J)\rangle. \nonumber
\end{eqnarray}
The double sum can be split into terms where $I=J$ and those where $I\neq J$. We will neglect the terms where $I\neq J$, as their sum is suppressed by a factor $\gamma_{\rm PT}$ relative to the sum over $I=J$ (see Appendix~\ref{app:two bubble terms}). We can replace the ensemble average of the $I=J$ terms with an integral over bubble coordinates:
\begin{widetext}
\begin{equation}\label{eq:2pt raw}
    \langle \delta_\chi(\bk)\delta_\chi(\bkp)\rangle = e^{2t_e/\tau_{PT}}\frac{(4\pi)^2}{k^3{k'}^3}
    N\int d^4x\, p_1(x) e^{-i(\bk+\bkp)\cdot \boldsymbol{x}}\mathcal{A}(kr(t))\mathcal{A}(k'r(t))\,.
\end{equation}
\end{widetext}
Here, $p_1(x)$ is the single-bubble probability distribution function over nucleation coordinates $x =(\bs{x}, t)$, and is given by
\begin{equation}\label{eq:single bubble pdf}
    p_1(x) = \frac{1}{\mathcal{V}}\left(\frac{1}{N}\frac{dN}{dt}\right),
\end{equation}
where $N$ is the total number of bubbles nucleated in a comoving volume $\mathcal{V}$ before $t_e$ and $dN/dt$ is given by Eq.~\eqref{eq:rate of bubble nucleation} with
\begin{equation}\label{eq:V_false theoretical}
    \mathcal{V}_{\rm false}(t) = \mathcal{V}p_{\rm false}(t) = \mathcal{V}e^{-t/\tau_{\rm PT}}.
\end{equation}
%%%%%%%%%%%%%%%%%%%%%%%%%%%%%%%%%%%%%%%%%%%%%%%%%%%%%%%%%%%%%%
\begin{figure}
    \centering
    \includegraphics[width=\columnwidth]{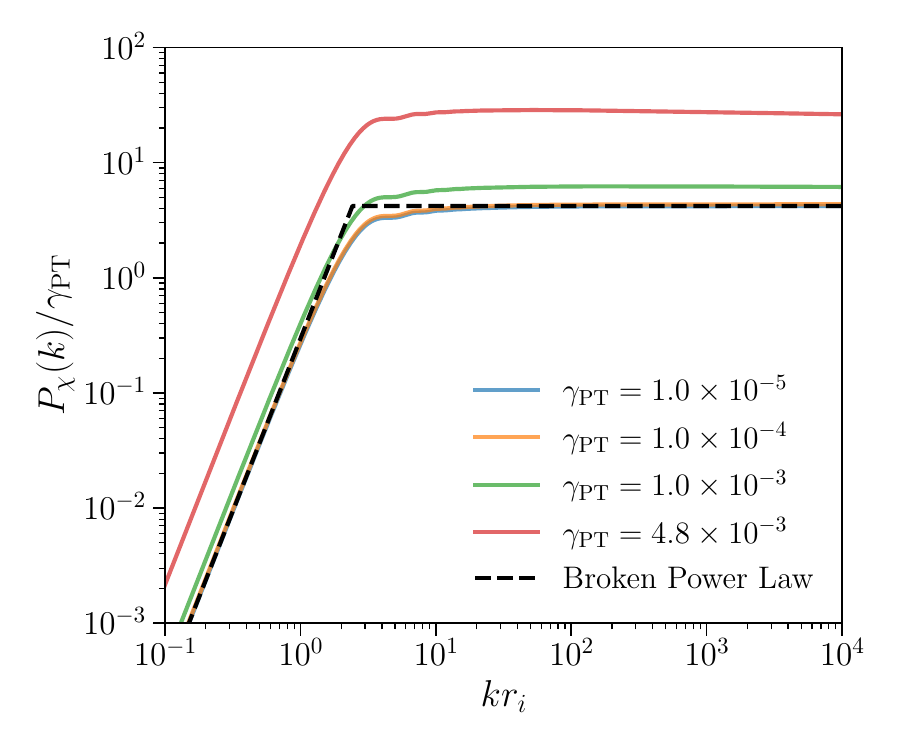}
    \caption{Power Spectra of the density contrast of $\chi$ normalized by $\gamma_{\rm PT}$ for different values of $\gamma_{\rm PT}$ and $H_{\rm inf} t_e = 50$. The largest value of $\gamma_{\rm PT}$ is chosen such that the incomplete PT condition $4\pi \gamma_{\rm PT}H_{\rm inf}t_e/3<1$ is barely satisfied. The dashed line is a broken power law with the same asymptotic behavior as the power spectrum in the limit of $\gamma_{\rm PT}H_{\rm inf}t_e\ll 1$ (see Eq.~\eqref{eq:chi power})\,. }\label{fig:power_spec}
\end{figure}
%%%%%%%%%%%%%%%%%%%%%%%%%%%%%%%%%%%%%%%%%%%%%%%%%%%%%%%%%%%%%%
After performing the $\bs{x}$ integral of Eq.~\eqref{eq:2pt raw}, the two point function can be written in terms of the dimensionless power spectrum $P_\chi(k)$ as 
\begin{equation} \label{eq:power spectrum conventions}
    \langle \delta_\chi(\bk) \delta_\chi(\bkp)\rangle \equiv 2\pi^2\frac{(2\pi)^3 \delta^3(\bk +\bkp)}{k^3}P_\chi(k).
\end{equation}
where 
\begin{eqnarray}\label{eq:chi power}
    P_\chi(k) & = & \, 8 \gamma_{\rm PT}e^{(8\pi \gamma_{\rm PT} H_{\rm inf}t_e)/3}(kr_i)^{-4\pi \gamma_{\rm PT}/3} \nonumber \\
   & & \times \int\displaylimits_{kr_e}^{kr_i} du\, u^{-4+4\pi\gamma_{\rm PT}/3} \mathcal{A}(u)^2\\
   & \approx & \, \gamma_{\rm PT}\left\{
   \begin{array}{cc}
        \frac{8}{27}(k r_i)^3 & \quad k r_i \ll 1 \\
        4.2 & \quad k r_i \gg 1 
    \end{array}\right.  \nonumber
\end{eqnarray}
to leading order of $\gamma_{\rm PT}$ in the limit $\gamma_{\rm PT}H_{\rm inf} t_e \ll 1$.
The result for small $kr_i$ is calculated analytically by Taylor-expanding the integrand and the result for large $kr_i$ is found by numerically calculating the limit as $kr_i\to \infty$.\footnote{We numerically integrate Eq.~\eqref{eq:chi power} for a series of values of $kr_i$ until the integral converges to a constant asymptotic value at 1\% precision.} As there are no bubbles larger than $r_i$, the $k^3$ dependence when $kr_i\ll 1$ is a consequence of the fact that the two-point function of $\delta_{\chi}$ is constant for modes larger than $r_i$. The flat power spectrum when $kr_i\gg 1$ can be understood as follows: when the nucleation rate is low, the majority of the Universe is in the false vacuum at all times $0<t<t_e$, and thus the rate of bubble nucleation in any physical volume of fixed size is the same at all times resulting in a scale invariant distribution of bubbles.
In Figure~\ref{fig:power_spec}, we show numerical results for the power spectrum as a function of $k$ for various values of $\gamma_{\rm PT}$ along with a broken power law with the same asymptotic behavior as Eq.~\eqref{eq:chi power}. The broken power law provides a good approximation to the power spectrum provided $\gamma_{\rm PT}H_{\rm inf}t_e\ll 1$. Therefore, assuming $H_{\rm inf}t_e \sim 50$, the approximation is valid for $\gamma_{\rm PT} \lesssim 10^{-3}$.

%%%%%%%%%%%%%%%%%%%%%%%%%%%%%%%%%%%%%%%%%%%%%%%%%%%%%%%%%%%%%%
%%%%%%%%%%%%%%%%%%%%%%%%%%%%%%%%%%%%%%%%%%%%%%%%%%%%%%%%%%%%%%
\section{Imprints of a non-thermal FOPT}\label{sec:cosmological perturbation theory}

After inflation and reheating, we assume the Universe is radiation dominated. As the Hubble parameter drops as $\propto T^2$, $\Gamma_{\rm PT}/H(T)^4$ will increase rapidly and reach unity at the nucleation temperature $T_*$, when the phase transition completes. We assume the energy density in the $\chi$ field will be converted to DR at $T_*$ by couplings between DR and the $\chi$ field. As a result, the DR will inherit the $\chi$ field's inhomogeneities, which, on large scales, follow the stochastic distribution of bubbles nucleated during inflation. This process will generate a relative entropy perturbation between DR and photons as well as a curvature perturbation which is distinct from the curvature perturbation produced by quantum fluctuations of the inflaton. The relative entropy perturbation will contribute to the DR isocurvature mode, while both curvature perturbations will contribute to the adiabatic mode. The isocurvature mode also has a non-negligible 3-point function and therefore has non-Gaussian statistics.

We begin this section by calculating the angular power spectrum of CMB temperature anisotropies in terms of the adiabatic and isocurvature power spectra. We then identify the super-horizon initial conditions for the DR isocurvature and adiabatic modes and calculate the power spectra. We show in Appendix~\ref{app:curvature FOPT} that -- for the class of models considered in this work -- the FOPT-induced curvature~\cite{Barir:2022kzo} has a negligible impact on cosmological observables compared to the DR isocurvature.
The DR isocurvature therefore sets the relevant constraints on these models of FOPTs. Finally, we calculate the isocurvature bispectrum from the non-Gaussian statistics.

In the synchronous gauge, the perturbed FRW metric is written as
\begin{equation}
    ds^2 = a(\tau)^2 \left[-d\tau^2 + (\delta_{ij} + h_{ij})dx^idx^j\right],
\end{equation}
where $\tau$ is the conformal time.\footnote{While it was natural to use physical time $t$ during inflation, the initial conditions for CLASS are most conveniently written in terms of $\tau$. Therefore, all functions that directly enter the CLASS calculation will be written in terms of $\tau$ rather than $t$.} The metric perturbation $h_{ij}$ can be written in Fourier space as
\begin{equation}
   h_{ij}(\bs{k}, \tau) = \left[\hat{k}_i\hat{k}_jh(\bk, \tau) + \left(\hat{k}_i\hat{k}_j - \frac{1}{3}\delta_{ij}\right)6\eta(\bk, \tau)\right],
\end{equation}
where $h$ and $\eta$ denote the trace and traceless longitudinal part of $h_{ij}$ respectively. 

In general, we can split a perturbation $X\in \{h,\eta,\delta_a,\theta_a,\sigma_a,...\}$ into a sum of isocurvature and adiabatic modes.  Here $\delta_a$, $\theta_a$, and $\sigma_a$ correspond to the $\ell=0,1$, and $2$ moments of the phase-space perturbation of species $a$. In particular, $\delta\equiv \delta \rho/\bar\rho$. For radiation species, moments $\ell>2$ are needed, and it is useful to define $F_{a, \ell}$ as the $\ell^{\rm \, th}$ moment of the phase-space perturbation of relativistic species $a$. 

For the FOPT we study in this paper, there are two linearly independent modes: the adiabatic mode that contains the curvature perturbation, and the isocurvature mode that contains the relative entropy perturbation. Therefore each perturbation variable can be written as a sum over these two modes using the decomposition 
\begin{eqnarray}\label{eq:split_modes}
    X(\bk,\tau)=c^{\rm ad}(\bk) X^{\rm ad}(k,\tau)+c^{\rm iso}(\bk) X^{\rm iso}(k,\tau).
\end{eqnarray}
Here $c^A(\bk)$ is a time-independent coefficient (where $A\in \{\rm ad, iso\}$) which is common for all perturbation variables, and $X^A(k, \tau)$ is perturbation-specific but independent of the direction $\bhat{k}{}$. The initial conditions for the modes of $X^A(k,\tau)$ are obtained by solving the Boltzmann and Einstein equations in the super-horizon limit $(k\tau\ll1)$.

The initial conditions for the adiabatic mode with DR are a straightforward extension of the adiabatic initial conditions for radiation in the standard cosmology~\cite{Ma:1995ey}: for example,  $\delta^{\rm ad}_{\rm dr} = \delta^{\rm ad}_\gamma=\delta^{\rm ad}_\nu$.
The initial conditions for the DR isocurvature mode can be derived in analogy to neutrino density isocurvature~\cite{Bucher:1999re}. As in the standard procedure, we choose a basis such that $\sum_{a}\delta\rho^{\rm iso}_a=0$, where the sum runs over the isocurvature energy density perturbations of all radiation species ($a=\gamma$, $\nu$, ${\rm dr}$). This ensures that the total energy density of radiation in the isocurvature mode is homogeneous at the initial time $\tau\to0$ in the radiation-dominated Universe and thus does not source curvature. Since there is no isocurvature in neutrinos in our models, their density perturbations are equal to that of photons $(\delta_{\gamma}^{\rm iso}= \delta_{\nu}^{\rm iso}$). We also choose the normalization such that $\delta_{\rm dr}^{\rm iso}=1$ when $k\tau\to 0$.  These requirements fix the initial conditions for the DR isocurvature mode:
\begin{eqnarray}\label{eq:iso_relation}
  \delta_{\rm dr}^{\rm iso}=1~,~  \delta_{\gamma}^{\rm iso}= \delta_{\nu}^{\rm iso}=\frac{-R_{\rm dr}}{1-R_{\rm dr}}~~~(k\tau \to 0),
\end{eqnarray}
where $R_{\rm dr}\equiv \bar\rho_{\rm dr}/(\bar\rho_{\gamma}+\bar\rho_{\nu}+\bar\rho_{\rm dr})$.
 The complete initial conditions for the DR isocurvature mode  in the synchronous gauge are given in Appendix~\ref{app:DR_iso_ic}. Notably, in the limit $R_{\rm dr}\ll 1$, all moments of the photon and neutrino phase-space perturbations in the isocurvature mode ($F^{\rm iso}_{\gamma,\ell}$ and $F_{\nu,\ell}^{\rm iso}$) are proportional to $R_{\rm dr}$, which can be seen based on their initial conditions and time evolution controlled by the coupled Boltzmann hierarchy and Einstein equations.

The coefficients $c^{A}(\bk) = \{c^{\rm ad}(\bk),c^{\rm iso}(\bk) \}$ encode the initial random configuration of each mode. For the FOPT models of interest, the modes are approximately uncorrelated so the statistics of Gaussian observables are determined by
\begin{equation}\label{eq:C power spectrum}
        \langle c^A(\bk) c^A(\bkp)\rangle = 2\pi^2\frac{(2\pi)^3 \delta^3(\bk +\bkp)}{k^3}P_A(k),
\end{equation}
where $P_A(k) = \{P_{\rm ad}(k),P_{\rm iso}(k) \}$ are the power spectra of the adiabatic and isocurvature modes. 

One relevant observable in the CMB is the photon temperature anisotropy $\Delta \equiv \Delta T/\bar{T}$, which can be written in a Fourier-Legendre series as 
\begin{equation}\label{eq:photon temp anisotropy}
    \Delta(\bs{x}, \bs{\hat{n}}, \tau) = \sum_{\ell=0}^\infty (-i)^\ell (2\ell + 1)\int \frac{d^3k}{(2\pi)^3} e^{i\bk\cdot\bs{x}} \Delta_\ell(\bk, \tau) P_\ell(\bs{\hat{k}} \cdot \bs{\hat{n}}),
\end{equation}
where $\bs{\hat n}$ is the direction of the momentum and $\Delta_\ell$ is proportional to the multipole moment of the photon phase-space density perturbation $\Delta_\ell= F_{\gamma, \ell}/4$. Its angular two-point function is
\begin{equation}\label{eq:two point generic}
    \langle\Delta ( \bs{x}, \bs{\hat{n}}, 
    \tau_0)\Delta (\bs{x}, \bs{\hat{n}'}, 
    \tau_0)\rangle = \frac{1}{4\pi}\sum_\ell (2\ell+1)C^{TT}_\ell P_\ell(\bs{\hat{n}} \cdot\bs{\hat{n}'}).
\end{equation}
Here $C^{TT}_\ell$ is the temperature anisotropy angular power spectrum:
\begin{equation}\label{eq:C_l moments}
\begin{split}
    C^{TT}_\ell = 4\pi\int d(\ln{k}) \left(P_{\rm ad}(k) |\Delta_\ell^{\rm ad}(k, \tau_0)|^2 + \right.\\
    \left. P_{\rm iso}(k) |\Delta_\ell^{\rm iso}(k, \tau_0)|^2\right).
\end{split}
\end{equation}
In this expression, $\Delta_\ell^A(k,\tau)$ is the photon transfer function, which is obtained by solving the Boltzmann and Einstein equations for $F_{\gamma,\ell}^A(k,\tau)$ with the initial conditions for mode $A=\{\rm ad, iso\}$.
The dependence on the particle physics models is therefore encapsulated in $P_{\rm ad}(k)$ and $P_{\rm iso}(k)$.  

We focus on a class of models for which the contribution of the FOPT to $P_{\rm ad}(k)$ is small compared to the adiabatic modes arising from inflation (see Appendix~\ref{app:curvature FOPT}). For this class of models, the dominant effects on CMB observables is from $P_{\rm iso}(k)$ so we can set $P_{\rm ad} (k)$ to be the standard $\Lambda$CDM adiabatic power spectrum
\begin{eqnarray}\label{eq:P_ab}
    P_{\rm ad}(k)=A_s \left(\frac{k}{k_{\rm pivot}}\right)^{n_s-1},
\end{eqnarray}
where $A_s$ and $n_s$ are the scalar amplitude and spectral index. $k_{\rm pivot}$ is the pivot scale, conventionally defined as $k_{\rm pivot} \equiv 0.05\,{\rm Mpc}^{-1}$ \cite{Planck:2018vyg}. The photon transfer function in the isocurvature mode scales as $\Delta^{\rm iso}_\ell=\frac{1}{4}F^{\rm iso}_{\gamma, \ell}\propto R_{\rm dr}$ for $R_{\rm dr}\ll 1$ (see the discussion below Eq.~(\ref{eq:iso_relation})). Therefore, the effect of isocurvature on the angular power spectrum is determined by $R_{\rm dr}^2 P_{\rm iso}(k)$. This behavior will be manifest in our results.

\subsection{Isocurvature \label{sec:isocurvature}}

To obtain $P_{\rm iso}(k)$ in Eq.~\eqref{eq:C_l moments}, we consider the gauge-invariant relative entropy perturbation between the DR and the photons:
\begin{eqnarray}
    \mathcal S_{{\rm{dr}},\gamma}\equiv -3{\cal H}\left(\frac{\delta\rho_{\rm dr}}{\rho_{\rm dr}'}-\frac{\delta\rho_{\gamma}}{\rho_{\gamma}'}\right).
\end{eqnarray}
Here $'$ denotes $\partial/\partial \tau$, and ${\cal H}=a H$ is the conformal Hubble parameter. With the definition in Eq.~(\ref{eq:split_modes}) and the isocurvature mode relations in Eq.~(\ref{eq:iso_relation}), as $k\tau\to0$ this reduces to
\begin{eqnarray}\label{eq:S_general}
    \mathcal S_{{\rm{dr}},\gamma}= \frac{3}{4}\frac{c^{\rm iso}(\bk)}{1-R_{\rm dr}}.
\end{eqnarray}
Therefore, the isocurvature power spectrum $P_{\rm iso}$ can be directly related to $\langle \mathcal S_{{\rm{dr}},\gamma} \, \mathcal S_{{\rm{dr}},\gamma} \rangle$.

To calculate $\mathcal S_{{\rm{dr}},\gamma}$,  we first determine the relations between $\delta_{\rm dr}$ and $\delta_{\chi}$. As the DR we are interested in was generated by rapid bubble percolation around a temperature $T_*$ (given by $H(T_*)^4 = \Gamma_{\rm PT}$), we can assume an instantaneous transfer of energy from the $\chi$ field into DR when the temperature of the Universe reaches $T_*$. 
We therefore work in the gauge where the temperature of the Universe is uniform just before the phase transition completes. This can be achieved by the following gauge transformation:
\begin{equation}
    \delta \tau(\bs{x},\tau)=-\frac{\delta \rho_{\rm SM}}{\rho_{\rm SM}'},
\end{equation}
where SM means a total over particle species in the Standard Model. In this gauge, the standard model density perturbation (defined as $\delta\hat{\rho}_{\rm SM}\equiv\delta \rho_{\rm SM} +\rho_{\rm SM}'\, \delta \tau$) vanishes and the density perturbation $\chi$ remains unchanged ($\delta \hat{\rho}_\chi=\delta \rho_\chi$).

Since the temperature is uniform in the new gauge, the phase transition completes at the same time $\tau_*$ everywhere and the large-scale perturbations (super-horizon at $T_*$) in $\hat{\rho}_\chi$ will be directly converted to those in $\hat{\rho}_{\rm dr}$. The sub-horizon perturbations of DR around $T_*$  are more complicated, and depend on the dynamics of the FOPT. However, as we are only interested in modes that are accessible to CMB, we can safely neglect contributions from sub-horizon dynamics at $T_*$, assuming $T_*\gg T_{\rm CMB}$. In this paper, we can therefore approximate
\begin{eqnarray}\label{eq:chi_DR_relation}
    \hat\delta_{\rm dr}(\bk,\tau_*)\approx \delta_\chi(\bk,t_*)\approx\delta_\chi(\bk,t_e),
\end{eqnarray}
where $t=t_*$ when the phase transition completes (at proper time $\tau_*$). The last approximation in Eq.~\eqref{eq:chi_DR_relation} holds because bubbles nucleated at times $t\in [t_e, t_*]$ have radii that are subhorizon at $T_*$. 

The Standard Model density perturbations prior to the end of the FOPT are generated from inflaton fluctuations and thus are all adiabatic, satisfying
\begin{equation}
    \frac{\delta \rho_{a}}{\rho_{a}'} = \frac{\delta \rho_{\rm SM}}{\rho_{\rm SM}'}
\end{equation}
for any standard model particle species $a$. Therefore density perturbations of each individual Standard Model species vanish in the new gauge as well ($\delta \hat{\rho}_a = 0$) and the relative entropy perturbation is simply
\begin{eqnarray}\label{eq:S_hat_gauge}
     \mathcal S_{{\mathrm{dr}},\gamma}= \frac{3}{4}\hat\delta_{\rm dr}.
\end{eqnarray}
As $\mathcal S_{{\mathrm{dr}},\gamma}$ is gauge invariant, we can directly relate Eq.~(\ref{eq:S_general}) and Eq.~(\ref{eq:S_hat_gauge}). Together with Eq.~(\ref{eq:chi_DR_relation}), we find
\begin{eqnarray}\label{eq:c delta relation}
    c^{\rm iso}(\bk)\approx (1-R_{\rm dr})\delta_{\chi}(\bk,\tau_e).
\end{eqnarray}
Therefore, $P_{\rm iso}$ is proportional to $P_\chi$:
\begin{eqnarray}
        P_{\rm iso}(k)\approx (1-R_{\rm dr})^2 P_\chi(k).
\end{eqnarray}
As $P_\chi$ exhibits the broken power law behavior in Eq.~\eqref{eq:chi power}, for the remainder of our work, we take $P_{\rm iso}$ as
\begin{eqnarray}\label{eq:P_iso_class}
    P_{\rm iso}(k)= f_{\rm iso}^2 A_s \left\{
    \begin{array}{cc}
        (k/k_i)^3 & \quad k \leq k_i \\
        1 & \quad k > k_i 
    \end{array}\right.,
\end{eqnarray}
where $A_s$ is the amplitude of the scalar curvature power spectrum and $f_{\rm iso}$ is a coefficient that sets the amplitude of the isocurvature power spectrum. In our model, $f_{\rm iso}^2 A_s \approx 4.2\gamma_{\rm PT}$. %for $R_{\rm dr}\ll 1$.
We note that we use $k_i$ as the scale of the transition of the broken power law, which is related to $r_i^{-1}$ by an ${\cal O}(1)$ factor.

\subsection{Non-Gaussianity}\label{sec:general_NG}

The perturbations generated by the FOPT have non-Gaussian statistics that may be observable in the CMB. The CMB power spectra discussed before are based on two-point functions, which only contain Gaussian information. To quantify non-Gaussianity in the CMB, we need to examine higher point functions of the photon temperature anisotropy $\Delta$. %, defined in Eq.~\eqref{eq:photon temp anisotropy}. 
One example is the 
three-point function, which 
comes predominantly from the isocurvature mode for the models considered, given by %given by
\begin{widetext}
\begin{equation}\label{eq:photon 3 point}
\begin{split}
    \langle \Delta(\bs{\hat{n}_1})\Delta(\bs{\hat{n}_2})\Delta(\bs{\hat{n}_3})\rangle = & \left(\frac{20}{3}\right)^3\sum_{\ell_1, \ell_2, \ell_3} (2 \ell_1 +1)(2 \ell_2 +1) (2\ell_3 + 1)(-i)^{\ell_1 + \ell_2+\ell_3}\\
    \times & \int\frac{d^3k_1}{(2\pi)^3}\int\frac{d^3k_2}{(2\pi)^3}\int\frac{d^3k_3}{(2\pi)^3}(2\pi)^3 \delta^3(\bs{k_1} + \bs{k_2} + \bs{k_3}) P_{\ell_1}(\bhat{k}{1}\cdot \bhat{n}{1})P_{\ell_2}(\bhat{k}{2}\cdot \bhat{n}{2})P_{\ell_3}(\bhat{k}{3}\cdot \bhat{n}{3})\\
    \times & B_{\rm iso}(k_1, k_2, k_3)\Delta_{\ell_1}^{\rm iso}(k_1, \tau_0)\Delta_{\ell_2}^{\rm iso}(k_2, \tau_0)\Delta_{\ell_3}^{\rm iso}(k_3, \tau_0) .
\end{split}
\end{equation}
\end{widetext}
We have introduced the isocurvature bispectrum $B_{\rm iso}$, which we define using the convention \cite{Planck:2019kim}: 
\begin{equation}
\begin{split}
    &\langle c^{\rm iso}(\bs{k_1})c^{\rm iso}(\bs{k_2}) c^{\rm iso}(\bs{k_3})\rangle \\
    &\equiv (2\pi)^3 \delta^3(\bs{k_1} + \bs{k_2} + \bs{k_3}) \left(\frac{20}{3}\right)^3 B_{\rm iso}(k_1, k_2, k_3).
\end{split}
\end{equation}

To calculate the bispectrum for our model, we use Eq.~\eqref{eq:c delta relation} to relate $c^{\rm iso}(\bk)$ to $\delta_\chi(\bk, t_e)$ and perform steps similar to those for the calculation of $P_\chi$, leading to
\begin{equation}\label{eq:bispectrum}
\begin{split}
    B_{\rm iso}(k_1, k_2, k_3) = & - \left(\frac{3\pi}{5}\right)^3(1-R_{\rm dr})^3 e^{4\pi\gamma_{\rm PT}H_{\rm inf}t_e}\frac{\gamma_{\rm PT}}{k_1^3 k_2^3} \\
    \times & \int\displaylimits_{k_3 r_e}^{k_3r_i}du u^{-4}\mathcal{A}(\alpha_1 u)\mathcal{A}(\alpha_2 u)\mathcal{A}(u) ,
\end{split}
\end{equation}
where $\alpha_1 \equiv k_1/k_3$, $\alpha_2 \equiv k_2/k_3$, and $\mathcal A(u)$ is defined below Eq.~\eqref{eq:density parameter chi}. 
In Section~\ref{sec:non_Gaussianity constraints} we will estimate constraints by comparing this result to the equilateral and local bispectra which have been studied in the literature \cite{Planck:2019kim}. To simplify the comparison, we will write down expressions for $B_{\rm iso}$ evaluated at wavenumbers in the squeezed $k_1\ll k_2\approx k_3$) and equilateral ($k_1\approx k_2\approx k_3$) configurations. For $R_{\rm dr}\ll 1$ and $\gamma_{\rm PT} H_{\rm inf} t_e \ll 1$ these expressions are
\begin{equation} \label{eq:bispectrum sq eq}
\begin{split}
        B_{\rm iso}(k_1, k, k) =& -\left(\frac{3\pi}{5}\right)^3\frac{\gamma_{\rm PT}}{k^6}  \\
        \times &\left\{
        \begin{array}{cc}
            \frac{1}{3}\int\displaylimits_0^{kr_i}du u^{-1} \mathcal{A}(u)^2 \quad & k_1\ll k \\
            \int\displaylimits_0^{kr_i} du u^{-4} \mathcal{A}(u)^3 \quad & k_1=k 
        \end{array}\right. \, .
\end{split}
\end{equation}
Restricting to the case where $kr_i\gg 1$, the asymptotic behavior of the integrals in these expressions is
\begin{equation}
\begin{split}
    \frac{1}{3}\int\displaylimits_0^{kr_i}du u^{-1} \mathcal{A}(u)^2 \propto (kr_i)^2 \\
    \int\displaylimits_0^{kr_i} du u^{-4} \mathcal{A}(u)^3\propto {\rm constant}, %(kr_i)^0,
\end{split}
\end{equation}
so 
\begin{equation}\label{eq:FOPT_bispectrum_scaling}
    B_{\rm iso}(k_1, k, k) \propto 
    \left\{
        \begin{array}{cr}
            k^{-4} & \quad k_1\ll k \\
            k^{-6} & \quad k_1=k
        \end{array}\right..
\end{equation}

%%%%%%%%%%%%%%%%%%%%%%%%%%%%%%%%%%%%%%%%%%%%%%
%%%%%%%%%%%%%%%%%%%%%%%%%%%%%%%%%%%%%%%%%%%%%%
\section{Effects on CMB observations}\label{sec:observables} 

In the previous section, we have calculated the DR isocurvature power spectrum from non-thermal FOPTs.  In this section, we implement the model in CLASS, and simulate the CMB angular power spectra. As mentioned before, we can approximate the adiabatic power spectrum as unchanged compared to the standard $\Lambda$CDM model. The presence of DR (parameterized by $\Delta N_{\rm eff}$) shifts CMB observables as a result of the change in the background energy density -- such shifts create the standard limit on $\Delta N_{\rm eff} \lesssim 0.3$.
In addition, DR isocurvature creates additional shifts in CMB observables which are the main focus of this work.

As explained in the introduction to Section~\ref{sec:cosmological perturbation theory}, the effect of DR isocurvature on the CMB depends on $R_{\rm dr}^2 P_{\rm iso}(k)$.
To simplify comparisons with limits on DR with adiabatic perturbations, we write $R_{\rm dr}$ in terms of the equivalent number of effective neutrino species, $\Delta N_{\rm eff}\equiv 3.044 \times (\bar \rho_{\rm dr}/\bar\rho_\nu)$. Around  photon decoupling, $R_{\rm dr}\approx 0.13\times\Delta N_{\rm eff}$.
The isocurvature power spectrum (defined in Eq.~(\ref{eq:P_iso_class})) depends on two additional parameters: $f_{\rm iso}$ determines the amplitude, and $k_i$ is the wavenumber at which the broken power law transitions. This transition wavenumber also corresponds to the comoving scale at the start of the FOPT. Given the form of $P_{\rm iso}$, the effect of DR isocurvature on the CMB is proportional to $\Delta N_{\rm eff}^2 f_{\rm iso}^2$ for fixed $k_i$.  We express $k_i$ in terms of multiples of $k_{\rm pivot} \equiv 0.05\,{\rm Mpc}^{-1}$, the pivot scale of the adiabatic power spectrum given in Eq.~\eqref{eq:P_ab}.

\begin{figure}[ht]
    \centering
    \includegraphics[width=\columnwidth]{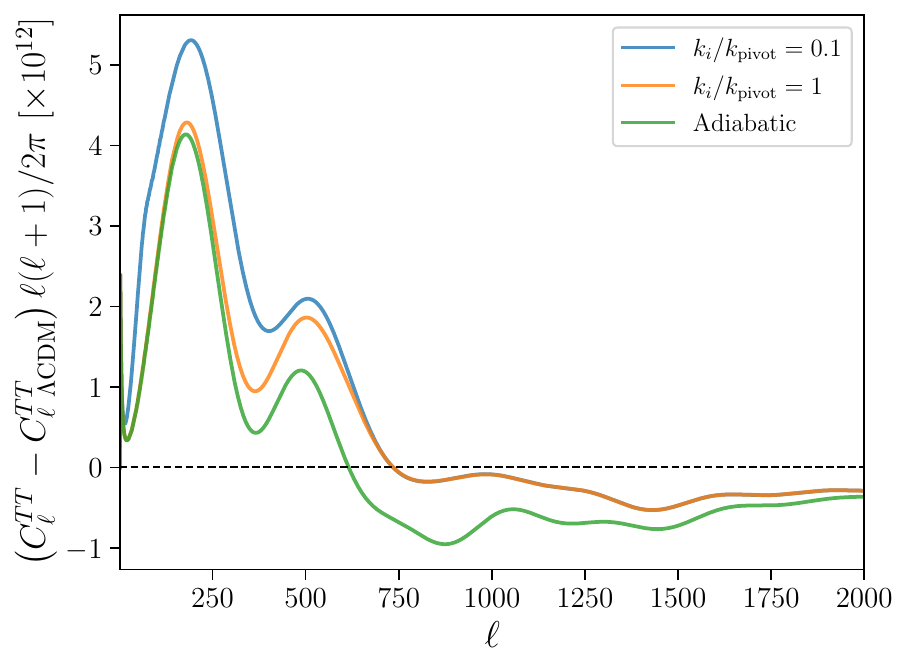}
    \caption{The difference of temperature anisotropy angular power spectra between cases with dark radiation ($\Delta N_{\rm eff}=0.1$) and the $\Lambda$CDM model. The blue and orange lines denote DR isocurvature with different $k_i$ and fixed $f_{\rm iso}=10$. The green line shows the result with adiabatic initial conditions. The pivot scale is chosen to be $k_{\rm pivot}=0.05 \, \textrm{Mpc}^{-1}$.}
    \label{fig:CMB_ISO_LCDM}
\end{figure}

\begin{figure*}[ht]
    \centering
    \includegraphics[width=\columnwidth]{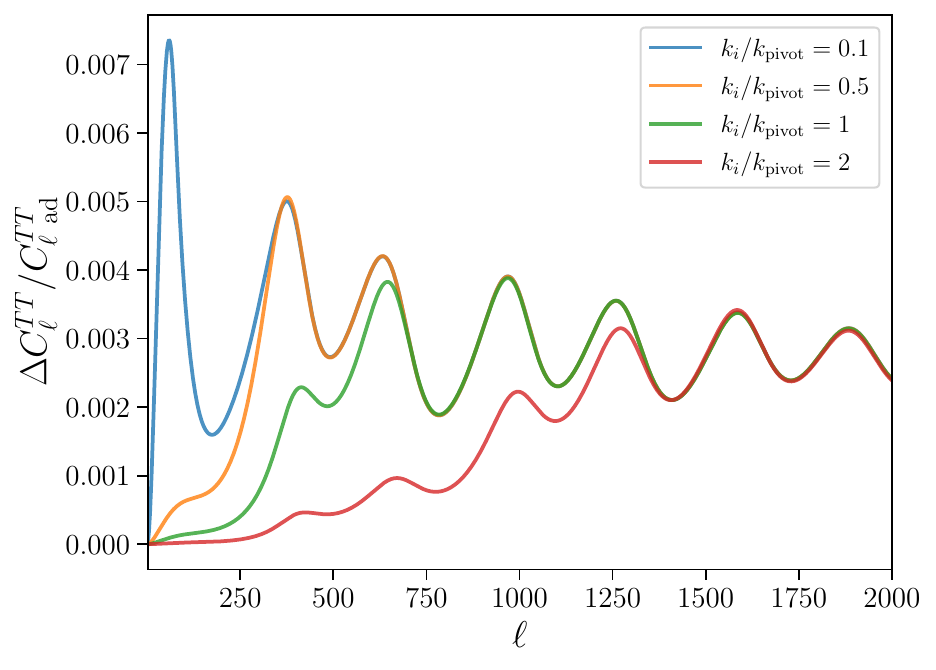}
    \includegraphics[width=\columnwidth]{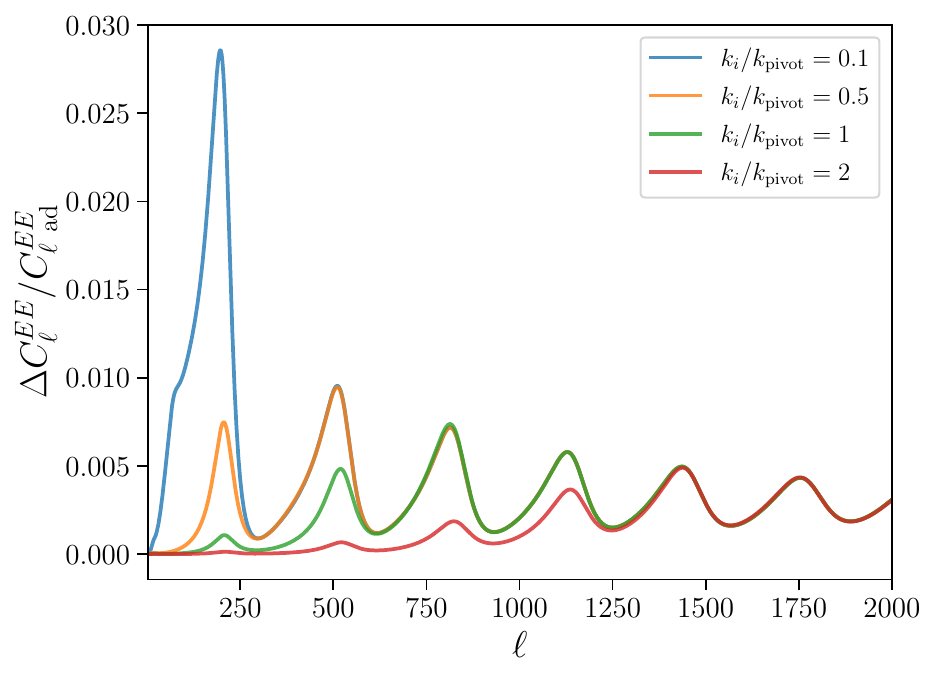}
    \caption{The fractional difference of CMB power spectra ($C^{TT}_\ell$ in the left panel, $C^{EE}_\ell$ in the right panel) between cases with dark radiation isocurvature and adiabatic initial conditions for different $k_i$. We set $f_{\rm iso}=10$ for DR isocurvature, while $\Delta N_{\rm eff}=0.1$ for both adiabatic and isocurvature cases. The pivot scale is chosen to be $k_{\rm pivot}=0.05\,\textrm{Mpc}^{-1}$.  }
    \label{fig:CMB_ISO}
\end{figure*}

To show the effects of DR isocurvature on the CMB observables, we simulate the CMB power spectra with difference choices of $k_i$, and compare them to the case with purely adiabatic initial conditions. In Figure~\ref{fig:CMB_ISO_LCDM}, we show the difference of the temperature angular power spectrum $C_\ell^{TT}$ between $\Lambda$CDM and three models with DR, all with $\Delta N_{\rm eff} = 0.1$. In one case, we consider only the adiabatic effects of DR (setting $f_{\rm iso} = 0$). In the other two we include isocurvature with $k_i = k_{\rm pivot}$ and $0.1\, k_{\rm pivot}$ (with $f_{\rm iso} = 10$ in both cases).  
The acoustic peaks in models with isocurvature shift towards higher $\ell$ as compared to the adiabatic case with the same $\Delta N_{\rm eff}$. This result is consistent with  the literature (see for example~\cite{Savelainen:2013iwa}). 
In Figure~\ref{fig:CMB_ISO}, we show the fractional difference of angular power spectra between the DR isocurvature and adiabatic cases (again holding $\Delta N_{\rm eff}=0.1$) for the $\rm TT$ (left) and $\rm EE$ (right) modes. For the DR isocurvature simulations again we fix $f_{\rm iso} = 10$ and vary $k_i$. The oscillations seen in both panels are a consequence of the shift in the acoustic peaks due to isocurvature, as discussed.

For all DR isocurvature models shown in Figures~\ref{fig:CMB_ISO_LCDM} and \ref{fig:CMB_ISO}, differences between the CMB power spectra disappear at large $\ell$. 
This occurs because CMB spherical harmonics with $\ell\gtrsim \ell_i \sim k_i d_{\rm SLS}$ mostly receive contributions from wavenumbers $k\gtrsim k_i$, where the isocurvature power spectrum is flat and independent of $k_i$. Here $d_{\rm SLS} \sim 10^4\,\rm{Mpc}$ is the comoving distance to the surface of last scattering.\footnote{Thus, for $k_i=k_{\rm pivot}$, $\ell_i\sim 500$.}
For $k_i\lesssim 2\times 10^{-3} \,k_{\rm pivot}$, the corresponding transition spherical harmonic is $\ell_i\lesssim 1$ and the entire CMB angular power spectrum can only probe the plateau of the DR isocurvature power spectrum. For such $k_i$, the effect of DR isocurvature saturates and the CMB becomes insensitive to further reductions in $k_i$.

%%%%%%%%%%%%%%%%%%%%%%%%%%%%%%%%%%%%%%%%%%%%%%%%%%%%%%%%%%%%%%
%%%%%%%%%%%%%%%%%%%%%%%%%%%%%%%%%%%%%%%%%%%%%%%%%%%%%%%%%%%%%%

%%%%%%%%%%%%%%%%%%%%%%%%%%%%%%%%%%%%%%%%%%%%%%
%%%%%%%%%%%%%%%%%%%%%%%%%%%%%%%%%%%%%%%%%%%%%%
\section{Methodology and Data Sets}\label{sec:MCMC}
To evaluate the cosmological constraints on the FOPT model parameters, we perform a full likelihood analysis using cosmological datasets. To compute the observables within these datasets, we employ a modified version of the \texttt{CLASS} code \cite{Ghosh:2021axu,Blas:2011rf} along with \texttt{MontePython} \cite{Audren:2012wb,Brinckmann:2018cvx} to perform the MCMC analysis.

We use the datasets from the following experiments to set limits on the FOPT model:
\begin{itemize}
    \item Planck 2018 high-$\ell$ TTTEEE and low-$\ell$ TT, EE anisotropy angular power spectra and the gravitational lensing data \cite{Planck:2018vyg}.
    
    \item The BAO measurements from the Six-degree Field Galaxy Survey (6dFGS) \cite{Beutler:2011hx}, the Sloan Digital Sky Survey (SDSS) DR7 MGS \cite{Ross:2014qpa}, and the LOWZ galaxy samples of BOSS DR12 \cite{BOSS:2016wmc}.
\end{itemize}

Our baseline cosmology consists of the usual six cosmological parameters of $\Lambda$CDM. To this, we add the three isocurvature parameters discussed previously: $\Delta N_{\rm eff}$, $k_{i}$, $f_{\rm iso}$. We assume three active neutrinos species, one with mass $0.06 \ev$ and two massless. As explained in Section~\ref{sec:observables}, the effects of DR isocurvature are sensitive to the product $\Delta N_{\rm eff} f_{\rm iso}$ while the effects of the background cosmology are sensitive to $\Delta N_{\rm eff}$. It is also convenient to express the transition wavenumber $k_i$ in terms of the CMB pivot wavenumber $k_{\rm pivot}$. Therefore we use $\Delta N_{\rm eff}$, $k_{i}/k_{\mathrm{pivot}}$ and $\Delta N_{\rm eff} f_{\rm iso}$ as the primary parameters for our MCMC runs with a log-flat prior for each parameter. The boundaries of these priors are: $\log_{10}(\Delta{}N_\mathrm{eff}) \in [-5, -0.3]$, $\log_{10}(k_{i}/k_{\mathrm{pivot}})\in [-4, 3]$ and $\log_{10}(\Delta{}N_\mathrm{eff}f_\mathrm{iso})\in [-2.3, 2.3]$. 
The $\Lambda$CDM parameters have standard wide flat priors. The FOPT isocurvature power spectrum of our simplified model (calculated in Section~\ref{sec:isocurvature}) is a good approximation provided $\gamma_{\rm PT} \lesssim 10^{-3}$. Therefore, the particular FOPT model studied in this work maps onto the priors used in the likelihood analysis for $f_{\rm iso} \lesssim 10^3$. We note that our cosmological model and priors are sufficiently general that other isocurvature generation mechanisms may lead to cosmologies that map onto subsets of our parameter space. With this in mind, we show the resulting posteriors for $f_{\rm iso}$ outside of the strict domain of validity for this class of FOPTs.

We use the Metropolis-Hastings algorithm with the Gelman-Rubin convergence criterion $R-1<0.02$ \cite{Gelman:1992zz} for the MCMC chains. To analyze MCMC chains and calculate Bayesian posteriors, we use the \texttt{GetDist} package \cite{Lewis:2019xzd}.
The best-fit values are obtained using simulated annealing for the minimization of $\chi^2$ values \cite{Schoneberg:2021qvd}. The procedure of simulated annealing systematically lowers the temperature during an MCMC chain run, effectively navigating and optimizing against noisy likelihood functions with numerous local maxima.

%%%%%%%%%%%%%%%%%%%%%%%%%%%%%%%%%%%%%%%%%%%%%%
%%%%%%%%%%%%%%%%%%%%%%%%%%%%%%%%%%%%%%%%%%%%%%

\section{Results}\label{sec:results}
\begin{figure}
\includegraphics[width=\columnwidth]{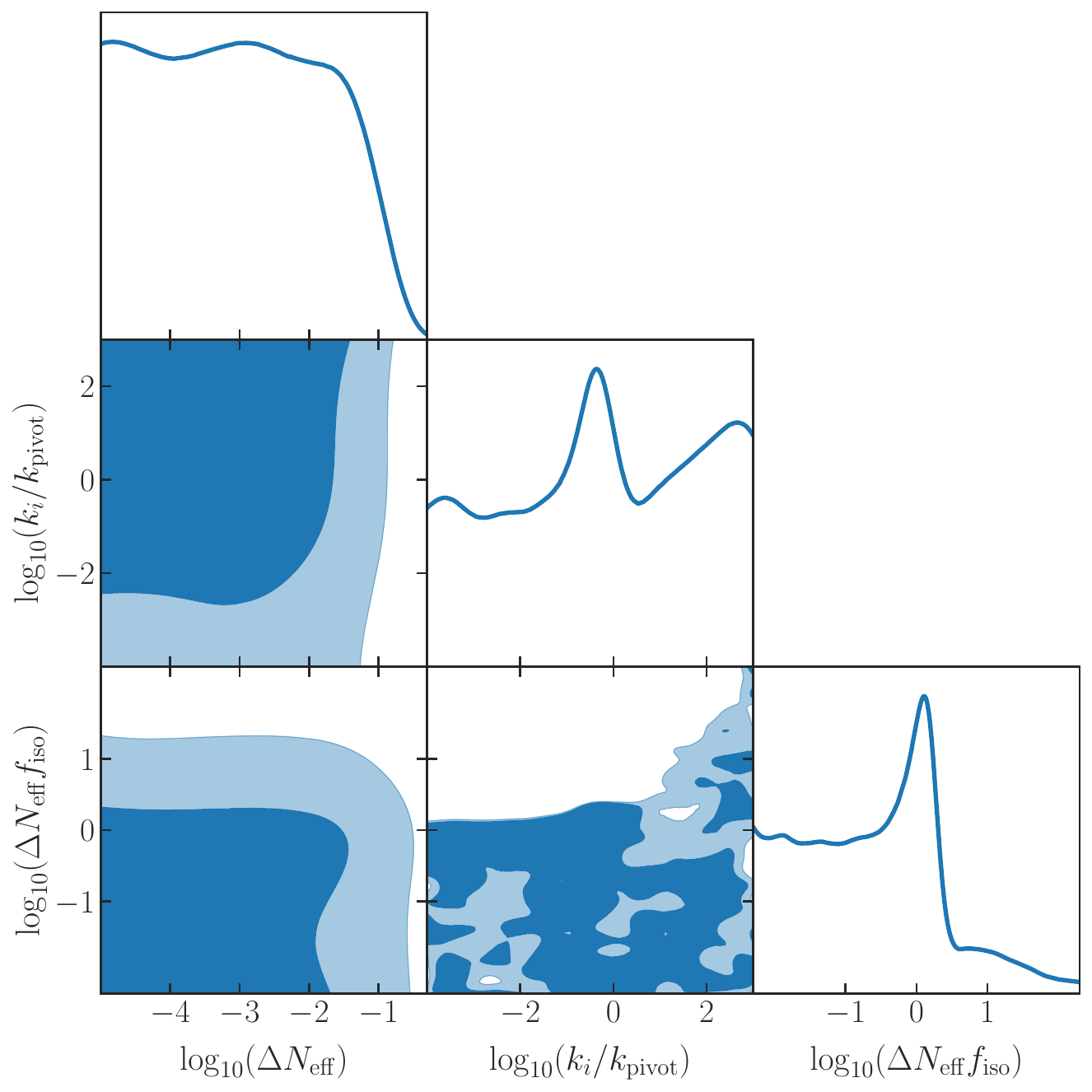}% 
\caption{\label{fig:Posteriors} The two dimensional marginalized posterior distributions for the FOPT new physics parameters $\Delta N_{\rm eff}$, $\Delta N_{\rm eff}f_{\rm iso}$ and $k_i/k_{\rm pivot}$, when analyzing Planck+BAO (see text for details). We omit the posteriors of the six $\Lambda$CDM parameters for clarity.}
\end{figure}

Having discussed our theoretical and statistical methodology as well as the datasets that we use in our analysis, we are ready to calculate posteriors for the parameters of DR isocurvature model.
The mean and best-fit values of each parameter in our fit are shown in Table~\ref{tab:summary}.  
The posterior distributions ($65\%$ and $95\%$ CL) for these new physics parameters are shown in Figure~\ref{fig:Posteriors}. 
The limits on $\Delta N_{\rm eff}f_{\rm iso}$ become approximately independent of $\Delta N_{\rm eff}$ for $\Delta N_{\rm eff}\ll 0.1$ and are also independent of $k_i$ for $k_i\ll k_{\rm pivot}$.

%%%%%%%%%%%%%%%%%%%%%%%%%%%%%%%%%%%%%%%%%%%%%%%%%%%%%%%%%%%%%%
%%%%%%%%%%%%%%%%%%%%%%%%%%%%%%%%%%%%%%%%%%%%%%%%%%%%%%%%%%%%%%

\begin{figure}
\includegraphics[width=\columnwidth]{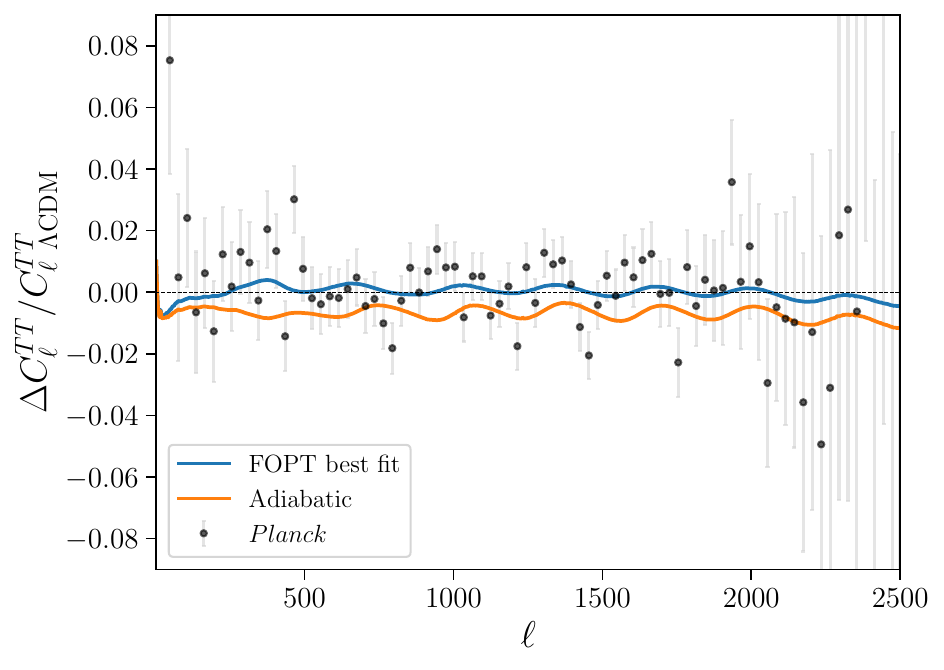}% 
\caption{The change (relative to $\Lambda$CDM) in $C_{\ell}^{TT}$  power spectrum normalized to $C_{\ell,\Lambda{\rm CDM}}^{TT}$ for the FOPT best-fit model (see Table~\ref{tab:summary}), an adiabatic model and the Planck 2018 data \cite{Planck:2018vyg}. For the adiabatic case we use the same parameters as in the FOPT best-fit except we turn off isocurvature ($f_{\rm iso}=0$).
\label{fig:residuals}}
\end{figure}

As seen in Table~\ref{tab:summary}, there is a slight preference (compared to $\Lambda$CDM) in the data for non-zero isocurvature, with best-fit parameters $k_i/k_{\mathrm{pivot}}=0.36$, $\Delta N_{\rm eff}=7.5\times 10^{-3}$, and $\Delta N_{\rm eff}f_{\rm iso}=1.58$. We show in Figure~\ref{fig:residuals} the fractional difference in $C_{\ell}^{TT}$ between this best-fit model and the best-fit $\Lambda$CDM model compared to the Planck 2018 residuals (also normalized to the $\Lambda$CDM result). We also show the fractional difference compared to $\Lambda$CDM of an adiabatic-only model with the same parameters as the DR isocurvature best-fit other than setting $f_{\rm iso}=0$.

To better understand the features of our posteriors, it is useful to hold one of the three new physics parameters fixed and consider posteriors over the other two. We do this in the next two subsections, holding $f_{\rm iso}$ fixed in Section~\ref{sec:fixed f_iso} and $k_i$ fixed in Section~\ref{sec:fixed ki}. Finally, we estimate limits from non-Gaussianities in Section~\ref{sec:non_Gaussianity constraints}, though a dedicated search is needed to derive a robust bound.

%%%%%%%%%%%%%%%%%%%%%%%%%%%%%%%
\begin{table}
\begin{tabular}{|l|c|} 
 \hline 
Parameters & FOPT \\ \hline \hline
$\Omega_{\mathrm{b} } h^{2}$  & $0.0225$ ($0.0223 \pm  0.0002$)  \\
$\Omega_{\mathrm{c} } h^{2}$  & $0.1199$ ($0.01954 \pm  0.0011$)  \\
$100*\theta{}_{s }$ & $1.0421$ ($1.0420 \pm  0.0003$)  \\ 
$\ln (10^{10}A_{s })$  & $3.0459$ ($3.0474 \pm 0.0146$)  \\
$n_{s}$  & $0.9678$ ($0.9679 \pm  0.0398$) \\
$\tau_{\mathrm{reio}}$  &$0.0580$ ($0.0564 \pm  0.0073$)  \\
$\log_{10}(\Delta{}N_\mathrm{eff})$ & $-2.1244$ ($ < -1.0741$)  \\
$\log_{10}(k_{i}/k_{\mathrm{pivot}})$ & $-0.4496$ (unconstrained)  \\
$\log_{10}(\Delta{}N_\mathrm{eff}f_\mathrm{iso})$ & $0.1974$ ($<1.1453$)  \\ \hline
$\Delta \chi^2_{\mathrm{tot}}$ & $-0.68$  \\
\hline 
 \end{tabular} \\
 \caption{The best-fit parameters of the FOPT isocurvature model resulting from fits to the Planck+BAO datasets. The mean and $1\sigma$ variation for each parameter when fit to the data are shown in parenthesis or the 95\% CL upper bound.}
 \label{tab:summary}
\end{table}
%%%%%%%%%%%%%%%%%%%%%%%%%%%%%%%

%%%%%%%%%%%%%%%%%%%%%%%%%%%%%%%%%%%%%%%%%%%%%%%%%%%%%%%%%%%%%%
%%%%%%%%%%%%%%%%%%%%%%%%%%%%%%%%%%%%%%%%%%%%%%%%%%%%%%%%%%%%%%

%%%%%%%%%%%%%%%%%%%%%%%%%%%%%%%
%%%%%%%%%%%%%%%%%%%%%%%%%%%%%%%

\subsection{$\Delta N_{\rm eff}$ constraints with fixed $f_{\rm iso}$}\label{sec:fixed f_iso}

%%%%%%%%%%%%%%%%%%%%%%%%%%%%%%%%%%%%%%%%%%%%%%%%%%%%%%%%%%%%%%
\begin{figure}
\includegraphics[width=\columnwidth]{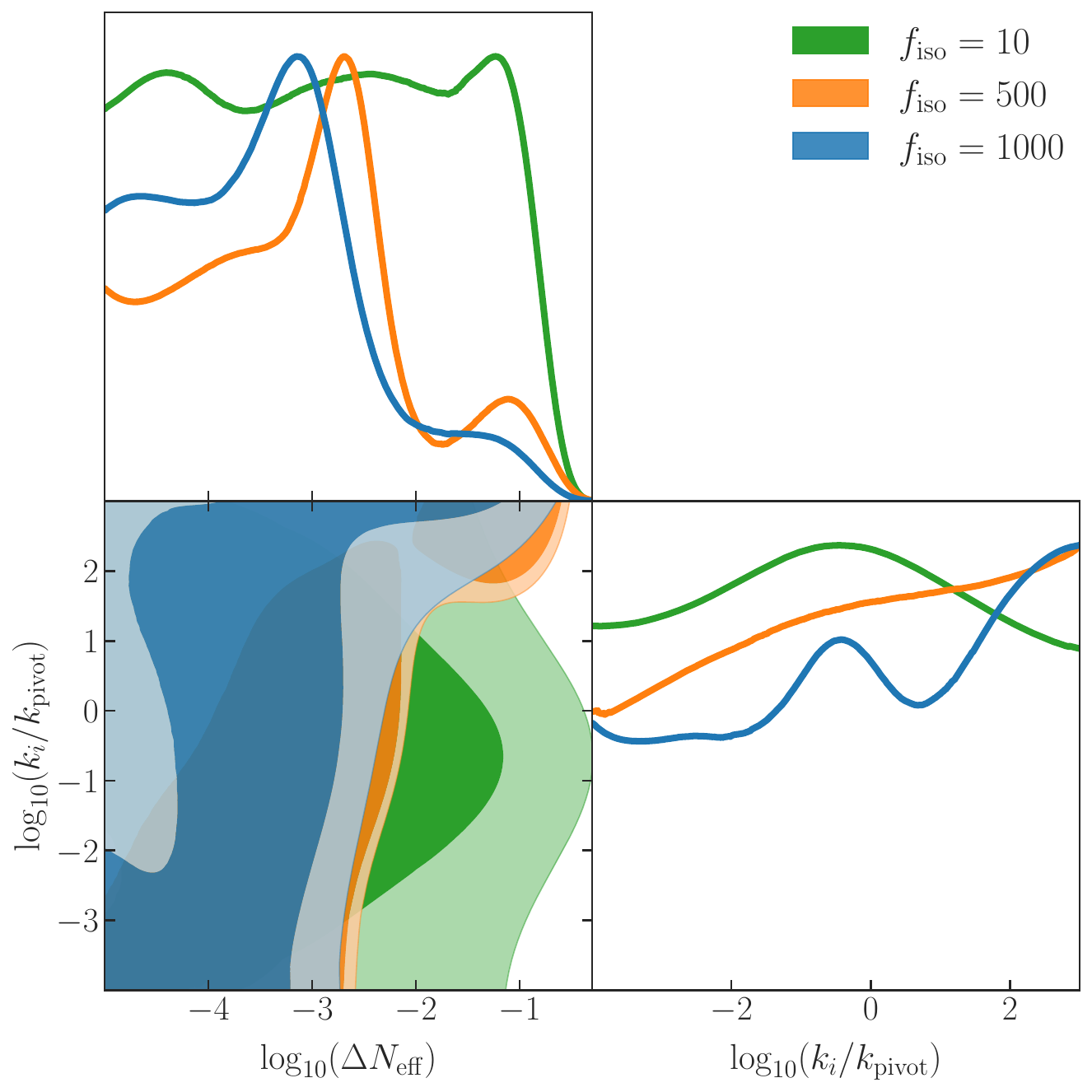}% 
\caption{The two dimensional marginalized posterior distributions for the FOPT new physics parameters $\Delta N_{\rm eff}$ and $k_i/k_{\rm pivot}$, holding $f_{\rm iso}$ fixed.}
\label{fig:N_eff_f_iso}
\end{figure}
%%%%%%%%%%%%%%%%%%%%%%%%%%%%%%%%%%%%%%%%%%%%%%%%%%%%%%%%%%%%%%

In this subsection, we fix $f_{\rm iso}$ and perform the scan over $\Delta N_{\rm eff}$, $k_i$ and the other six $\rm \Lambda$CDM parameters. This corresponds to a scenario where the phase transition rate $\gamma_{\rm PT}\sim f_{\rm iso}^2\,A_s$ is held fixed while the horizon size at the beginning of the transition $r_i\sim k_i^{-1}$ varies. In Figure~\ref{fig:N_eff_f_iso}, we show posteriors for this scan for three values of $f_{\rm iso}$: 10, 500 and 1000.

For the largest two values of $f_{\rm iso}$ we find that the 2$\sigma$ contours approach the adiabatic constraint $\Delta N_{\rm eff}<0.3$ for large $k_i$. This is expected as the DR isocurvature power spectrum modifies the CMB power spectrum primarily at $\ell$ above the range currently probed by Planck when $k_i\gtrsim 10^2\, k_{\rm pivot}$, making the CMB power spectrum observationally indistinguishable from the adiabatic-only result.  
As $k_i$ decreases, the limit on $\Delta N_{\rm eff}$ strengthens rapidly. For sufficiently small $k_i$, CMB data is mainly sensitive to the plateau of the isocurvature power spectrum (see Eq.~(\ref{eq:P_iso_class})) and the constraint approaches an asymptotic value which depends on $\Delta N_{\rm eff}f_{\rm iso}$.

For $f_{\rm iso} = 10$ we find a preference for the range $k_i\sim (0.1 - 1)\times k_{\rm pivot}$. That is, models with this combination of $k_i$ and $f_{\rm iso}$ fit the data better than models with large $k_i$ (which are themselves indistinguishable from the adiabatic model).

\subsection{$\Delta N_{\rm eff}$ constraints with fixed $k_i$}\label{sec:fixed ki}

%%%%%%%%%%%%%%%%%%%%%%%%%%%%%%%%%%%%%%%%%%%%%%%%%%%%%%%%%%%%%%
\begin{figure}
\includegraphics[width=\columnwidth]{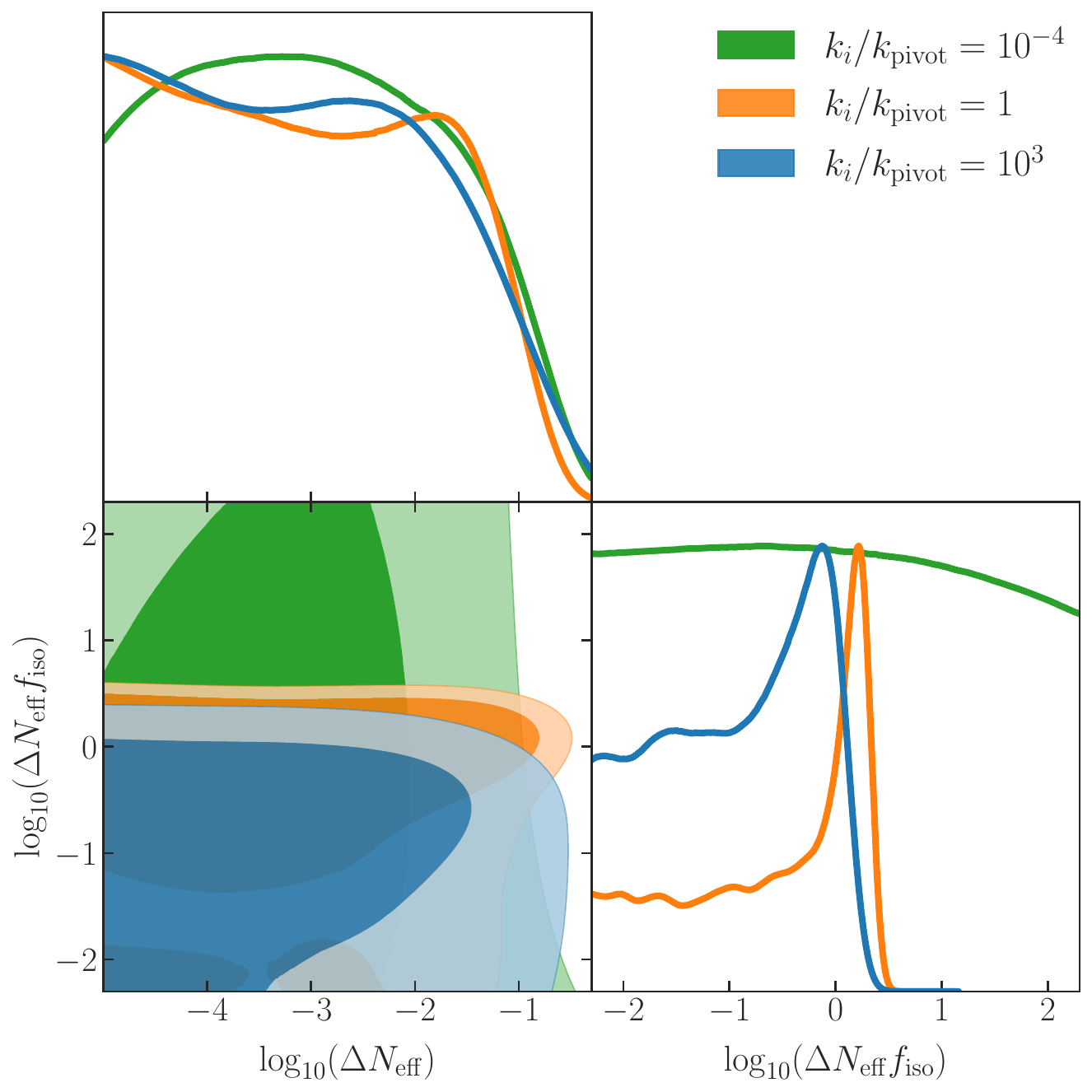}% 
\caption{The two dimensional marginalized posterior distributions for the FOPT new physics parameters $\Delta N_{\rm eff}$ and $\Delta N_{\rm eff}f_{\rm iso}$, holding $k_i$ fixed. \label{fig:N_eff_k_i}}
\end{figure}
%%%%%%%%%%%%%%%%%%%%%%%%%%%%%%%%%%%%%%%%%%%%%%%%%%%%%%%%%%%%%%

Next, we fix $k_i$ and calculate the posteriors in $\Delta N_{\rm eff}$ and $\Delta N_{\rm eff} f_{\rm iso}$. This corresponds to the scenario where the horizon size at the beginning of the PT is fixed, and the PT rate varies. The results of our MCMC fit to the data are shown in Figure~\ref{fig:N_eff_k_i} for three choices of $(k_i/k_{\mathrm{pivot}})$:  $10^{3}, 1$, and $10^{-4}$.

Again, we see that when $k_i$ is large ($k_i=10^3\, k_{\rm pivot}$) the constraints on $\Delta N_{\rm eff}$ approximate the adiabatic result for all values of $f_{\rm iso}$ within our priors. For the other two values of $k_i$, isocurvature becomes important and the model can be constrained even for very small values of $\Delta N_{\rm eff}$. For $\Delta N_{\rm eff} \ll 0.1$, the adiabatic and background effects are negligible and the CMB data mainly has sensitivity to the combination $\Delta N_{\rm eff} f_{\rm iso}$ which parameterizes the strength of isocurvature effects. In this regime, we find the $95\%$ CL constraints  $\Delta N_{\rm eff} f_{\rm iso}< 1.25$ for $k_i = 10^{-4}\,k_{\rm pivot}$ and $\Delta N_{\rm eff} f_{\rm iso}< 2.09$ for $k_i = k_{\rm pivot}$. We have confirmed through explicit calculation using linear-flat priors on $\Delta N_{\rm eff}$ that these results are robust against our choice of priors.

The limits on $\Delta N_{\rm eff} f_{\rm iso}$ are stronger for the smallest value of $k_i$ as the CMB only probes the plateau (the maximum) of the isocurvature power spectrum. For the larger value, the CMB is sensitive to the portion of the power spectrum which is proportional to $k^3$ and suppressed relative to the plateau. 
As $\Delta N_{\rm eff}$ becomes $\mathcal{O}(0.1)$, the adiabatic effects become important. At this point, the limits deviate from $\Delta N_{\rm eff}f_{\rm iso}=\mathrm{const.}$ and we (approximately) recover the adiabatic result: $\Delta N_{\rm eff} < 0.3$. 

The constraint on $\Delta N_{\rm eff}f_{\rm iso}$ for $\Delta N_{\rm eff}\ll 0.1$ can be rewritten as a constraint on $\Delta N_{\rm eff}$ in terms of other quantities directly related to FOPTs. As shown below Eq.~\eqref{eq:P_iso_class}, $A_sf_{\rm iso}^2\approx 4.2\gamma_{\rm PT}$ and $\gamma_{\rm PT}=\Gamma_{\rm PT}/H_{\rm inf}^4$. Using the definitions of nucleation temperature $T_*$ ($ \Gamma_{\rm PT}= H^4(T_*)$) and reheating temperature $T_{\rm rh}$  ($H_{\rm inf}=H(T_{\rm rh})$) for instantaneous reheating, we can rewrite $\gamma_{\rm PT}=(T_*/T_{\rm rh})^8$ in the radiation-dominated Universe. Therefore, the constraint $\Delta N_{\rm eff}f_{\rm iso}<\beta(k_i)$ is equivalent to 
\begin{eqnarray}
    \Delta N_{\rm eff}< 2.8\times 10^{-5} \left(\frac{T_{\rm rh}}{T_*}\right)^{4} \left(\frac{\beta(k_i)}{1.25}\right).
\end{eqnarray}
for $T_*/T_{\rm rh} \gtrsim 0.1$. $\beta(k_i)$ is mildly sensitive to $k_i$ for $k_i \lesssim k_{\rm pivot}$ as demonstrated by our limits on $\Delta N_{\rm eff} f_{\rm iso}$ for small $\Delta N_{\rm eff}$ discussed earlier this subsection.

For the two smallest values of $k_i$ shown in Figure~\ref{fig:N_eff_k_i}, there is a preference for models with isocurvature ($\Delta N_{\rm eff} f_{\rm iso} \sim 0.1-1$) compared to the adiabatic-only model which is well approximated by the smallest values of $\Delta N_{\rm eff} f_{\rm iso}$ allowed by our prior. This effect is more prominent for $k_i=k_{\rm pivot}$ which is close to the best-fit value of $k_i$. 

\subsection{$\Delta N_{\rm eff}$ constraints from non-Gaussianity} \label{sec:non_Gaussianity constraints}
In this section, we will estimate constraints on this class of FOPTs from non-Gaussianity in the CMB. Currently, there is no dedicated search for non-Gaussianity with DR isocurvature produced in a FOPT. The most relevant study is on non-Gaussianity with neutrino density isocurvature \cite{Planck:2019kim}. 

Based on Eq.~\eqref{eq:photon 3 point}, the effect of non-Gaussianity in our study will be comparable in size to that of neutrino density isocurvature when
\begin{equation}\label{eq:simple mapping}
\begin{split}
    & B_{\rm iso}(k_1, k_2, k_3)\Delta_{\ell_1}^{\rm iso}(k_1, \tau_0)\Delta_{\ell_2}^{\rm iso}(k_2, \tau_0)\Delta_{\ell_3}^{\rm iso}(k_3, \tau_0) \sim \\
     & B_{\nu, \rm iso}(k_1, k_2, k_3)\Delta_{\ell_1}^{\nu, \rm iso}(k_1, \tau_0)\Delta_{\ell_2}^{\nu, \rm iso}(k_2, \tau_0)\Delta_{\ell_3}^{\nu, \rm iso}(k_3, \tau_0).
\end{split}
\end{equation}
where $B_{\nu, \rm iso}$ and $\Delta_{\ell}^{\nu, \rm iso}$ are the bispectrum and transfer function for the neutrino density isocurvature model. 
This transfer function can be simply related to the transfer function for DR isocurvature by accounting for the different energy fraction of DR and neutrinos. As discussed in Section~\ref{sec:cosmological perturbation theory}, the transfer function for DR isocurvature is proportional to $R_{\rm dr}$. Therefore, the transfer function for neutrino isocurvature can be well-approximated by 
\begin{equation}\label{eq:transfer function relation}
    \Delta^{\nu, \rm iso}_\ell(k, \tau) \sim \frac{R_\nu}{R_{\rm dr}}\Delta^{\rm iso}_\ell(k, \tau),
\end{equation}
where $R_{i} \equiv \bar{\rho}_{i}/(\bar{\rho}_\gamma+\bar{\rho}_\nu+\bar{\rho}_{\rm dr})$.
Subsituting this result into Eq.~\eqref{eq:simple mapping}, we find that the effect of non-Gaussianity will be similar in the two models when
\begin{equation}\label{eq:simple mapping bispec}
    B_{\rm iso}(k_1, k_2, k_3) \sim \left(\frac{R_\nu}{R_{\rm dr}}\right)^3 B_{\nu, \rm iso}(k_1, k_2, k_3).
\end{equation}

The available constraints on neutrino isocurvature non-Gaussianity \cite{Planck:2019kim} assume a local bispectrum which peaks in the squeeze configuration:
\begin{equation}\label{eq:local bispectrum}
    B_{\rm local}(k_1, k, k) \propto 
    \frac{f_{\rm NL}^{\rm local}}{k_1^{4-n_s} k^{4-n_s}} \quad k_1\ll k
\end{equation}
where $f_{\rm NL}^{\rm local}$ is a dimensionless parameter that quantifies the amount of non-Gaussianity present. The FOPT bispectrum, on the other hand, is independent of $k_1$ in the squeeze configuration (as shown in Eq.~\eqref{eq:FOPT_bispectrum_scaling}) so the bound on $f^{\rm local}_{\rm NL}$ does not apply to our case. 

The bispectrum from the FOPT is similar to the equilateral template for which the dominant effects come from the equilateral configuration \cite{Planck:2019kim}
\begin{equation}\label{eq:equilateral_bispectrum}
    B_{\rm equil}(k, k, k) = 6\left(\frac{2\pi^2}{25}\right)^2A_s^2 f_{\rm NL}^{\rm equil}\frac{1}{k^6}\left(\frac{k_{\rm pivot}}{k}\right)^{2(1-n_s)}.
\end{equation}
However, measurements of $f_{\rm NL}^{\rm equil}$ for isocurvature non-Gaussianity do not exist in the literature. Based on measurements of equilateral and local non-Gaussianity for curvature \cite{Planck:2019kim}, we assume that a search for equilateral non-Gaussianity for isocurvature would find a value of $f_{\rm NL}^{\rm equil}$ within an order of magnitude of $f_{\rm NL}^{\rm local}$. Under this assumption and given the 2$\sigma$ upper bound on $|f_{\rm NL}^{\rm local}|$ for neutrino density isocurvature~\cite{Planck:2019kim}, we can estimate an approximate bound $|f_{\rm NL}^{\rm equil}|\lesssim 10^3$ for the same cosmological model. This result can be translated into a bound on FOPT non-Gaussianity by evaluating Eq.~\eqref{eq:simple mapping bispec} in the equilateral configuration ($k_1\approx k_2 \approx k_3$) near the pivot scale $k_{\rm pivot}$ and taking $B_{\rm \nu, iso}=B_{\rm equil}$.
After numerical integration of the second line of Eq.~\eqref{eq:bispectrum sq eq}, the FOPT bispectrum can be written as
\begin{equation}\label{eq:FOPT_bispectrum}
    B_{\rm iso}(k, k, k) = -0.97  A_s f_{\rm iso}^2\frac{1}{k^6}.
\end{equation}
Combining Eqs.~\eqref{eq:simple mapping bispec}, \eqref{eq:equilateral_bispectrum} and \eqref{eq:FOPT_bispectrum} leads to the approximate constraint
\begin{equation}
    \Delta N_{\rm eff}^3 f_{\rm iso}^2 \lesssim 2\times 10^{-4}.
\end{equation}
In Section~\ref{sec:fixed ki} we found $\Delta N_{\rm eff} f_{\rm iso}\lesssim 1$ in the regime $k_i\lesssim k_{\rm pivot}$ and $\Delta N_{\rm eff}\ll 0.3$. Therefore the dominant constraint on FOPT may come from non-Gaussianity for $\Delta N_{\rm eff}\gtrsim 2\times 10^{-4}$, though we again emphasize that a dedicated study is required to set a robust bound.

%%%%%%%%%%%%%%%%%%%%%%%%%%%%%%%%%%%%%%%%%%%%%%%%%%%%%%%%%%%%%%
%%%%%%%%%%%%%%%%%%%%%%%%%%%%%%%%%%%%%%%%%%%%%%%%%%%%%%%%%%%%%%

\section{Conclusions}\label{sec:conclusions}
We have showed that for a broad class of non-thermal FOPTs, the scalar field undergoes a phase transition that nucleates bubbles during inflation. If the PT rate is small during inflation, the bubbles do not collide and their size is determined by the horizon size at the time of nucleation. The FOPT will eventually complete after reheating, at which point we assume most of the remaining energy density in the scalar field converts into dark radiation. The DR inherits the inhomogeneity of the scalar field, leading to a DR isocurvature mode. 

If the PT starts sufficiently early during inflation, isocurvature in the DR can create measurable perturbations in the CMB. In this work, we performed a detailed calculation of the DR isocurvature power spectrum and implemented it in CLASS code. Using Planck 2018 and BAO data, we calculated the constraint on the energy density of DR (in terms of the effective relativistic degrees of freedom $\Delta N_{\rm eff}$). We demonstrated that the constraint on $\Delta N_{\rm eff}$ can be much stronger than the limit derived assuming adiabatic initial conditions. In particular, when the temperature $T_*$ at which the phase transition completes, is close to the reheating temperature $T_{\rm rh}$, the isocurvature perturbations set a limit of $\Delta N_{\rm eff} \lesssim 10^{-5}(T_*/T_{\rm rh})^{-4}$. This limit weakens as $T_*$ decreases and approaches the adiabatic constraint of $\Delta N_{\rm eff}<0.3$ when $T_*$ becomes much less than $T_{\rm rh}$.

Since the distribution of the perturbation caused by bubbles is intrinsically non-Gaussian, it could leave sizable non-Gaussianity signals in the CMB. We have also estimated the strength of the non-Gaussianity constraint on FOPTs from the CMB by comparing to neutrino density isocurvature. However, we showed that the bispectrum calculated for this class of FOPTs is different than those studied previously. Therefore, a dedicated search is needed to derive a robust bound on FOPTs from non-Gaussianity in the CMB.

Moreover, our study can also put a lower bound on the scale of inflation if we can determine $\Delta N_{\rm eff}$ and $T_*$. These quantities can be derived if we observe the direct gravitational wave spectrum from the FOPT. We will show in a follow-up paper that the recent nano-Hz gravitational wave data could already set a lower limit on the scale of the reheating temperature. In the future, it will be interesting to explore the correlation of gravitational wave signals and cosmological data to further probe FOPTs.

\section{Acknowledgements}

We thank Subhajit Ghosh, Soubhik Kumar and Chen Sun for useful discussions.
This work was supported by DOE grant DOE-SC0010008. Key parts of this work were motivated by discussions at the Aspen Center for Physics, which is supported by National Science Foundation grant PHY-2210452.

%%%%%%%%
\appendix

\section{Bubble Wall Dynamics}\label{app:bubble wall}

As the $\chi$ field tunnels to the true vacuum in a region of spacetime, a wall forms as $\chi$ smoothly varies from $\chi_-$ inside the region to $\chi_+$ outside. 
We follow the approach of Ref.~\cite{Megevand:2023nin} to calculate the field profile and dynamics inside the wall, working in the thin wall approximation which assumes that the wall thickness is smaller than any other length scale in the problem.

For this calculation, it is convenient to work in Gaussian-normal coordinates adapted to a hypersurface $\Sigma$ along the wall. On the hypersurface, the field takes on a constant value $\chi(x^\mu) = \chi_\Sigma \in (\chi_-, \chi_+)$. Points on $\Sigma$ are parameterized as $x^\mu = X^\mu(\xi^a)$ for three coordinates  $\xi^a = (\xi^0, \xi^1, \xi^2)$. The fourth coordinate $n$ measures the proper distance along a geodesic that originates from $X^\mu(\xi^a)$ with a tangent vector that is orthogonal to $\Sigma$. In this coordinate system, an arbitrary point near $\Sigma$ has Gaussian-normal coordinates which can be related to the original coordinates by
\begin{equation}
x^\mu = X^\mu(\xi^a) + n N^\mu(\xi^a) + \mathcal{O}(n^2)
\end{equation}
where $N^\mu(\xi^a)$ is a unit normal vector satisfying
\begin{equation}
\begin{split}
    N_\mu N^\mu =& 1 \\
    N_\mu \partial_a X^\mu =& 0 .\\
\end{split}
\end{equation}
On the hypersurface $\Sigma$, $n=0$. 

The scalar field $\chi$ obeys the standard equation of motion:
\begin{equation}
    g^{\mu \nu} \left[\partial_\mu\partial_\nu \chi -\Gamma^\rho_{\mu\nu} \partial_\rho \chi\right] = V'(\chi).
\end{equation}
In the thin-wall approximation, $\partial_n\chi \gg \partial_a \chi$, allowing us to drop derivatives other than those with respect to $n$. Thus, in Gaussian-normal coordinates, the equations of motion are approximately\footnote{This equation is exact at $n=0$ as, in the Gaussian normal coordinates, $\left.\partial_a \chi\right|_{n=0} = 0$.}
\begin{equation}\label{eq:differential equation for chi(n)}
    \partial^2_n \chi - K(n) \partial_n \chi = V'(\chi(n)),
\end{equation}
where $K(n)$ is the mean extrinsic curvature of the hypersurface at $n$. In the thin wall approximation, $K(n)$ does not vary considerably over the width of the wall, so we can replace $K(n)\to K(0) \equiv K$. Multiplying Eq.~\eqref{eq:differential equation for chi(n)} by $\partial_n \chi$ and integrating, we obtain
\begin{equation}\label{eq:integrated EOM}
    \frac{1}{2} (\partial_n\chi)^2 - K\int\displaylimits_{-\infty}^n dn (\partial_n \chi)^2 = V(\chi(n)),
\end{equation}
where $V(\chi)$ is the potential defined in Eq.~\eqref{eq:bare_potential_chi} with an appropriate constant added such that $V$ is positive definite. 
Here we have defined $n$ such that $\chi(n)\to\chi_-$ as $n\to -\infty$.
Taking $n\to \infty$ in Eq.~\eqref{eq:integrated EOM} gives
\begin{equation} \label{eq:integrated wall simple}
-K\sigma = \Delta V
\end{equation}
where $\sigma$ is the surface tension, defined as
\begin{equation}
\sigma \equiv \int dn (\partial_n \chi)^2.
\end{equation}
This is equivalent to the radial bounce action $S_1$, introduced in Section~\ref{sec:models_PT}.
Since $K$ can be written in terms of the shape and dynamics of the wall, Eq.~\eqref{eq:integrated wall simple} can be used as a dynamic equation for the wall after we calculate $\sigma$. 

To determine the surface tension $\sigma$, we solve Eq.~\eqref{eq:integrated EOM}. To do so, it is useful to consider the order of magnitude of the terms in the equation for values of $n$ inside the wall:
\begin{equation}\label{eq: orders of mag}
\begin{split}
K\int\displaylimits_{-\infty}^{n} dn (\partial_n \chi)^2 \sim & \mathcal{O}\left(\Delta V\right)  \\
V(\chi) \sim & \mathcal{O}\left(V_{\rm max} \right),
\end{split}
\end{equation}
where $V_{\rm max}$ is the maximum potential value between the two minima.
The thin-wall approximation requires that $\Delta V \ll V_{\rm max}$.
Therefore, an approximate solution for the profile of the bubble wall can be found by setting $K = 0$ in Eq.~\eqref{eq:integrated EOM}. However, from Eq.~\eqref{eq:integrated wall simple} this solution is only consistent if we set $\Delta V=0$ as well. 
We find a self-consistent approximate solution by splitting the potential into two parts:
\begin{equation}
\begin{split}
V = & V_0 + V_1\\
V_0\equiv &\frac{\lambda}{4} \left(\chi^2-\frac{m^2}{\lambda}\right)^2\\
V_1\equiv & \frac{\mu}{3}\chi^3 + \rm const.
\end{split}
\end{equation} 
where $V_0$ is the $Z_2$ symmetric part of the potential and the constant on the last line makes $V_1(\chi_-)\approx 0$.

Therefore, an approximate solution for the bubble wall profile can be found by solving
\begin{equation}\label{eq:symmetrized wall equation}
\frac{1}{2}(\partial_n \chi)^2 \approx V_0(\chi(n)).
\end{equation} 
subject to boundary conditions $\chi(\pm \infty) = \pm m/\sqrt{\lambda}$. The solution is
\begin{equation}\label{eq:thin wall solution}
\chi = \frac{m}{\sqrt{\lambda}} \tanh\left(\frac{n}{\ell}\right),
\end{equation}
where $\ell \equiv \sqrt{2}/m$. This allows us to calculate $\sigma$ explicitly:
\begin{equation}
    \sigma = \frac{2\sqrt{2}}{3}\frac{m^3}{\lambda} \,.
\end{equation}

Having calculated the surface tension, we next consider the mean extrinsic curvature of the surface $\Sigma$:
\begin{equation}\label{eq:mean curvature}
K = -(g^{\mu \nu} \nabla_\mu \tilde{N}_\nu)|_{n=0}.
\end{equation} 
Here $\tilde{N}_\nu(n, \xi^a)$ is an arbitrary extension of $N_\nu$ to a unit vector field that agrees with $N_\nu$ on $\Sigma$. 
In our case, $\Sigma$ is the surface of a spherical bubble nucleated at some time $t_I$ with a radius 
\begin{equation}
    r(t, t_I) = \int\displaylimits_{t_I}^t \frac{dt' v_w(t')}{a(t')} + \frac{r_c}{a(t_I)}
\end{equation}
where the critical radius $r_c$ is the initial physical radius of the bubble as mentioned in Section~\ref{sec:models_PT}. Points $x^\mu = (t, r, \theta, \phi)$ on $\Sigma$ satisfy
\begin{equation}
    0 = F(x^\mu)\equiv r-r(t, t_I)
\end{equation}
The unit normal vector $N_\mu$ is then given by
\begin{equation}\label{eq:unit normal}
    N_\mu = \frac{\partial_\mu F}{\sqrt{|\partial_\mu F \partial^\mu F|}} = \gamma_w (-v_w, a(t) \bhat{r}{}).
\end{equation}
and
\begin{equation}\label{eq:normal coordinate}
    n = (x^\mu -X^\mu)N_\mu = \frac{F(x^\mu)}{\sqrt{|\partial_\mu F \partial^\mu F|}} = a(t)\gamma_w(r-r(t, t_I)).
\end{equation}

Using Eqs.~\eqref{eq:mean curvature} and \eqref{eq:unit normal}, we calculate the mean curvature of $\Sigma$. Since $N_\mu$ does not depend explicitly on $r$ but $n$ depends on $t$ and $r$, we can extend the domain of $N_\mu$ away from $n=0$ and take $\tilde{N}_\mu\equiv N_\mu$.  
Then the mean curvature can be written as
\begin{equation}
    K = -\partial_t(\gamma_w v_w) - 3H \gamma_w v_w - \frac{2\gamma_w}{a(t)r(t, t_I)}.
\end{equation}
Substituting this result into Eq.~\eqref{eq:integrated wall simple} leads to a dynamic equation for the expansion of the bubble wall
\begin{equation}\label{eq:bubble wall dynamic equation}
    \partial_t(\gamma_w v_w) + 3H \gamma_w v_w + \frac{2\gamma_w}{a(t)r(t, t_I)} = \frac{\Delta V}{\sigma}.
\end{equation}

When a bubble first nucleates its physical size is the critical radius $r_c\equiv 3 S_1/\Delta V=3 \sigma/\Delta V$ and its wall is at rest \cite{Coleman:1977py,Linde:1981zj}.
Putting $a(t_I)r(t_I,t_I)=r_c$ and $v_w=0$ in Eq.~\eqref{eq:bubble wall dynamic equation}, we can see that, the bubble wall will accelerate and expand until it reaches the terminal velocity given by 
\begin{equation} \label{eq:terminal velocity}
    \left. (\gamma_w v_w) \right|_\infty = \frac{\Delta V}{3\sigma H_{\rm inf}} = \frac{1}{r_cH_{\rm inf}} \gg 1,
\end{equation}
where we have used the condition in Eq.~\eqref{eq:condition_Hinf}.
This implies the terminal wall velocity is $v_w \approx 1$, with $\gamma_w \approx 1/(r_cH_{\rm inf})$, after a time $t-t_I \sim H_{\rm inf}^{-1}$.

\bigskip
%%%%%%%%%%%%%%%%%%%%%%%%%%%%%%%%%%%%%%%%%%%%%%%%%%%%%%%%%%%%%%%%%%%%%%%%%%%%%%%%%

\section{Two-Bubble Terms}\label{app:two bubble terms}
Here we quantify the errors from our calculation of the power spectrum of $\delta_{\rm \chi}$ in Section~\ref{sec:bubbles}. The power spectrum is determined by the two-point function which can be written as the double sum over bubbles given in Eq.~\eqref{eq:chi correlation sum}. The double sum can be split up as
\begin{equation}
    \langle \delta_\chi(\bk)\delta_\chi(\bkp)\rangle = \langle \delta_\chi(\bk)\delta_\chi(\bkp)\rangle^{(1)} + \langle \delta_\chi(\bk)\delta_\chi(\bkp)\rangle^{(2)}
\end{equation}
where the first term on the right is the contribution from terms where $I=J$ and the second term is the contribution from terms where $I\neq J$. In Section~\ref{sec:bubbles}, we calculated the power spectrum from the $I=J$ terms only and here we will approximate the correction from including the $I\neq J$ terms.

The correction to the 2-point function from the $I\neq J$ terms can be written as
\begin{widetext}
\begin{equation}\label{eq:full 2 point}
     \langle \delta_\chi(\bk)\delta_\chi(\bkp)\rangle^{(2)} = e^{2t_e/\tau_{\rm PT}}\frac{(4\pi)^2}{k^3{k'}^3} (N^2-N)\int d^4x d^4x'p_2(x, x')e^{-i(\bk\cdot \bs{x}+\bkp\cdot \bs{x'})}\mathcal{A}(kr(t))\mathcal{A}(k'r(t'))
\end{equation}
\end{widetext}
where $p_2(x, x')$ is the joint probability density function for bubbles nucleated at space-time coordinates $x$ and $x'$. For $t<t'$, we will write the joint probability density function as
\begin{equation} \label{eq:joint pdf}
    p_2(x, x') = p(x|x')p_1(x') 
\end{equation}
where 
\begin{equation}
    p_1(x') = \frac{1}{N}\Gamma a(t')^3p_{\rm false}(t')
\end{equation}
and the conditional probability can be written as
\begin{equation}\label{eq:conditional pdf}
    p(x|x') = \frac{1}{N}\Gamma a(t)^3\Theta(\Delta x - r(t))p_{\rm false}(t|t', \Delta x)\quad \quad t<t' .
\end{equation}
Here, $\Delta x \equiv |\bs{x}-\bs{x'}|$ and 
\begin{equation}\label{eq:p_false conditional}
    \begin{split}
        p_{\rm false}(t|t', \Delta x) = & e^{-J(t, \Delta x)} \quad \quad t<t'\\
        J(t, \Delta x) = & \int\displaylimits_0^t dt'' \Gamma a(t'')^3\left[\VH(t'') - \VO(t'', \Delta x)\right]
    \end{split}
\end{equation}
where $\mathcal{V_H}(t'')$ is the comoving Hubble volume at $t''$ and $\mathcal{V_O}(t'', \Delta x)$ is the overlap volume between two spheres of comoving radius $r(t'')$ at $\bs{x}$ and $\bs{x'}$. $J(t, \Delta x) \leq t/\tau_{\rm PT}$ where equality is acheived if $\mathcal{V_O}(t'', \Delta x)=0$ for $t''\in [0, t]$.
The $\Theta$ function in Eq.~\eqref{eq:conditional pdf} ensures that the probability goes to zero for a bubble nucleated at coordinates $\bs{x}, t$ that encloses the point $\bs{x'}$. For $t>t'$ we use $p_2(x,x') = p(x'|x)p_1(x)$ rather than Eq.~\eqref{eq:joint pdf} and exchange $x\leftrightarrow x'$ in Eq.~\eqref{eq:conditional pdf}.

Since $p_2(x, x')$ only depends on $\bs{x}$ and $\bs{x'}$ through $\Delta x$, we change integration variables to
\begin{equation}
    \begin{split}
        \bs{X} \equiv & \frac{\bs{x} + \bs{x'}}{2} \\
        \bs{\Delta x} \equiv & \bs{x}-\bs{x'}
    \end{split}
\end{equation}
and write $\bk \cdot\bs{x} + \bkp\cdot\bs{x'} = (\bk + \bkp)\cdot \bs{X} + (\bk-\bkp)\cdot \bs{\Delta x}/2$. Integration over $\bs{X}$ is then trivial. If $t_e/\tau_{\rm PT} \ll 1$, we can safely set $\exp{\left[-J(t, \Delta x)\right]} \to 1$ and $\exp{(-t/\tau_{\rm PT})}\to 1$, leading to the expression
\begin{widetext}
\begin{equation}
    \langle \delta_\chi(\bk)\delta_\chi(\bkp)\rangle^{(2)} = 2\frac{(4\pi)^2}{k^6} (2\pi)^3 \delta^3(\bk + \bkp) \Gamma^2\int\displaylimits_0^{t_e} dt \int\displaylimits_t^{t_e} dt'a(t)^3a(t')^3\mathcal{A}(kr(t))\mathcal{A}(kr(t'))\int d^3(\Delta x)e^{-i\bk \cdot \bs{\Delta x}} \Theta(\Delta x - r(t))
\end{equation}
\end{widetext}
The $\bs{\Delta x}$ integral leads to a term proportional to $\delta^3(\bk)$ and a finite term. We ignore the term proportional to the delta function and change to dimensionless integration variables $u \equiv kr(t)$ and $u'\equiv kr(t')$ to obtain
\begin{widetext}
\begin{equation}
    \langle \delta_\chi(\bk)\delta_\chi(\bkp)\rangle^{(2)} =-\frac{128\pi^3}{k^3}(2\pi)^3 \delta^3(\bk +\bkp) \gamma_{\rm PT}^2\int\displaylimits_{kr_e}^{kr_i} du u^{-4}\mathcal{A}(u)^2\int\displaylimits_{kr_e}^u du' {u'}^{-4}\mathcal{A}(u') + (\propto \delta^3(\bk))
\end{equation}
\end{widetext}
This contribution to the two point function is proportional to $\gamma_{\rm PT}^2$ so it is suppressed by an extra factor of $\gamma_{\rm PT}$ compared to the result found in Section~\ref{sec:bubbles}. We checked numerically that this term can be neglected for all values of $\gamma_{\rm PT}$ considered in this work.
%%%%%%%%%%%%%%%%%%%%%%%%%%%%%%%%%%%%%%%%%%%%%%%%%%%%%%%%%%%%%%
%%%%%%%%%%%%%%%%%%%%%%%%%%%%%%%%%%%%%%%%%%%%%%%%%%%%%%%%%%%%%%
\section{Curvature Produced by FOPT}\label{app:curvature FOPT}

In the main text we calculated the contribution of the FOPT to the isocurvature power spectrum.  In this section we will show that the FOPT can also source curvature perturbations during inflation due to non-adiabatic pressure in bubble walls. We will start by calculating the components of the stress-energy tensor for the $\chi$ field and extract the non-adiabatic pressure. We will then use our expression for the non-adiabatic pressure to calculate the evolution of the comoving curvature perturbation during inflation.

\subsection{Stress-Energy Tensor} \label{app:stress energy}
The stress-energy tensor for the $\chi$ field is given by the standard expression for a scalar field
\begin{equation}
    {T^\mu}_\nu = g^{\mu\rho}\partial_\rho\chi \partial_\nu\chi - \delta^\mu_\nu\left(\frac{1}{2}g^{\rho\sigma}\partial_\rho\chi \partial_\sigma \chi + V(\chi)\right)
\end{equation}
As discussed in Appendix \ref{app:bubble wall}, when a bubble is nucleated, $\chi$ only depends on the coordinate $n$ orthogonal to the wall (with unit vector $N_\nu$). As a result, the stress-energy tensor for a single bubble can be written as
\begin{equation}
    {T^\mu}_\nu = \left[N^\mu N_\nu-\frac{1}{2}\delta^\mu_\nu\right](\partial_n\chi)^2 - \delta^\mu_\nu V(\chi).
\end{equation}
The definition of $n$ for a bubble nucleated at the origin at time $t_I$ is given in Eq.~\eqref{eq:normal coordinate}. We split the potential into its two components $V=V_0+V_1$ and substitute Eq.~\eqref{eq:symmetrized wall equation} for $V_0$. Then the stress energy tensor can be written as
\begin{equation}
    {T^\mu}_\nu = \left[N^\mu N_\nu-\delta^\mu_\nu\right](\partial_n\chi)^2 - \delta^\mu_\nu V_1(\chi)
\end{equation}
The term proportional to $(\partial_n\chi)^2$ vanishes everywhere except on the wall. The term proportional to $V_1$ is negligible inside the wall compared to the first term but switches from $0$ for $n\lesssim 0$ to $+\Delta V$ for $n\gtrsim 0$. This motivates us to define ``wall" and ``bulk" contributions to the stress energy tensor
\begin{equation}
\begin{split}
    {{(T_{\rm wall})}^\mu}_\nu = & \left[N^\mu N_\nu-\delta^\mu_\nu\right](\partial_n \chi)^2\\
    {{(T_{\rm bulk})}^\mu}_\nu = & V_1(\chi(n))
\end{split}
\end{equation}
where the full tensor is the sum of the two:
\begin{equation}
    {T^\mu}_\nu = {{(T_{\rm wall})}^\mu}_\nu + {{(T_{\rm bulk})}^\mu}_\nu
\end{equation}
In this work, the most important properties of the stress tensor are the energy density $\rho$ and the pressure $p$. Using Eq.~\eqref{eq:unit normal}, we can write the ``wall" and ``bulk" components of the single bubble energy density as
\begin{equation}\label{eq:bulk and wall energy}
\begin{split}
    \rho_{\rm wall} =& -{{(T_{\rm wall})}^0}_0 = \gamma_w^2 (\partial_n \chi)^2 \\ 
    \rho_{\rm bulk} =& -{{(T_{\rm bulk})}^0}_0 = V_1(\chi(n))
\end{split}
\end{equation}
and of the pressure as
\begin{equation}
\begin{split}
    p_{\rm wall} = & \frac{1}{3}{{(T_{\rm wall})}^i}_i = \gamma_w^2(v_w^2 - 2/3)(\partial_n\chi)^2 = (v_w^2 - 2/3)\rho_{\rm wall}\\
    p_{\rm bulk} = & \frac{1}{3}{{(T_{\rm bulk})}^i}_i = -V_1(\chi(n)) = - \rho_{\rm bulk}
\end{split}
\end{equation}

\subsubsection{Wall Energy Density}

We next calculate $\bar{\rho}_{\rm wall}$ and $\delta \rho_{\rm wall}(\bs{k}, t)$ for the whole Universe at time $t$. To do so, we will sum the first line of Eq.~\eqref{eq:bulk and wall energy} over bubbles nucleated before $t$
\begin{equation}
    \rho_{\rm wall}(\bs{x}, t) = \gamma_w^2\sum_{I=1}^N\Theta(t-t_I)\left.(\partial_n\chi)^2\right|_{n=n_I}
\end{equation}
where $n_I \equiv a(t)\gamma_w(|\bs{x}-\bs{x_I}|-r(t, t_I))$ is the normal coordinate for bubble $I$, $\bs{x_I}$ is the center position of the bubble, and for $\gamma_w$ we take the terminal value for all bubbles given in Eq.~\eqref{eq:terminal velocity}. Using the solution for the profile of the bubble wall in Eq.~\eqref{eq:thin wall solution}, the wall energy density can be written as
\begin{equation}\label{eq:gradient energy 1 bubble}
    \rho_{\rm wall} = \frac{\gamma_w \Delta V}{4\ell} \sum_I\Theta(t-t_I)\sech^4{(n_I/\ell)}
\end{equation}

The spatial average of the wall energy is
\begin{equation}\label{eq:average gradient energy density}
\begin{split}
    \bar{\rho}_{\rm wall} = & \frac{1}{\mathcal{V}}\int\displaylimits_\mathcal{V} d^3 x \rho_g\\
    =& \frac{4\pi}{3}\frac{\Delta V}{\mathcal{V} a(t)} \sum_{I=1}^N \Theta(t-t_I) r(t, t_I)^2.
\end{split}
\end{equation}
We substitute Eq.~\eqref{eq:comoving radius} for $r(t, t_I)$ and use the probability density in Eq.~\eqref{eq:single bubble pdf} to convert the sum over bubbles into an integral over time
\begin{equation}\label{eq:average gradient energy density 2}
\bar{\rho}_{\rm wall} = \frac{4\pi}{3}\gamma_{\rm PT}\Delta Ve^{-t/\tau_{PT}}.
\end{equation}

The perturbation $\rho_{\rm wall} - \bar{\rho}_{\rm wall}$ of the wall's energy density in Fourier space is 
\begin{equation}
\label{eq:delta_rho_wall}
    \delta \rho_{\rm wall}   = \int d^3x \rho_{\rm wall}(\bk) e^{-i\bk\cdot\bs{x}} + {\rm const.}\times \delta^3(\bk),
\end{equation}
where the second term is only non-zero for $\bk = 0$. This term can be dropped, as only the non-zero wavenumbers enter our analysis. Assuming that $r_I\gg \ell/(\gamma_w a)$ for all relevant bubbles, and that we are interested in modes such that $k\ell/(\gamma_w a) \ll 1$, the integral over $\bs{x}$ results in the closed form solution
\begin{equation}
    \delta \rho_{\rm wall}=  \frac{4\pi}{3}\frac{\Delta V}{Ha(t) v_w k^2}\sum_I \Theta(t-t_I)\mathcal{B}(kr(t_I)) e^{-i\bk\cdot\bs{x_I}},
\end{equation}
where 
\begin{equation}
    \mathcal{B}(y) \equiv y\sin{y}.
\end{equation}

\subsubsection{Bulk Energy Density}
The bulk energy density is given by the point-wise function of $\chi$ in the second line of Eq.~\eqref{eq:bulk and wall energy}. In a region of space where bubble $I$ is the only bubble, the behavior of $\chi$ is given by Eq.~\eqref{eq:thin wall solution} with $n\to n_I$. Since we only care about the bulk energy density on scales much larger than the thickness of the wall, we can approximate
\begin{equation}
\begin{split}
    \chi(n_I) \approx & \frac{2m}{\sqrt{\lambda}}\left[\Theta(n_I)-\frac{1}{2}\right]\\
    =& \frac{2m}{\sqrt{\lambda}} \left[\Theta(|\bs{x}-\bs{x_I}|-r(t, t_I))-\frac{1}{2}\right],
\end{split}
\end{equation}
leading to
\begin{equation}
 \rho_{\mathrm{bulk}, I} = V_1(\chi(n_I)) \approx  \Delta V\left[1 - \Theta(r(t, t_I)-|\bs{x}-\bs{x_I}|)\right].
\end{equation}
where $\Delta V\equiv 2\mu m^3/(3\lambda^{3/2})$.
The fraction of space in true vacuum by the end of inflation is $\sim \gamma_{\rm PT}H_{\rm inf}t_e\ll 1$. As a result, most bubbles do not overlap and the bulk energy density for the whole Universe can be approximated as
\begin{equation}
    \rho_{\rm bulk} = \Delta V\left[1 - \sum_I \Theta(t-t_I)\Theta(r(t, t_I)-|\bs{x}-\bs{x_I}|)\right].
\end{equation}

Spatially averaging this bulk density results in
\begin{equation}\label{eq:average wall}
    \bar{\rho}_{\rm bulk} = \Delta V p_{\rm false}(t) = \Delta V e^{-t/\tau_{PT}},
\end{equation}
which dominates over the wall contribution, Eq.~\eqref{eq:average wall}. The density perturbation for the bulk contribution is
\begin{equation} 
\begin{split}
    \delta\rho_{\rm bulk}(\bk, t) = & -\Delta V\sum_I \Theta(t-t_I)\\
    \times & \int d^3x e^{-i\bk\cdot\bs{x}}\Theta(r(t, t_I)-|\bs{x}-\bs{x_I}|) \\
    + & {\rm const.}\times\delta^3(\bk).
\end{split}
\end{equation}
Again dropping the second term and performing the integral over $\bs{x}$
\begin{equation}
     \delta\rho_{\rm bulk}(\bk, t) = \frac{4\pi\Delta V}{k^3} \sum_I \Theta(t-t_I)\mathcal{A}(kr(t, t_I))e^{-i\bk\cdot\bs{x_I}},
\end{equation}
where we have defined
\begin{equation}
    \mathcal{A}(y) \equiv y \cos{y} - \sin{y}.
\end{equation}
\subsection{Curvature Power Spectrum}\label{app:curvature power}

In the standard inflationary model, the inflaton field produces subhorizon quantum fluctuations in the comoving curvature perturbation (denoted as $\mathcal{R}$) with a power spectrum given by \cite{Malik:2008im}
\begin{equation}\label{eq:curvature power q fluct}
  P_\mathcal{R_\phi}(k) \equiv P_\mathcal{R}(k, t_k) = \left.\frac{H^4}{4\pi^2 \dot{\bar{\phi}}^2}\right|_{t_k},
\end{equation}
where $t_k$ is the time that wave-number $k$ exits the horizon ($a(t_k)H_{\rm inf}\sim k$). In the class of models of FOPT considered in this paper, bubbles from the FOPT quickly become larger than the horizon. As a result, their effect on the subhorizon dynamics is mostly from modifying the background energy density $\bar\rho_{\chi}$, which can be neglected since we work in the region $\bar\rho_{\chi}\ll(\bar\rho_{\chi}+\bar\rho_{\phi})$. Thus, the subhorizon dynamics are almost unaltered from standard inflation. 

Outside of the horizon, the curvature perturbation evolves according to~\cite{Gordon:2000hv,Malik:2004tf}
\begin{equation}\label{eq:zeta_dot_general}
    \dot{\mathcal{R}} = \frac{H}{\bar{\rho}+\bar{p}}\delta p_{\rm nad} + \mathcal{O}\left(\frac{k^2}{a^2H^2}\right),
\end{equation}
where $\delta p_{\rm nad}$ is the non-adiabatic pressure perturbation
\begin{equation}\label{eq:non adiabatic pressure general}
    \delta p_{\rm nad} \equiv \delta p - \frac{\dot{\bar{p}}}{\dot{\bar{\rho}}}\delta \rho.
\end{equation}
The non-adiabatic pressure of the inflaton is suppressed outside the horizon \cite{Wands:2007bd}, while for the $\chi$ field it is sourced by the energy of the bubble wall.
Setting $v_w=1$ and using results in Appendix~\ref{app:stress energy}, the superhorizon evolution of $\mathcal{R}$ is
\begin{equation}\label{eq:zeta_dot_superhorizon}
    \dot{\mathcal{R}} \approx \frac{4}{3}\frac{H}{\dot{\bar{\phi}}^2} \delta \rho_{\rm wall}~~~~(t>t_k),
\end{equation}
where $\delta \rho_{\rm wall}$ is defined in Eq.~(\ref{eq:delta_rho_wall}). The curvature $\mathcal{R}$ at the end of inflation ($t=t_e$) is then
\begin{equation}
\begin{split}
   &\mathcal{R}(\bk, t_e) = \mathcal{R}(\bk, t_k) +  \frac{16\pi}{9}\frac{\Delta V}{\dot{\bar{\phi}}^2}\frac{1}{k^2}\\
    \times & \sum_I kr(t_I)\sin{(kr(t_I))}e^{-i\bk\cdot\bs{x_I}}\int_{t_k}^{t_e} dt \frac{1}{a(t)} \Theta(t-t_I) ,   
\end{split}
\end{equation}
where $I$ runs over all nucleated bubbles.

 To calculate the power spectrum of $\mathcal{R}$ at the end of inflation, we follow the same procedure as in Eq.~\eqref{eq:chi correlation sum}: we ignore cross-terms in the double sum, write the ensemble average as an integral, integrate over spatial coordinates, and change the $t$ integration variable to $u\equiv kr(t)$. The curvature power spectrum is then given as:
\begin{equation}\label{eq:curvature power 1}
\begin{split}
      & P_{\cal R}(k, t_e) -  P_{\cal R}(k, t_k)=\frac{128}{81}\left(\frac{\Delta V}{\dot{\bar{\phi}}^2}\right)^2 \gamma_{\rm PT}\\
      \times & \left[\frac{1}{2}\left(kr(t_k) - \frac{1}{2}\sin{(2kr(t_k))}\right) + \int\displaylimits_{kr(t_k)}^{kr_i}du\frac{\sin^2{u}}{u^2}\right].\\
      \quad &
\end{split}
\end{equation}
The first term in the square brackets comes from bubbles nucleated after $t_k$ while the second term comes from bubbles nucleated before $t_k$. If FOPT starts after $t_k$ ($t_i>t_k$), the second term in the brackets of Eq.~\eqref{eq:curvature power 1} is absent since there are no bubbles nucleated before $t_k$. The power spectrum in this case is simply given by the first term with $t_k\to t_i$. 
Since $r(t)$ is the radius of the horizon at $t$, then the definition of $t_k$ implies $r(t_k)\sim 1/k$ so the first term in the brackets is independent of $k$. 

Given our definition of $P_{\rm ad}(k)$ in Eq.~(\ref{eq:C power spectrum}), we find that $P_{\rm ad}(k)= P_{\cal R}(k, t_e)$. There is also a contribution to the curvature power spectrum from bubble percolation when the phase transition completes around $T_*$~\cite{Freese:2023fcr,Liu:2022lvz,Elor:2023xbz}. As mentioned in section~\ref{sec:isocurvature}, we can neglect those contributions when studying CMB observables since we consider scenarios where $T_*\gg T_
{\rm CMB}$. Bringing all of this together, we can obtain $P_{\rm ad}(k)$:
\begin{widetext}
    \begin{equation}\label{eq:curvature power spectrum final}
       P_{\rm ad}(k)\approx  P_{\cal R_\phi}(k)+\frac{128}{81}\left(\frac{\Delta V}{\dot{\bar{\phi}}^2}\right)^2 \gamma_{\rm PT}
    \begin{dcases}
        \frac{1}{3} (kr_i)^3 & kr_i\ll 1 \\
        0.95 & kr_i\gg 1
    \end{dcases}.
    \end{equation}
\end{widetext}
Here the inflaton contribution $P_{\cal{R}_{\phi}}(k)$ is usually parameterized as $P_{\cal R_\phi}(k)=A_s (k/k_{\rm pivot})^{n_s-1}$, where $A_s$ and $n_s$ are the amplitude and spectral index respectively. The contribution from FOPT has a distinct $k$ dependence: for $kr_i\ll 1$ it goes as $(kr_i)^3$ and for $kr_i\to \infty$ it approaches a constant. 

Since we focus on isocurvature effects of the FOPT in this work,  we choose the model parameters such that the contribution of FOPT to $ P_{\rm ad}(k)$ (second term of Eq.~\eqref{eq:curvature power spectrum final}) is negligible. This translates to the requirement that:
\begin{equation}
    \left(\frac{\Delta V}{\dot{\bar{\phi}}^2}\right)^2 \gamma_{\rm PT} \ll A_s .
\end{equation}
Moreover, we also require the effect of FOPT on CMB angular power spectrum (e.g., $C_{\ell}^{TT}$) is dominantly from isocurvature. Given the scaling of the isocurvature effects at the end of Section~\ref{sec:isocurvature}, this condition is given as:
\begin{equation}
    \left(\frac{\Delta V}{\dot{\bar{\phi}}^2}\right)^2 \ll  R_{\rm dr}^2 .
\end{equation}

%%%%%%%%%%%%%%%%
%%%%%%%%%%%%%%%%%%%%%%%%%%%%%%%%%%%%%%%%%%%%%%%%%%%%%%%%%%%%%%
%%%%%%%%%%%%%%%%%%%%%%%%%%%%%%%%%%%%%%%%%%%%%%%%%%%%%%%%%%%%%%
\section{Initial Conditions for the Dark Radiation isocurvature mode}\label{app:DR_iso_ic}
In this section, we will present the full set of initial conditions for perturbations in the DR isocurvature mode in the synchronous gauge and in terms of the conformal time $\tau$ (see also~\cite{Ghosh:2021axu}). We assume DR is free-streaming, analogous to Standard Model neutrinos. We keep terms up to ${\cal O}((k\tau)^2)$. For terms that are zero up to this order, we retain the leading non-vanishing term. The initial conditions are
\begin{eqnarray}
    &&\eta^{\rm iso}= \frac{-R_{\rm dr}+R_{\rm dr}^2+R_{\rm dr}R_{\nu}}{6(1-R_{\rm dr})(15+4R_{\rm dr}+4R_{\nu})}(k\tau)^2 \nonumber\\
    &&h^{\rm iso}=\frac{R_{\rm dr}}{40(1-R_{\rm dr})}\omega_b k^2\tau^3\nonumber\\
    &&\delta_\gamma^{\rm iso}=\delta_\nu^{\rm iso}=\frac{-R_{\rm dr}}{1-R_{\rm dr}}\left(1-\frac{1}{6}(k\tau)^2\right)\nonumber\\
     &&\delta_b^{\rm iso}=\frac{R_{\rm dr}}{8(1-R_{\rm dr})}(k\tau)^2\nonumber\\
      &&\delta_c^{\rm iso}=\frac{-R_{\rm dr}}{80(1-R_{\rm dr})}\omega_b k^2\tau^3\\
&&\delta_{\rm dr}^{\rm iso}=1-\frac{1}{6}(k\tau)^2\nonumber\\
    &&\theta_\gamma^{\rm iso}=\theta_b^{\rm iso}=\frac{-R_{\rm dr}}{4(1-R_{\rm dr})}k^2\tau\nonumber\\
    &&\theta_{\rm dr}^{\rm iso}=\frac{1}{4}k^2\tau\nonumber\\
    &&\sigma_{\nu}^{\rm iso}=\frac{-19R_{\rm dr}}{30(1-R_{\rm dr})(15+4R_{\rm dr}+4R_{\nu})}k^2\tau^2\nonumber\\
     &&\sigma_{\rm dr}^{\rm iso}=\frac{15-15R_{\rm dr}+4R_{\nu}}{30(1-R_{\rm dr})(15+4R_{\rm dr}+4R_{\nu})}k^2\tau^2,\nonumber
\end{eqnarray}
where $\omega_{b}\equiv \sqrt{8\pi G/3}\, a(\tau_{\rm ini})\bar\rho_b(\tau_{\rm ini})/\sqrt{\bar\rho_{r}(\tau_{\rm ini})}$ and $\bar\rho_b$,$\bar\rho_r$ is the background baryon and radiation density respectively. At the time initial conditions are set, $\omega_b\tau\ll 1$ can be treated as an expansion parameter.

\medskip
%%%%%%%%%%%%%%%%%%%%%%%%%%%%%%

\bibliography{references}

\begin{thebibliography}{105}%
\makeatletter
\providecommand \@ifxundefined [1]{%
 \@ifx{#1\undefined}
}%
\providecommand \@ifnum [1]{%
 \ifnum #1\expandafter \@firstoftwo
 \else \expandafter \@secondoftwo
 \fi
}%
\providecommand \@ifx [1]{%
 \ifx #1\expandafter \@firstoftwo
 \else \expandafter \@secondoftwo
 \fi
}%
\providecommand \natexlab [1]{#1}%
\providecommand \enquote  [1]{``#1''}%
\providecommand \bibnamefont  [1]{#1}%
\providecommand \bibfnamefont [1]{#1}%
\providecommand \citenamefont [1]{#1}%
\providecommand \href@noop [0]{\@secondoftwo}%
\providecommand \href [0]{\begingroup \@sanitize@url \@href}%
\providecommand \@href[1]{\@@startlink{#1}\@@href}%
\providecommand \@@href[1]{\endgroup#1\@@endlink}%
\providecommand \@sanitize@url [0]{\catcode `\\12\catcode `\$12\catcode
  `\&12\catcode `\#12\catcode `\^12\catcode `\_12\catcode `\%12\relax}%
\providecommand \@@startlink[1]{}%
\providecommand \@@endlink[0]{}%
\providecommand \url  [0]{\begingroup\@sanitize@url \@url }%
\providecommand \@url [1]{\endgroup\@href {#1}{\urlprefix }}%
\providecommand \urlprefix  [0]{URL }%
\providecommand \Eprint [0]{\href }%
\providecommand \doibase [0]{http://dx.doi.org/}%
\providecommand \selectlanguage [0]{\@gobble}%
\providecommand \bibinfo  [0]{\@secondoftwo}%
\providecommand \bibfield  [0]{\@secondoftwo}%
\providecommand \translation [1]{[#1]}%
\providecommand \BibitemOpen [0]{}%
\providecommand \bibitemStop [0]{}%
\providecommand \bibitemNoStop [0]{.\EOS\space}%
\providecommand \EOS [0]{\spacefactor3000\relax}%
\providecommand \BibitemShut  [1]{\csname bibitem#1\endcsname}%
\let\auto@bib@innerbib\@empty
%</preamble>
\bibitem [{\citenamefont {Kuzmin}\ \emph {et~al.}(1985)\citenamefont {Kuzmin},
  \citenamefont {Rubakov},\ and\ \citenamefont {Shaposhnikov}}]{KUZMIN198536}%
  \BibitemOpen
  \bibfield  {author} {\bibinfo {author} {\bibfnamefont {V.}~\bibnamefont
  {Kuzmin}}, \bibinfo {author} {\bibfnamefont {V.}~\bibnamefont {Rubakov}}, \
  and\ \bibinfo {author} {\bibfnamefont {M.}~\bibnamefont {Shaposhnikov}},\
  }\href {\doibase https://doi.org/10.1016/0370-2693(85)91028-7} {\bibfield
  {journal} {\bibinfo  {journal} {Physics Letters B}\ }\textbf {\bibinfo
  {volume} {155}},\ \bibinfo {pages} {36} (\bibinfo {year} {1985})}\BibitemShut
  {NoStop}%
\bibitem [{\citenamefont {Shaposhnikov}(1986)}]{Shaposhnikov:1986jp}%
  \BibitemOpen
  \bibfield  {author} {\bibinfo {author} {\bibfnamefont {M.~E.}\ \bibnamefont
  {Shaposhnikov}},\ }\href@noop {} {\bibfield  {journal} {\bibinfo  {journal}
  {JETP Lett.}\ }\textbf {\bibinfo {volume} {44}},\ \bibinfo {pages} {465}
  (\bibinfo {year} {1986})}\BibitemShut {NoStop}%
\bibitem [{\citenamefont {Shaposhnikov}(1987)}]{Shaposhnikov:1987tw}%
  \BibitemOpen
  \bibfield  {author} {\bibinfo {author} {\bibfnamefont {M.~E.}\ \bibnamefont
  {Shaposhnikov}},\ }\href {\doibase 10.1016/0550-3213(87)90127-1} {\bibfield
  {journal} {\bibinfo  {journal} {Nucl. Phys. B}\ }\textbf {\bibinfo {volume}
  {287}},\ \bibinfo {pages} {757} (\bibinfo {year} {1987})}\BibitemShut
  {NoStop}%
\bibitem [{\citenamefont {Cohen}\ \emph {et~al.}(1993)\citenamefont {Cohen},
  \citenamefont {Kaplan},\ and\ \citenamefont {Nelson}}]{Cohen:1993nk}%
  \BibitemOpen
  \bibfield  {author} {\bibinfo {author} {\bibfnamefont {A.~G.}\ \bibnamefont
  {Cohen}}, \bibinfo {author} {\bibfnamefont {D.~B.}\ \bibnamefont {Kaplan}}, \
  and\ \bibinfo {author} {\bibfnamefont {A.~E.}\ \bibnamefont {Nelson}},\
  }\href {\doibase 10.1146/annurev.ns.43.120193.000331} {\bibfield  {journal}
  {\bibinfo  {journal} {Ann. Rev. Nucl. Part. Sci.}\ }\textbf {\bibinfo
  {volume} {43}},\ \bibinfo {pages} {27} (\bibinfo {year} {1993})},\ \Eprint
  {http://arxiv.org/abs/hep-ph/9302210} {arXiv:hep-ph/9302210} \BibitemShut
  {NoStop}%
\bibitem [{\citenamefont {Trodden}(1999)}]{Trodden:1998ym}%
  \BibitemOpen
  \bibfield  {author} {\bibinfo {author} {\bibfnamefont {M.}~\bibnamefont
  {Trodden}},\ }\href {\doibase 10.1103/RevModPhys.71.1463} {\bibfield
  {journal} {\bibinfo  {journal} {Rev. Mod. Phys.}\ }\textbf {\bibinfo {volume}
  {71}},\ \bibinfo {pages} {1463} (\bibinfo {year} {1999})},\ \Eprint
  {http://arxiv.org/abs/hep-ph/9803479} {arXiv:hep-ph/9803479} \BibitemShut
  {NoStop}%
\bibitem [{\citenamefont {Riotto}\ and\ \citenamefont
  {Trodden}(1999)}]{Riotto:1999yt}%
  \BibitemOpen
  \bibfield  {author} {\bibinfo {author} {\bibfnamefont {A.}~\bibnamefont
  {Riotto}}\ and\ \bibinfo {author} {\bibfnamefont {M.}~\bibnamefont
  {Trodden}},\ }\href {\doibase 10.1146/annurev.nucl.49.1.35} {\bibfield
  {journal} {\bibinfo  {journal} {Ann. Rev. Nucl. Part. Sci.}\ }\textbf
  {\bibinfo {volume} {49}},\ \bibinfo {pages} {35} (\bibinfo {year} {1999})},\
  \Eprint {http://arxiv.org/abs/hep-ph/9901362} {arXiv:hep-ph/9901362}
  \BibitemShut {NoStop}%
\bibitem [{\citenamefont {Dine}\ and\ \citenamefont
  {Kusenko}(2003)}]{Dine:2003ax}%
  \BibitemOpen
  \bibfield  {author} {\bibinfo {author} {\bibfnamefont {M.}~\bibnamefont
  {Dine}}\ and\ \bibinfo {author} {\bibfnamefont {A.}~\bibnamefont {Kusenko}},\
  }\href {\doibase 10.1103/RevModPhys.76.1} {\bibfield  {journal} {\bibinfo
  {journal} {Rev. Mod. Phys.}\ }\textbf {\bibinfo {volume} {76}},\ \bibinfo
  {pages} {1} (\bibinfo {year} {2003})},\ \Eprint
  {http://arxiv.org/abs/hep-ph/0303065} {arXiv:hep-ph/0303065} \BibitemShut
  {NoStop}%
\bibitem [{\citenamefont {Morrissey}\ and\ \citenamefont
  {Ramsey-Musolf}(2012)}]{Morrissey:2012db}%
  \BibitemOpen
  \bibfield  {author} {\bibinfo {author} {\bibfnamefont {D.~E.}\ \bibnamefont
  {Morrissey}}\ and\ \bibinfo {author} {\bibfnamefont {M.~J.}\ \bibnamefont
  {Ramsey-Musolf}},\ }\href {\doibase 10.1088/1367-2630/14/12/125003}
  {\bibfield  {journal} {\bibinfo  {journal} {New J. Phys.}\ }\textbf {\bibinfo
  {volume} {14}},\ \bibinfo {pages} {125003} (\bibinfo {year} {2012})},\
  \Eprint {http://arxiv.org/abs/1206.2942} {arXiv:1206.2942 [hep-ph]}
  \BibitemShut {NoStop}%
\bibitem [{\citenamefont {Creminelli}\ \emph {et~al.}(2002)\citenamefont
  {Creminelli}, \citenamefont {Nicolis},\ and\ \citenamefont
  {Rattazzi}}]{Creminelli:2001th}%
  \BibitemOpen
  \bibfield  {author} {\bibinfo {author} {\bibfnamefont {P.}~\bibnamefont
  {Creminelli}}, \bibinfo {author} {\bibfnamefont {A.}~\bibnamefont {Nicolis}},
  \ and\ \bibinfo {author} {\bibfnamefont {R.}~\bibnamefont {Rattazzi}},\
  }\href {\doibase 10.1088/1126-6708/2002/03/051} {\bibfield  {journal}
  {\bibinfo  {journal} {JHEP}\ }\textbf {\bibinfo {volume} {03}},\ \bibinfo
  {pages} {051} (\bibinfo {year} {2002})},\ \Eprint
  {http://arxiv.org/abs/hep-th/0107141} {arXiv:hep-th/0107141} \BibitemShut
  {NoStop}%
\bibitem [{\citenamefont {Randall}\ and\ \citenamefont
  {Servant}(2007)}]{Randall:2006py}%
  \BibitemOpen
  \bibfield  {author} {\bibinfo {author} {\bibfnamefont {L.}~\bibnamefont
  {Randall}}\ and\ \bibinfo {author} {\bibfnamefont {G.}~\bibnamefont
  {Servant}},\ }\href {\doibase 10.1088/1126-6708/2007/05/054} {\bibfield
  {journal} {\bibinfo  {journal} {JHEP}\ }\textbf {\bibinfo {volume} {05}},\
  \bibinfo {pages} {054} (\bibinfo {year} {2007})},\ \Eprint
  {http://arxiv.org/abs/hep-ph/0607158} {arXiv:hep-ph/0607158} \BibitemShut
  {NoStop}%
\bibitem [{\citenamefont {Nardini}\ \emph {et~al.}(2007)\citenamefont
  {Nardini}, \citenamefont {Quiros},\ and\ \citenamefont
  {Wulzer}}]{Nardini:2007me}%
  \BibitemOpen
  \bibfield  {author} {\bibinfo {author} {\bibfnamefont {G.}~\bibnamefont
  {Nardini}}, \bibinfo {author} {\bibfnamefont {M.}~\bibnamefont {Quiros}}, \
  and\ \bibinfo {author} {\bibfnamefont {A.}~\bibnamefont {Wulzer}},\ }\href
  {\doibase 10.1088/1126-6708/2007/09/077} {\bibfield  {journal} {\bibinfo
  {journal} {JHEP}\ }\textbf {\bibinfo {volume} {09}},\ \bibinfo {pages} {077}
  (\bibinfo {year} {2007})},\ \Eprint {http://arxiv.org/abs/0706.3388}
  {arXiv:0706.3388 [hep-ph]} \BibitemShut {NoStop}%
\bibitem [{\citenamefont {Konstandin}\ \emph {et~al.}(2010)\citenamefont
  {Konstandin}, \citenamefont {Nardini},\ and\ \citenamefont
  {Quiros}}]{Konstandin:2010cd}%
  \BibitemOpen
  \bibfield  {author} {\bibinfo {author} {\bibfnamefont {T.}~\bibnamefont
  {Konstandin}}, \bibinfo {author} {\bibfnamefont {G.}~\bibnamefont {Nardini}},
  \ and\ \bibinfo {author} {\bibfnamefont {M.}~\bibnamefont {Quiros}},\ }\href
  {\doibase 10.1103/PhysRevD.82.083513} {\bibfield  {journal} {\bibinfo
  {journal} {Phys. Rev. D}\ }\textbf {\bibinfo {volume} {82}},\ \bibinfo
  {pages} {083513} (\bibinfo {year} {2010})},\ \Eprint
  {http://arxiv.org/abs/1007.1468} {arXiv:1007.1468 [hep-ph]} \BibitemShut
  {NoStop}%
\bibitem [{\citenamefont {Konstandin}\ and\ \citenamefont
  {Servant}(2011)}]{Konstandin:2011dr}%
  \BibitemOpen
  \bibfield  {author} {\bibinfo {author} {\bibfnamefont {T.}~\bibnamefont
  {Konstandin}}\ and\ \bibinfo {author} {\bibfnamefont {G.}~\bibnamefont
  {Servant}},\ }\href {\doibase 10.1088/1475-7516/2011/12/009} {\bibfield
  {journal} {\bibinfo  {journal} {JCAP}\ }\textbf {\bibinfo {volume} {12}},\
  \bibinfo {pages} {009} (\bibinfo {year} {2011})},\ \Eprint
  {http://arxiv.org/abs/1104.4791} {arXiv:1104.4791 [hep-ph]} \BibitemShut
  {NoStop}%
\bibitem [{\citenamefont {Bunk}\ \emph {et~al.}(2018)\citenamefont {Bunk},
  \citenamefont {Hubisz},\ and\ \citenamefont {Jain}}]{Bunk:2017fic}%
  \BibitemOpen
  \bibfield  {author} {\bibinfo {author} {\bibfnamefont {D.}~\bibnamefont
  {Bunk}}, \bibinfo {author} {\bibfnamefont {J.}~\bibnamefont {Hubisz}}, \ and\
  \bibinfo {author} {\bibfnamefont {B.}~\bibnamefont {Jain}},\ }\href {\doibase
  10.1140/epjc/s10052-018-5529-2} {\bibfield  {journal} {\bibinfo  {journal}
  {Eur. Phys. J. C}\ }\textbf {\bibinfo {volume} {78}},\ \bibinfo {pages} {78}
  (\bibinfo {year} {2018})},\ \Eprint {http://arxiv.org/abs/1705.00001}
  {arXiv:1705.00001 [hep-ph]} \BibitemShut {NoStop}%
\bibitem [{\citenamefont {Baratella}\ \emph {et~al.}(2019)\citenamefont
  {Baratella}, \citenamefont {Pomarol},\ and\ \citenamefont
  {Rompineve}}]{Baratella:2018pxi}%
  \BibitemOpen
  \bibfield  {author} {\bibinfo {author} {\bibfnamefont {P.}~\bibnamefont
  {Baratella}}, \bibinfo {author} {\bibfnamefont {A.}~\bibnamefont {Pomarol}},
  \ and\ \bibinfo {author} {\bibfnamefont {F.}~\bibnamefont {Rompineve}},\
  }\href {\doibase 10.1007/JHEP03(2019)100} {\bibfield  {journal} {\bibinfo
  {journal} {JHEP}\ }\textbf {\bibinfo {volume} {03}},\ \bibinfo {pages} {100}
  (\bibinfo {year} {2019})},\ \Eprint {http://arxiv.org/abs/1812.06996}
  {arXiv:1812.06996 [hep-ph]} \BibitemShut {NoStop}%
\bibitem [{\citenamefont {Bruggisser}\ \emph {et~al.}(2018)\citenamefont
  {Bruggisser}, \citenamefont {Von~Harling}, \citenamefont {Matsedonskyi},\
  and\ \citenamefont {Servant}}]{Bruggisser:2018mrt}%
  \BibitemOpen
  \bibfield  {author} {\bibinfo {author} {\bibfnamefont {S.}~\bibnamefont
  {Bruggisser}}, \bibinfo {author} {\bibfnamefont {B.}~\bibnamefont
  {Von~Harling}}, \bibinfo {author} {\bibfnamefont {O.}~\bibnamefont
  {Matsedonskyi}}, \ and\ \bibinfo {author} {\bibfnamefont {G.}~\bibnamefont
  {Servant}},\ }\href {\doibase 10.1007/JHEP12(2018)099} {\bibfield  {journal}
  {\bibinfo  {journal} {JHEP}\ }\textbf {\bibinfo {volume} {12}},\ \bibinfo
  {pages} {099} (\bibinfo {year} {2018})},\ \Eprint
  {http://arxiv.org/abs/1804.07314} {arXiv:1804.07314 [hep-ph]} \BibitemShut
  {NoStop}%
\bibitem [{\citenamefont {Meg\'\i{}as}\ \emph {et~al.}(2018)\citenamefont
  {Meg\'\i{}as}, \citenamefont {Nardini},\ and\ \citenamefont
  {Quir\'os}}]{Megias:2018sxv}%
  \BibitemOpen
  \bibfield  {author} {\bibinfo {author} {\bibfnamefont {E.}~\bibnamefont
  {Meg\'\i{}as}}, \bibinfo {author} {\bibfnamefont {G.}~\bibnamefont
  {Nardini}}, \ and\ \bibinfo {author} {\bibfnamefont {M.}~\bibnamefont
  {Quir\'os}},\ }\href {\doibase 10.1007/JHEP09(2018)095} {\bibfield  {journal}
  {\bibinfo  {journal} {JHEP}\ }\textbf {\bibinfo {volume} {09}},\ \bibinfo
  {pages} {095} (\bibinfo {year} {2018})},\ \Eprint
  {http://arxiv.org/abs/1806.04877} {arXiv:1806.04877 [hep-ph]} \BibitemShut
  {NoStop}%
\bibitem [{\citenamefont {Agashe}\ \emph {et~al.}(2020)\citenamefont {Agashe},
  \citenamefont {Du}, \citenamefont {Ekhterachian}, \citenamefont {Kumar},\
  and\ \citenamefont {Sundrum}}]{Agashe:2019lhy}%
  \BibitemOpen
  \bibfield  {author} {\bibinfo {author} {\bibfnamefont {K.}~\bibnamefont
  {Agashe}}, \bibinfo {author} {\bibfnamefont {P.}~\bibnamefont {Du}}, \bibinfo
  {author} {\bibfnamefont {M.}~\bibnamefont {Ekhterachian}}, \bibinfo {author}
  {\bibfnamefont {S.}~\bibnamefont {Kumar}}, \ and\ \bibinfo {author}
  {\bibfnamefont {R.}~\bibnamefont {Sundrum}},\ }\href {\doibase
  10.1007/JHEP05(2020)086} {\bibfield  {journal} {\bibinfo  {journal} {JHEP}\
  }\textbf {\bibinfo {volume} {05}},\ \bibinfo {pages} {086} (\bibinfo {year}
  {2020})},\ \Eprint {http://arxiv.org/abs/1910.06238} {arXiv:1910.06238
  [hep-ph]} \BibitemShut {NoStop}%
\bibitem [{\citenamefont {Fujikura}\ \emph {et~al.}(2020)\citenamefont
  {Fujikura}, \citenamefont {Nakai},\ and\ \citenamefont
  {Yamada}}]{Fujikura:2019oyi}%
  \BibitemOpen
  \bibfield  {author} {\bibinfo {author} {\bibfnamefont {K.}~\bibnamefont
  {Fujikura}}, \bibinfo {author} {\bibfnamefont {Y.}~\bibnamefont {Nakai}}, \
  and\ \bibinfo {author} {\bibfnamefont {M.}~\bibnamefont {Yamada}},\ }\href
  {\doibase 10.1007/JHEP02(2020)111} {\bibfield  {journal} {\bibinfo  {journal}
  {JHEP}\ }\textbf {\bibinfo {volume} {02}},\ \bibinfo {pages} {111} (\bibinfo
  {year} {2020})},\ \Eprint {http://arxiv.org/abs/1910.07546} {arXiv:1910.07546
  [hep-ph]} \BibitemShut {NoStop}%
\bibitem [{\citenamefont {Agashe}\ \emph {et~al.}(2021)\citenamefont {Agashe},
  \citenamefont {Du}, \citenamefont {Ekhterachian}, \citenamefont {Kumar},\
  and\ \citenamefont {Sundrum}}]{Agashe:2020lfz}%
  \BibitemOpen
  \bibfield  {author} {\bibinfo {author} {\bibfnamefont {K.}~\bibnamefont
  {Agashe}}, \bibinfo {author} {\bibfnamefont {P.}~\bibnamefont {Du}}, \bibinfo
  {author} {\bibfnamefont {M.}~\bibnamefont {Ekhterachian}}, \bibinfo {author}
  {\bibfnamefont {S.}~\bibnamefont {Kumar}}, \ and\ \bibinfo {author}
  {\bibfnamefont {R.}~\bibnamefont {Sundrum}},\ }\href {\doibase
  10.1007/JHEP02(2021)051} {\bibfield  {journal} {\bibinfo  {journal} {JHEP}\
  }\textbf {\bibinfo {volume} {02}},\ \bibinfo {pages} {051} (\bibinfo {year}
  {2021})},\ \Eprint {http://arxiv.org/abs/2010.04083} {arXiv:2010.04083
  [hep-th]} \BibitemShut {NoStop}%
\bibitem [{\citenamefont {Ares}\ \emph {et~al.}(2020)\citenamefont {Ares},
  \citenamefont {Hindmarsh}, \citenamefont {Hoyos},\ and\ \citenamefont
  {Jokela}}]{Ares:2020lbt}%
  \BibitemOpen
  \bibfield  {author} {\bibinfo {author} {\bibfnamefont {F.~R.}\ \bibnamefont
  {Ares}}, \bibinfo {author} {\bibfnamefont {M.}~\bibnamefont {Hindmarsh}},
  \bibinfo {author} {\bibfnamefont {C.}~\bibnamefont {Hoyos}}, \ and\ \bibinfo
  {author} {\bibfnamefont {N.}~\bibnamefont {Jokela}},\ }\href {\doibase
  10.1007/JHEP04(2021)100} {\bibfield  {journal} {\bibinfo  {journal} {JHEP}\
  }\textbf {\bibinfo {volume} {21}},\ \bibinfo {pages} {100} (\bibinfo {year}
  {2020})},\ \Eprint {http://arxiv.org/abs/2011.12878} {arXiv:2011.12878
  [hep-th]} \BibitemShut {NoStop}%
\bibitem [{\citenamefont {Agrawal}\ and\ \citenamefont
  {Nee}(2021)}]{Agrawal:2021alq}%
  \BibitemOpen
  \bibfield  {author} {\bibinfo {author} {\bibfnamefont {P.}~\bibnamefont
  {Agrawal}}\ and\ \bibinfo {author} {\bibfnamefont {M.}~\bibnamefont {Nee}},\
  }\href {\doibase 10.1007/JHEP10(2021)105} {\bibfield  {journal} {\bibinfo
  {journal} {JHEP}\ }\textbf {\bibinfo {volume} {10}},\ \bibinfo {pages} {105}
  (\bibinfo {year} {2021})},\ \Eprint {http://arxiv.org/abs/2103.05646}
  {arXiv:2103.05646 [hep-ph]} \BibitemShut {NoStop}%
\bibitem [{\citenamefont {Levi}\ \emph {et~al.}(2023)\citenamefont {Levi},
  \citenamefont {Opferkuch},\ and\ \citenamefont {Redigolo}}]{Levi:2022bzt}%
  \BibitemOpen
  \bibfield  {author} {\bibinfo {author} {\bibfnamefont {N.}~\bibnamefont
  {Levi}}, \bibinfo {author} {\bibfnamefont {T.}~\bibnamefont {Opferkuch}}, \
  and\ \bibinfo {author} {\bibfnamefont {D.}~\bibnamefont {Redigolo}},\ }\href
  {\doibase 10.1007/JHEP02(2023)125} {\bibfield  {journal} {\bibinfo  {journal}
  {JHEP}\ }\textbf {\bibinfo {volume} {02}},\ \bibinfo {pages} {125} (\bibinfo
  {year} {2023})},\ \Eprint {http://arxiv.org/abs/2212.08085} {arXiv:2212.08085
  [hep-ph]} \BibitemShut {NoStop}%
\bibitem [{\citenamefont {Cs\'aki}\ \emph {et~al.}(2023)\citenamefont
  {Cs\'aki}, \citenamefont {Geller}, \citenamefont {Heller-Algazi},\ and\
  \citenamefont {Ismail}}]{Csaki:2023pwy}%
  \BibitemOpen
  \bibfield  {author} {\bibinfo {author} {\bibfnamefont {C.}~\bibnamefont
  {Cs\'aki}}, \bibinfo {author} {\bibfnamefont {M.}~\bibnamefont {Geller}},
  \bibinfo {author} {\bibfnamefont {Z.}~\bibnamefont {Heller-Algazi}}, \ and\
  \bibinfo {author} {\bibfnamefont {A.}~\bibnamefont {Ismail}},\ }\href
  {\doibase 10.1007/JHEP06(2023)202} {\bibfield  {journal} {\bibinfo  {journal}
  {JHEP}\ }\textbf {\bibinfo {volume} {06}},\ \bibinfo {pages} {202} (\bibinfo
  {year} {2023})},\ \Eprint {http://arxiv.org/abs/2301.10247} {arXiv:2301.10247
  [hep-ph]} \BibitemShut {NoStop}%
\bibitem [{\citenamefont {Er\"oncel}\ \emph {et~al.}(2023)\citenamefont
  {Er\"oncel}, \citenamefont {Hubisz}, \citenamefont {Lee}, \citenamefont
  {Rigo},\ and\ \citenamefont {Sambasivam}}]{Eroncel:2023uqf}%
  \BibitemOpen
  \bibfield  {author} {\bibinfo {author} {\bibfnamefont {C.}~\bibnamefont
  {Er\"oncel}}, \bibinfo {author} {\bibfnamefont {J.}~\bibnamefont {Hubisz}},
  \bibinfo {author} {\bibfnamefont {S.~J.}\ \bibnamefont {Lee}}, \bibinfo
  {author} {\bibfnamefont {G.}~\bibnamefont {Rigo}}, \ and\ \bibinfo {author}
  {\bibfnamefont {B.}~\bibnamefont {Sambasivam}},\ }\href@noop {} {\  (\bibinfo
  {year} {2023})},\ \Eprint {http://arxiv.org/abs/2305.03773} {arXiv:2305.03773
  [hep-ph]} \BibitemShut {NoStop}%
\bibitem [{\citenamefont {Mishra}\ and\ \citenamefont
  {Randall}(2023)}]{Mishra:2023kiu}%
  \BibitemOpen
  \bibfield  {author} {\bibinfo {author} {\bibfnamefont {R.~K.}\ \bibnamefont
  {Mishra}}\ and\ \bibinfo {author} {\bibfnamefont {L.}~\bibnamefont
  {Randall}},\ }\href@noop {} {\  (\bibinfo {year} {2023})},\ \Eprint
  {http://arxiv.org/abs/2309.10090} {arXiv:2309.10090 [hep-ph]} \BibitemShut
  {NoStop}%
\bibitem [{\citenamefont {Mishra}\ and\ \citenamefont
  {Randall}(2024)}]{Mishra:2024ehr}%
  \BibitemOpen
  \bibfield  {author} {\bibinfo {author} {\bibfnamefont {R.~K.}\ \bibnamefont
  {Mishra}}\ and\ \bibinfo {author} {\bibfnamefont {L.}~\bibnamefont
  {Randall}},\ }\href@noop {} {\  (\bibinfo {year} {2024})},\ \Eprint
  {http://arxiv.org/abs/2401.09633} {arXiv:2401.09633 [hep-ph]} \BibitemShut
  {NoStop}%
\bibitem [{\citenamefont {Bai}\ and\ \citenamefont
  {Schwaller}(2014)}]{Bai:2013xga}%
  \BibitemOpen
  \bibfield  {author} {\bibinfo {author} {\bibfnamefont {Y.}~\bibnamefont
  {Bai}}\ and\ \bibinfo {author} {\bibfnamefont {P.}~\bibnamefont
  {Schwaller}},\ }\href {\doibase 10.1103/PhysRevD.89.063522} {\bibfield
  {journal} {\bibinfo  {journal} {Phys. Rev. D}\ }\textbf {\bibinfo {volume}
  {89}},\ \bibinfo {pages} {063522} (\bibinfo {year} {2014})},\ \Eprint
  {http://arxiv.org/abs/1306.4676} {arXiv:1306.4676 [hep-ph]} \BibitemShut
  {NoStop}%
\bibitem [{\citenamefont {Kribs}\ and\ \citenamefont
  {Neil}(2016)}]{Kribs:2016cew}%
  \BibitemOpen
  \bibfield  {author} {\bibinfo {author} {\bibfnamefont {G.~D.}\ \bibnamefont
  {Kribs}}\ and\ \bibinfo {author} {\bibfnamefont {E.~T.}\ \bibnamefont
  {Neil}},\ }\href {\doibase 10.1142/S0217751X16430041} {\bibfield  {journal}
  {\bibinfo  {journal} {Int. J. Mod. Phys. A}\ }\textbf {\bibinfo {volume}
  {31}},\ \bibinfo {pages} {1643004} (\bibinfo {year} {2016})},\ \Eprint
  {http://arxiv.org/abs/1604.04627} {arXiv:1604.04627 [hep-ph]} \BibitemShut
  {NoStop}%
\bibitem [{\citenamefont {Cline}(2022)}]{Cline:2021itd}%
  \BibitemOpen
  \bibfield  {author} {\bibinfo {author} {\bibfnamefont {J.~M.}\ \bibnamefont
  {Cline}},\ }\href {\doibase 10.21468/SciPostPhysLectNotes.52} {\bibfield
  {journal} {\bibinfo  {journal} {SciPost Phys. Lect. Notes}\ }\textbf
  {\bibinfo {volume} {52}},\ \bibinfo {pages} {1} (\bibinfo {year} {2022})},\
  \Eprint {http://arxiv.org/abs/2108.10314} {arXiv:2108.10314 [hep-ph]}
  \BibitemShut {NoStop}%
\bibitem [{\citenamefont {Shelton}\ and\ \citenamefont
  {Zurek}(2010)}]{Shelton:2010ta}%
  \BibitemOpen
  \bibfield  {author} {\bibinfo {author} {\bibfnamefont {J.}~\bibnamefont
  {Shelton}}\ and\ \bibinfo {author} {\bibfnamefont {K.~M.}\ \bibnamefont
  {Zurek}},\ }\href {\doibase 10.1103/PhysRevD.82.123512} {\bibfield  {journal}
  {\bibinfo  {journal} {Phys. Rev. D}\ }\textbf {\bibinfo {volume} {82}},\
  \bibinfo {pages} {123512} (\bibinfo {year} {2010})},\ \Eprint
  {http://arxiv.org/abs/1008.1997} {arXiv:1008.1997 [hep-ph]} \BibitemShut
  {NoStop}%
\bibitem [{\citenamefont {Dutta}\ and\ \citenamefont
  {Kumar}(2011)}]{DUTTA2011364}%
  \BibitemOpen
  \bibfield  {author} {\bibinfo {author} {\bibfnamefont {B.}~\bibnamefont
  {Dutta}}\ and\ \bibinfo {author} {\bibfnamefont {J.}~\bibnamefont {Kumar}},\
  }\href {\doibase https://doi.org/10.1016/j.physletb.2011.04.036} {\bibfield
  {journal} {\bibinfo  {journal} {Physics Letters B}\ }\textbf {\bibinfo
  {volume} {699}},\ \bibinfo {pages} {364} (\bibinfo {year}
  {2011})}\BibitemShut {NoStop}%
\bibitem [{\citenamefont {Holthausen}\ \emph {et~al.}(2013)\citenamefont
  {Holthausen}, \citenamefont {Kubo}, \citenamefont {Lim},\ and\ \citenamefont
  {Lindner}}]{Holthausen:2013ota}%
  \BibitemOpen
  \bibfield  {author} {\bibinfo {author} {\bibfnamefont {M.}~\bibnamefont
  {Holthausen}}, \bibinfo {author} {\bibfnamefont {J.}~\bibnamefont {Kubo}},
  \bibinfo {author} {\bibfnamefont {K.~S.}\ \bibnamefont {Lim}}, \ and\
  \bibinfo {author} {\bibfnamefont {M.}~\bibnamefont {Lindner}},\ }\href
  {\doibase 10.1007/JHEP12(2013)076} {\bibfield  {journal} {\bibinfo  {journal}
  {JHEP}\ }\textbf {\bibinfo {volume} {12}},\ \bibinfo {pages} {076} (\bibinfo
  {year} {2013})},\ \Eprint {http://arxiv.org/abs/1310.4423} {arXiv:1310.4423
  [hep-ph]} \BibitemShut {NoStop}%
\bibitem [{\citenamefont {Schwaller}(2015)}]{Schwaller:2015tja}%
  \BibitemOpen
  \bibfield  {author} {\bibinfo {author} {\bibfnamefont {P.}~\bibnamefont
  {Schwaller}},\ }\href {\doibase 10.1103/PhysRevLett.115.181101} {\bibfield
  {journal} {\bibinfo  {journal} {Phys. Rev. Lett.}\ }\textbf {\bibinfo
  {volume} {115}},\ \bibinfo {pages} {181101} (\bibinfo {year} {2015})},\
  \Eprint {http://arxiv.org/abs/1504.07263} {arXiv:1504.07263 [hep-ph]}
  \BibitemShut {NoStop}%
\bibitem [{\citenamefont {Hall}\ \emph {et~al.}(2023)\citenamefont {Hall},
  \citenamefont {Konstandin}, \citenamefont {McGehee},\ and\ \citenamefont
  {Murayama}}]{Hall:2019rld}%
  \BibitemOpen
  \bibfield  {author} {\bibinfo {author} {\bibfnamefont {E.}~\bibnamefont
  {Hall}}, \bibinfo {author} {\bibfnamefont {T.}~\bibnamefont {Konstandin}},
  \bibinfo {author} {\bibfnamefont {R.}~\bibnamefont {McGehee}}, \ and\
  \bibinfo {author} {\bibfnamefont {H.}~\bibnamefont {Murayama}},\ }\href
  {\doibase 10.1103/PhysRevD.107.055011} {\bibfield  {journal} {\bibinfo
  {journal} {Phys. Rev. D}\ }\textbf {\bibinfo {volume} {107}},\ \bibinfo
  {pages} {055011} (\bibinfo {year} {2023})},\ \Eprint
  {http://arxiv.org/abs/1911.12342} {arXiv:1911.12342 [hep-ph]} \BibitemShut
  {NoStop}%
\bibitem [{\citenamefont {Bai}\ and\ \citenamefont
  {Korwar}(2022)}]{Bai:2021ibt}%
  \BibitemOpen
  \bibfield  {author} {\bibinfo {author} {\bibfnamefont {Y.}~\bibnamefont
  {Bai}}\ and\ \bibinfo {author} {\bibfnamefont {M.}~\bibnamefont {Korwar}},\
  }\href {\doibase 10.1103/PhysRevD.105.095015} {\bibfield  {journal} {\bibinfo
   {journal} {Phys. Rev. D}\ }\textbf {\bibinfo {volume} {105}},\ \bibinfo
  {pages} {095015} (\bibinfo {year} {2022})},\ \Eprint
  {http://arxiv.org/abs/2109.14765} {arXiv:2109.14765 [hep-ph]} \BibitemShut
  {NoStop}%
\bibitem [{\citenamefont {Bottaro}\ \emph {et~al.}(2021)\citenamefont
  {Bottaro}, \citenamefont {Costa},\ and\ \citenamefont
  {Popov}}]{Bottaro:2021aal}%
  \BibitemOpen
  \bibfield  {author} {\bibinfo {author} {\bibfnamefont {S.}~\bibnamefont
  {Bottaro}}, \bibinfo {author} {\bibfnamefont {M.}~\bibnamefont {Costa}}, \
  and\ \bibinfo {author} {\bibfnamefont {O.}~\bibnamefont {Popov}},\ }\href
  {\doibase 10.1007/JHEP11(2021)055} {\bibfield  {journal} {\bibinfo  {journal}
  {JHEP}\ }\textbf {\bibinfo {volume} {11}},\ \bibinfo {pages} {055} (\bibinfo
  {year} {2021})},\ \Eprint {http://arxiv.org/abs/2104.14244} {arXiv:2104.14244
  [hep-ph]} \BibitemShut {NoStop}%
\bibitem [{\citenamefont {Pasechnik}\ \emph {et~al.}(2023)\citenamefont
  {Pasechnik}, \citenamefont {Reichert}, \citenamefont {Sannino},\ and\
  \citenamefont {Wang}}]{Pasechnik:2023hwv}%
  \BibitemOpen
  \bibfield  {author} {\bibinfo {author} {\bibfnamefont {R.}~\bibnamefont
  {Pasechnik}}, \bibinfo {author} {\bibfnamefont {M.}~\bibnamefont {Reichert}},
  \bibinfo {author} {\bibfnamefont {F.}~\bibnamefont {Sannino}}, \ and\
  \bibinfo {author} {\bibfnamefont {Z.-W.}\ \bibnamefont {Wang}},\ }\href@noop
  {} {\  (\bibinfo {year} {2023})},\ \Eprint {http://arxiv.org/abs/2309.16755}
  {arXiv:2309.16755 [hep-ph]} \BibitemShut {NoStop}%
\bibitem [{\citenamefont {Karwal}\ and\ \citenamefont
  {Kamionkowski}(2016)}]{Karwal:2016vyq}%
  \BibitemOpen
  \bibfield  {author} {\bibinfo {author} {\bibfnamefont {T.}~\bibnamefont
  {Karwal}}\ and\ \bibinfo {author} {\bibfnamefont {M.}~\bibnamefont
  {Kamionkowski}},\ }\href {\doibase 10.1103/PhysRevD.94.103523} {\bibfield
  {journal} {\bibinfo  {journal} {Phys. Rev. D}\ }\textbf {\bibinfo {volume}
  {94}},\ \bibinfo {pages} {103523} (\bibinfo {year} {2016})},\ \Eprint
  {http://arxiv.org/abs/1608.01309} {arXiv:1608.01309 [astro-ph.CO]}
  \BibitemShut {NoStop}%
\bibitem [{\citenamefont {M\"ortsell}\ and\ \citenamefont
  {Dhawan}(2018)}]{Mortsell:2018mfj}%
  \BibitemOpen
  \bibfield  {author} {\bibinfo {author} {\bibfnamefont {E.}~\bibnamefont
  {M\"ortsell}}\ and\ \bibinfo {author} {\bibfnamefont {S.}~\bibnamefont
  {Dhawan}},\ }\href {\doibase 10.1088/1475-7516/2018/09/025} {\bibfield
  {journal} {\bibinfo  {journal} {JCAP}\ }\textbf {\bibinfo {volume} {09}},\
  \bibinfo {pages} {025} (\bibinfo {year} {2018})},\ \Eprint
  {http://arxiv.org/abs/1801.07260} {arXiv:1801.07260 [astro-ph.CO]}
  \BibitemShut {NoStop}%
\bibitem [{\citenamefont {Poulin}\ \emph {et~al.}(2019)\citenamefont {Poulin},
  \citenamefont {Smith}, \citenamefont {Karwal},\ and\ \citenamefont
  {Kamionkowski}}]{Poulin:2018cxd}%
  \BibitemOpen
  \bibfield  {author} {\bibinfo {author} {\bibfnamefont {V.}~\bibnamefont
  {Poulin}}, \bibinfo {author} {\bibfnamefont {T.~L.}\ \bibnamefont {Smith}},
  \bibinfo {author} {\bibfnamefont {T.}~\bibnamefont {Karwal}}, \ and\ \bibinfo
  {author} {\bibfnamefont {M.}~\bibnamefont {Kamionkowski}},\ }\href {\doibase
  10.1103/PhysRevLett.122.221301} {\bibfield  {journal} {\bibinfo  {journal}
  {Phys. Rev. Lett.}\ }\textbf {\bibinfo {volume} {122}},\ \bibinfo {pages}
  {221301} (\bibinfo {year} {2019})},\ \Eprint
  {http://arxiv.org/abs/1811.04083} {arXiv:1811.04083 [astro-ph.CO]}
  \BibitemShut {NoStop}%
\bibitem [{\citenamefont {Aghanim}\ \emph {et~al.}(2020)\citenamefont {Aghanim}
  \emph {et~al.}}]{Planck:2018vyg}%
  \BibitemOpen
  \bibfield  {author} {\bibinfo {author} {\bibfnamefont {N.}~\bibnamefont
  {Aghanim}} \emph {et~al.} (\bibinfo {collaboration} {Planck}),\ }\href
  {\doibase 10.1051/0004-6361/201833910} {\bibfield  {journal} {\bibinfo
  {journal} {Astron. Astrophys.}\ }\textbf {\bibinfo {volume} {641}},\ \bibinfo
  {pages} {A6} (\bibinfo {year} {2020})},\ \bibinfo {note} {[Erratum:
  Astron.Astrophys. 652, C4 (2021)]},\ \Eprint
  {http://arxiv.org/abs/1807.06209} {arXiv:1807.06209 [astro-ph.CO]}
  \BibitemShut {NoStop}%
\bibitem [{\citenamefont {Riess}\ \emph {et~al.}(2022)\citenamefont {Riess}
  \emph {et~al.}}]{Riess:2021jrx}%
  \BibitemOpen
  \bibfield  {author} {\bibinfo {author} {\bibfnamefont {A.~G.}\ \bibnamefont
  {Riess}} \emph {et~al.},\ }\href {\doibase 10.3847/2041-8213/ac5c5b}
  {\bibfield  {journal} {\bibinfo  {journal} {Astrophys. J. Lett.}\ }\textbf
  {\bibinfo {volume} {934}},\ \bibinfo {pages} {L7} (\bibinfo {year} {2022})},\
  \Eprint {http://arxiv.org/abs/2112.04510} {arXiv:2112.04510 [astro-ph.CO]}
  \BibitemShut {NoStop}%
\bibitem [{\citenamefont {Niedermann}\ and\ \citenamefont
  {Sloth}(2021)}]{Niedermann:2019olb}%
  \BibitemOpen
  \bibfield  {author} {\bibinfo {author} {\bibfnamefont {F.}~\bibnamefont
  {Niedermann}}\ and\ \bibinfo {author} {\bibfnamefont {M.~S.}\ \bibnamefont
  {Sloth}},\ }\href {\doibase 10.1103/PhysRevD.103.L041303} {\bibfield
  {journal} {\bibinfo  {journal} {Phys. Rev. D}\ }\textbf {\bibinfo {volume}
  {103}},\ \bibinfo {pages} {L041303} (\bibinfo {year} {2021})},\ \Eprint
  {http://arxiv.org/abs/1910.10739} {arXiv:1910.10739 [astro-ph.CO]}
  \BibitemShut {NoStop}%
\bibitem [{\citenamefont {Niedermann}\ and\ \citenamefont
  {Sloth}(2020)}]{Niedermann:2020dwg}%
  \BibitemOpen
  \bibfield  {author} {\bibinfo {author} {\bibfnamefont {F.}~\bibnamefont
  {Niedermann}}\ and\ \bibinfo {author} {\bibfnamefont {M.~S.}\ \bibnamefont
  {Sloth}},\ }\href {\doibase 10.1103/PhysRevD.102.063527} {\bibfield
  {journal} {\bibinfo  {journal} {Phys. Rev. D}\ }\textbf {\bibinfo {volume}
  {102}},\ \bibinfo {pages} {063527} (\bibinfo {year} {2020})},\ \Eprint
  {http://arxiv.org/abs/2006.06686} {arXiv:2006.06686 [astro-ph.CO]}
  \BibitemShut {NoStop}%
\bibitem [{\citenamefont {Niedermann}\ and\ \citenamefont
  {Sloth}(2022)}]{Niedermann:2021vgd}%
  \BibitemOpen
  \bibfield  {author} {\bibinfo {author} {\bibfnamefont {F.}~\bibnamefont
  {Niedermann}}\ and\ \bibinfo {author} {\bibfnamefont {M.~S.}\ \bibnamefont
  {Sloth}},\ }\href {\doibase 10.1103/PhysRevD.105.063509} {\bibfield
  {journal} {\bibinfo  {journal} {Phys. Rev. D}\ }\textbf {\bibinfo {volume}
  {105}},\ \bibinfo {pages} {063509} (\bibinfo {year} {2022})},\ \Eprint
  {http://arxiv.org/abs/2112.00770} {arXiv:2112.00770 [hep-ph]} \BibitemShut
  {NoStop}%
\bibitem [{\citenamefont {Kosowsky}\ \emph
  {et~al.}(1992{\natexlab{a}})\citenamefont {Kosowsky}, \citenamefont
  {Turner},\ and\ \citenamefont {Watkins}}]{Kosowsky:1992rz}%
  \BibitemOpen
  \bibfield  {author} {\bibinfo {author} {\bibfnamefont {A.}~\bibnamefont
  {Kosowsky}}, \bibinfo {author} {\bibfnamefont {M.~S.}\ \bibnamefont
  {Turner}}, \ and\ \bibinfo {author} {\bibfnamefont {R.}~\bibnamefont
  {Watkins}},\ }\href {\doibase 10.1103/PhysRevLett.69.2026} {\bibfield
  {journal} {\bibinfo  {journal} {Phys. Rev. Lett.}\ }\textbf {\bibinfo
  {volume} {69}},\ \bibinfo {pages} {2026} (\bibinfo {year}
  {1992}{\natexlab{a}})}\BibitemShut {NoStop}%
\bibitem [{\citenamefont {Kosowsky}\ \emph
  {et~al.}(1992{\natexlab{b}})\citenamefont {Kosowsky}, \citenamefont
  {Turner},\ and\ \citenamefont {Watkins}}]{Kosowsky:1991ua}%
  \BibitemOpen
  \bibfield  {author} {\bibinfo {author} {\bibfnamefont {A.}~\bibnamefont
  {Kosowsky}}, \bibinfo {author} {\bibfnamefont {M.~S.}\ \bibnamefont
  {Turner}}, \ and\ \bibinfo {author} {\bibfnamefont {R.}~\bibnamefont
  {Watkins}},\ }\href {\doibase 10.1103/PhysRevD.45.4514} {\bibfield  {journal}
  {\bibinfo  {journal} {Phys. Rev. D}\ }\textbf {\bibinfo {volume} {45}},\
  \bibinfo {pages} {4514} (\bibinfo {year} {1992}{\natexlab{b}})}\BibitemShut
  {NoStop}%
\bibitem [{\citenamefont {Kosowsky}\ and\ \citenamefont
  {Turner}(1993)}]{Kosowsky:1992vn}%
  \BibitemOpen
  \bibfield  {author} {\bibinfo {author} {\bibfnamefont {A.}~\bibnamefont
  {Kosowsky}}\ and\ \bibinfo {author} {\bibfnamefont {M.~S.}\ \bibnamefont
  {Turner}},\ }\href {\doibase 10.1103/PhysRevD.47.4372} {\bibfield  {journal}
  {\bibinfo  {journal} {Phys. Rev. D}\ }\textbf {\bibinfo {volume} {47}},\
  \bibinfo {pages} {4372} (\bibinfo {year} {1993})},\ \Eprint
  {http://arxiv.org/abs/astro-ph/9211004} {arXiv:astro-ph/9211004} \BibitemShut
  {NoStop}%
\bibitem [{\citenamefont {Kamionkowski}\ \emph {et~al.}(1994)\citenamefont
  {Kamionkowski}, \citenamefont {Kosowsky},\ and\ \citenamefont
  {Turner}}]{Kamionkowski:1993fg}%
  \BibitemOpen
  \bibfield  {author} {\bibinfo {author} {\bibfnamefont {M.}~\bibnamefont
  {Kamionkowski}}, \bibinfo {author} {\bibfnamefont {A.}~\bibnamefont
  {Kosowsky}}, \ and\ \bibinfo {author} {\bibfnamefont {M.~S.}\ \bibnamefont
  {Turner}},\ }\href {\doibase 10.1103/PhysRevD.49.2837} {\bibfield  {journal}
  {\bibinfo  {journal} {Phys. Rev. D}\ }\textbf {\bibinfo {volume} {49}},\
  \bibinfo {pages} {2837} (\bibinfo {year} {1994})},\ \Eprint
  {http://arxiv.org/abs/astro-ph/9310044} {arXiv:astro-ph/9310044} \BibitemShut
  {NoStop}%
\bibitem [{\citenamefont {Caprini}\ \emph {et~al.}(2016)\citenamefont {Caprini}
  \emph {et~al.}}]{Caprini:2015zlo}%
  \BibitemOpen
  \bibfield  {author} {\bibinfo {author} {\bibfnamefont {C.}~\bibnamefont
  {Caprini}} \emph {et~al.},\ }\href {\doibase 10.1088/1475-7516/2016/04/001}
  {\bibfield  {journal} {\bibinfo  {journal} {JCAP}\ }\textbf {\bibinfo
  {volume} {04}},\ \bibinfo {pages} {001} (\bibinfo {year} {2016})},\ \Eprint
  {http://arxiv.org/abs/1512.06239} {arXiv:1512.06239 [astro-ph.CO]}
  \BibitemShut {NoStop}%
\bibitem [{\citenamefont {Caprini}\ \emph {et~al.}(2020)\citenamefont {Caprini}
  \emph {et~al.}}]{Caprini:2019egz}%
  \BibitemOpen
  \bibfield  {author} {\bibinfo {author} {\bibfnamefont {C.}~\bibnamefont
  {Caprini}} \emph {et~al.},\ }\href {\doibase 10.1088/1475-7516/2020/03/024}
  {\bibfield  {journal} {\bibinfo  {journal} {JCAP}\ }\textbf {\bibinfo
  {volume} {03}},\ \bibinfo {pages} {024} (\bibinfo {year} {2020})},\ \Eprint
  {http://arxiv.org/abs/1910.13125} {arXiv:1910.13125 [astro-ph.CO]}
  \BibitemShut {NoStop}%
\bibitem [{\citenamefont {Caldwell}\ \emph {et~al.}(2022)\citenamefont
  {Caldwell} \emph {et~al.}}]{Caldwell:2022qsj}%
  \BibitemOpen
  \bibfield  {author} {\bibinfo {author} {\bibfnamefont {R.}~\bibnamefont
  {Caldwell}} \emph {et~al.},\ }\href {\doibase 10.1007/s10714-022-03027-x}
  {\bibfield  {journal} {\bibinfo  {journal} {Gen. Rel. Grav.}\ }\textbf
  {\bibinfo {volume} {54}},\ \bibinfo {pages} {156} (\bibinfo {year} {2022})},\
  \Eprint {http://arxiv.org/abs/2203.07972} {arXiv:2203.07972 [gr-qc]}
  \BibitemShut {NoStop}%
\bibitem [{\citenamefont {Agazie}\ \emph {et~al.}(2023)\citenamefont {Agazie}
  \emph {et~al.}}]{NANOGrav:2023gor}%
  \BibitemOpen
  \bibfield  {author} {\bibinfo {author} {\bibfnamefont {G.}~\bibnamefont
  {Agazie}} \emph {et~al.} (\bibinfo {collaboration} {NANOGrav}),\ }\href
  {\doibase 10.3847/2041-8213/acdac6} {\bibfield  {journal} {\bibinfo
  {journal} {Astrophys. J. Lett.}\ }\textbf {\bibinfo {volume} {951}},\
  \bibinfo {pages} {L8} (\bibinfo {year} {2023})},\ \Eprint
  {http://arxiv.org/abs/2306.16213} {arXiv:2306.16213 [astro-ph.HE]}
  \BibitemShut {NoStop}%
\bibitem [{\citenamefont {Afzal}\ \emph {et~al.}(2023)\citenamefont {Afzal}
  \emph {et~al.}}]{NANOGrav:2023hvm}%
  \BibitemOpen
  \bibfield  {author} {\bibinfo {author} {\bibfnamefont {A.}~\bibnamefont
  {Afzal}} \emph {et~al.} (\bibinfo {collaboration} {NANOGrav}),\ }\href
  {\doibase 10.3847/2041-8213/acdc91} {\bibfield  {journal} {\bibinfo
  {journal} {Astrophys. J. Lett.}\ }\textbf {\bibinfo {volume} {951}},\
  \bibinfo {pages} {L11} (\bibinfo {year} {2023})},\ \Eprint
  {http://arxiv.org/abs/2306.16219} {arXiv:2306.16219 [astro-ph.HE]}
  \BibitemShut {NoStop}%
\bibitem [{\citenamefont {Antoniadis}\ \emph {et~al.}(2023)\citenamefont
  {Antoniadis} \emph {et~al.}}]{EPTA:2023fyk}%
  \BibitemOpen
  \bibfield  {author} {\bibinfo {author} {\bibfnamefont {J.}~\bibnamefont
  {Antoniadis}} \emph {et~al.} (\bibinfo {collaboration} {EPTA}),\ }\href@noop
  {} {\  (\bibinfo {year} {2023})},\ \Eprint {http://arxiv.org/abs/2306.16214}
  {arXiv:2306.16214 [astro-ph.HE]} \BibitemShut {NoStop}%
\bibitem [{\citenamefont {Reardon}\ \emph {et~al.}(2023)\citenamefont {Reardon}
  \emph {et~al.}}]{Reardon:2023gzh}%
  \BibitemOpen
  \bibfield  {author} {\bibinfo {author} {\bibfnamefont {D.~J.}\ \bibnamefont
  {Reardon}} \emph {et~al.},\ }\href {\doibase 10.3847/2041-8213/acdd02}
  {\bibfield  {journal} {\bibinfo  {journal} {Astrophys. J. Lett.}\ }\textbf
  {\bibinfo {volume} {951}},\ \bibinfo {pages} {L6} (\bibinfo {year} {2023})},\
  \Eprint {http://arxiv.org/abs/2306.16215} {arXiv:2306.16215 [astro-ph.HE]}
  \BibitemShut {NoStop}%
\bibitem [{\citenamefont {Xu}\ \emph {et~al.}(2023)\citenamefont {Xu} \emph
  {et~al.}}]{Xu:2023wog}%
  \BibitemOpen
  \bibfield  {author} {\bibinfo {author} {\bibfnamefont {H.}~\bibnamefont {Xu}}
  \emph {et~al.},\ }\href {\doibase 10.1088/1674-4527/acdfa5} {\bibfield
  {journal} {\bibinfo  {journal} {Res. Astron. Astrophys.}\ }\textbf {\bibinfo
  {volume} {23}},\ \bibinfo {pages} {075024} (\bibinfo {year} {2023})},\
  \Eprint {http://arxiv.org/abs/2306.16216} {arXiv:2306.16216 [astro-ph.HE]}
  \BibitemShut {NoStop}%
\bibitem [{\citenamefont {Amaro-Seoane}\ \emph {et~al.}(2017)\citenamefont
  {Amaro-Seoane} \emph {et~al.}}]{LISA:2017pwj}%
  \BibitemOpen
  \bibfield  {author} {\bibinfo {author} {\bibfnamefont {P.}~\bibnamefont
  {Amaro-Seoane}} \emph {et~al.} (\bibinfo {collaboration} {LISA}),\
  }\href@noop {} {\  (\bibinfo {year} {2017})},\ \Eprint
  {http://arxiv.org/abs/1702.00786} {arXiv:1702.00786 [astro-ph.IM]}
  \BibitemShut {NoStop}%
\bibitem [{\citenamefont {Auclair}\ \emph {et~al.}(2023)\citenamefont {Auclair}
  \emph {et~al.}}]{LISACosmologyWorkingGroup:2022jok}%
  \BibitemOpen
  \bibfield  {author} {\bibinfo {author} {\bibfnamefont {P.}~\bibnamefont
  {Auclair}} \emph {et~al.} (\bibinfo {collaboration} {LISA Cosmology Working
  Group}),\ }\href {\doibase 10.1007/s41114-023-00045-2} {\bibfield  {journal}
  {\bibinfo  {journal} {Living Rev. Rel.}\ }\textbf {\bibinfo {volume} {26}},\
  \bibinfo {pages} {5} (\bibinfo {year} {2023})},\ \Eprint
  {http://arxiv.org/abs/2204.05434} {arXiv:2204.05434 [astro-ph.CO]}
  \BibitemShut {NoStop}%
\bibitem [{\citenamefont {Kawamura}\ \emph {et~al.}(2006)\citenamefont
  {Kawamura} \emph {et~al.}}]{Kawamura:2006up}%
  \BibitemOpen
  \bibfield  {author} {\bibinfo {author} {\bibfnamefont {S.}~\bibnamefont
  {Kawamura}} \emph {et~al.},\ }\href {\doibase 10.1088/0264-9381/23/8/S17}
  {\bibfield  {journal} {\bibinfo  {journal} {Class. Quant. Grav.}\ }\textbf
  {\bibinfo {volume} {23}},\ \bibinfo {pages} {S125} (\bibinfo {year}
  {2006})}\BibitemShut {NoStop}%
\bibitem [{\citenamefont {Kawamura}\ \emph {et~al.}(2021)\citenamefont
  {Kawamura} \emph {et~al.}}]{Kawamura:2020pcg}%
  \BibitemOpen
  \bibfield  {author} {\bibinfo {author} {\bibfnamefont {S.}~\bibnamefont
  {Kawamura}} \emph {et~al.},\ }\href {\doibase 10.1093/ptep/ptab019}
  {\bibfield  {journal} {\bibinfo  {journal} {PTEP}\ }\textbf {\bibinfo
  {volume} {2021}},\ \bibinfo {pages} {05A105} (\bibinfo {year} {2021})},\
  \Eprint {http://arxiv.org/abs/2006.13545} {arXiv:2006.13545 [gr-qc]}
  \BibitemShut {NoStop}%
\bibitem [{\citenamefont {Harry}\ \emph {et~al.}(2006)\citenamefont {Harry},
  \citenamefont {Fritschel}, \citenamefont {Shaddock}, \citenamefont
  {Folkner},\ and\ \citenamefont {Phinney}}]{Harry:2006fi}%
  \BibitemOpen
  \bibfield  {author} {\bibinfo {author} {\bibfnamefont {G.~M.}\ \bibnamefont
  {Harry}}, \bibinfo {author} {\bibfnamefont {P.}~\bibnamefont {Fritschel}},
  \bibinfo {author} {\bibfnamefont {D.~A.}\ \bibnamefont {Shaddock}}, \bibinfo
  {author} {\bibfnamefont {W.}~\bibnamefont {Folkner}}, \ and\ \bibinfo
  {author} {\bibfnamefont {E.~S.}\ \bibnamefont {Phinney}},\ }\href {\doibase
  10.1088/0264-9381/23/15/008} {\bibfield  {journal} {\bibinfo  {journal}
  {Class. Quant. Grav.}\ }\textbf {\bibinfo {volume} {23}},\ \bibinfo {pages}
  {4887} (\bibinfo {year} {2006})},\ \bibinfo {note} {[Erratum:
  Class.Quant.Grav. 23, 7361 (2006)]}\BibitemShut {NoStop}%
\bibitem [{\citenamefont {Froustey}\ \emph {et~al.}(2020)\citenamefont
  {Froustey}, \citenamefont {Pitrou},\ and\ \citenamefont
  {Volpe}}]{Froustey:2020mcq}%
  \BibitemOpen
  \bibfield  {author} {\bibinfo {author} {\bibfnamefont {J.}~\bibnamefont
  {Froustey}}, \bibinfo {author} {\bibfnamefont {C.}~\bibnamefont {Pitrou}}, \
  and\ \bibinfo {author} {\bibfnamefont {M.~C.}\ \bibnamefont {Volpe}},\ }\href
  {\doibase 10.1088/1475-7516/2020/12/015} {\bibfield  {journal} {\bibinfo
  {journal} {JCAP}\ }\textbf {\bibinfo {volume} {12}},\ \bibinfo {pages} {015}
  (\bibinfo {year} {2020})},\ \Eprint {http://arxiv.org/abs/2008.01074}
  {arXiv:2008.01074 [hep-ph]} \BibitemShut {NoStop}%
\bibitem [{\citenamefont {Bennett}\ \emph {et~al.}(2021)\citenamefont
  {Bennett}, \citenamefont {Buldgen}, \citenamefont {De~Salas}, \citenamefont
  {Drewes}, \citenamefont {Gariazzo}, \citenamefont {Pastor},\ and\
  \citenamefont {Wong}}]{Bennett:2020zkv}%
  \BibitemOpen
  \bibfield  {author} {\bibinfo {author} {\bibfnamefont {J.~J.}\ \bibnamefont
  {Bennett}}, \bibinfo {author} {\bibfnamefont {G.}~\bibnamefont {Buldgen}},
  \bibinfo {author} {\bibfnamefont {P.~F.}\ \bibnamefont {De~Salas}}, \bibinfo
  {author} {\bibfnamefont {M.}~\bibnamefont {Drewes}}, \bibinfo {author}
  {\bibfnamefont {S.}~\bibnamefont {Gariazzo}}, \bibinfo {author}
  {\bibfnamefont {S.}~\bibnamefont {Pastor}}, \ and\ \bibinfo {author}
  {\bibfnamefont {Y.~Y.~Y.}\ \bibnamefont {Wong}},\ }\href {\doibase
  10.1088/1475-7516/2021/04/073} {\bibfield  {journal} {\bibinfo  {journal}
  {JCAP}\ }\textbf {\bibinfo {volume} {04}},\ \bibinfo {pages} {073} (\bibinfo
  {year} {2021})},\ \Eprint {http://arxiv.org/abs/2012.02726} {arXiv:2012.02726
  [hep-ph]} \BibitemShut {NoStop}%
\bibitem [{\citenamefont {Akita}\ and\ \citenamefont
  {Yamaguchi}(2020)}]{Akita:2020szl}%
  \BibitemOpen
  \bibfield  {author} {\bibinfo {author} {\bibfnamefont {K.}~\bibnamefont
  {Akita}}\ and\ \bibinfo {author} {\bibfnamefont {M.}~\bibnamefont
  {Yamaguchi}},\ }\href {\doibase 10.1088/1475-7516/2020/08/012} {\bibfield
  {journal} {\bibinfo  {journal} {JCAP}\ }\textbf {\bibinfo {volume} {08}},\
  \bibinfo {pages} {012} (\bibinfo {year} {2020})},\ \Eprint
  {http://arxiv.org/abs/2005.07047} {arXiv:2005.07047 [hep-ph]} \BibitemShut
  {NoStop}%
\bibitem [{\citenamefont {Bucher}\ \emph {et~al.}(2000)\citenamefont {Bucher},
  \citenamefont {Moodley},\ and\ \citenamefont {Turok}}]{Bucher:1999re}%
  \BibitemOpen
  \bibfield  {author} {\bibinfo {author} {\bibfnamefont {M.}~\bibnamefont
  {Bucher}}, \bibinfo {author} {\bibfnamefont {K.}~\bibnamefont {Moodley}}, \
  and\ \bibinfo {author} {\bibfnamefont {N.}~\bibnamefont {Turok}},\ }\href
  {\doibase 10.1103/PhysRevD.62.083508} {\bibfield  {journal} {\bibinfo
  {journal} {Phys. Rev. D}\ }\textbf {\bibinfo {volume} {62}},\ \bibinfo
  {pages} {083508} (\bibinfo {year} {2000})},\ \Eprint
  {http://arxiv.org/abs/astro-ph/9904231} {arXiv:astro-ph/9904231} \BibitemShut
  {NoStop}%
\bibitem [{\citenamefont {Wands}\ \emph {et~al.}(2000)\citenamefont {Wands},
  \citenamefont {Malik}, \citenamefont {Lyth},\ and\ \citenamefont
  {Liddle}}]{Wands:2000dp}%
  \BibitemOpen
  \bibfield  {author} {\bibinfo {author} {\bibfnamefont {D.}~\bibnamefont
  {Wands}}, \bibinfo {author} {\bibfnamefont {K.~A.}\ \bibnamefont {Malik}},
  \bibinfo {author} {\bibfnamefont {D.~H.}\ \bibnamefont {Lyth}}, \ and\
  \bibinfo {author} {\bibfnamefont {A.~R.}\ \bibnamefont {Liddle}},\ }\href
  {\doibase 10.1103/PhysRevD.62.043527} {\bibfield  {journal} {\bibinfo
  {journal} {Phys. Rev. D}\ }\textbf {\bibinfo {volume} {62}},\ \bibinfo
  {pages} {043527} (\bibinfo {year} {2000})},\ \Eprint
  {http://arxiv.org/abs/astro-ph/0003278} {arXiv:astro-ph/0003278} \BibitemShut
  {NoStop}%
\bibitem [{\citenamefont {Gordon}\ \emph {et~al.}(2000)\citenamefont {Gordon},
  \citenamefont {Wands}, \citenamefont {Bassett},\ and\ \citenamefont
  {Maartens}}]{Gordon:2000hv}%
  \BibitemOpen
  \bibfield  {author} {\bibinfo {author} {\bibfnamefont {C.}~\bibnamefont
  {Gordon}}, \bibinfo {author} {\bibfnamefont {D.}~\bibnamefont {Wands}},
  \bibinfo {author} {\bibfnamefont {B.~A.}\ \bibnamefont {Bassett}}, \ and\
  \bibinfo {author} {\bibfnamefont {R.}~\bibnamefont {Maartens}},\ }\href
  {\doibase 10.1103/PhysRevD.63.023506} {\bibfield  {journal} {\bibinfo
  {journal} {Phys. Rev. D}\ }\textbf {\bibinfo {volume} {63}},\ \bibinfo
  {pages} {023506} (\bibinfo {year} {2000})},\ \Eprint
  {http://arxiv.org/abs/astro-ph/0009131} {arXiv:astro-ph/0009131} \BibitemShut
  {NoStop}%
\bibitem [{\citenamefont {Lyth}\ \emph {et~al.}(2003)\citenamefont {Lyth},
  \citenamefont {Ungarelli},\ and\ \citenamefont {Wands}}]{Lyth:2002my}%
  \BibitemOpen
  \bibfield  {author} {\bibinfo {author} {\bibfnamefont {D.~H.}\ \bibnamefont
  {Lyth}}, \bibinfo {author} {\bibfnamefont {C.}~\bibnamefont {Ungarelli}}, \
  and\ \bibinfo {author} {\bibfnamefont {D.}~\bibnamefont {Wands}},\ }\href
  {\doibase 10.1103/PhysRevD.67.023503} {\bibfield  {journal} {\bibinfo
  {journal} {Phys. Rev. D}\ }\textbf {\bibinfo {volume} {67}},\ \bibinfo
  {pages} {023503} (\bibinfo {year} {2003})},\ \Eprint
  {http://arxiv.org/abs/astro-ph/0208055} {arXiv:astro-ph/0208055} \BibitemShut
  {NoStop}%
\bibitem [{\citenamefont {Malik}\ and\ \citenamefont
  {Wands}(2005)}]{Malik:2004tf}%
  \BibitemOpen
  \bibfield  {author} {\bibinfo {author} {\bibfnamefont {K.~A.}\ \bibnamefont
  {Malik}}\ and\ \bibinfo {author} {\bibfnamefont {D.}~\bibnamefont {Wands}},\
  }\href {\doibase 10.1088/1475-7516/2005/02/007} {\bibfield  {journal}
  {\bibinfo  {journal} {JCAP}\ }\textbf {\bibinfo {volume} {02}},\ \bibinfo
  {pages} {007} (\bibinfo {year} {2005})},\ \Eprint
  {http://arxiv.org/abs/astro-ph/0411703} {arXiv:astro-ph/0411703} \BibitemShut
  {NoStop}%
\bibitem [{\citenamefont {Wands}(2008)}]{Wands:2007bd}%
  \BibitemOpen
  \bibfield  {author} {\bibinfo {author} {\bibfnamefont {D.}~\bibnamefont
  {Wands}},\ }\href {\doibase 10.1007/978-3-540-74353-8_8} {\bibfield
  {journal} {\bibinfo  {journal} {Lect. Notes Phys.}\ }\textbf {\bibinfo
  {volume} {738}},\ \bibinfo {pages} {275} (\bibinfo {year} {2008})},\ \Eprint
  {http://arxiv.org/abs/astro-ph/0702187} {arXiv:astro-ph/0702187} \BibitemShut
  {NoStop}%
\bibitem [{\citenamefont {Freese}\ and\ \citenamefont
  {Winkler}(2023)}]{Freese:2023fcr}%
  \BibitemOpen
  \bibfield  {author} {\bibinfo {author} {\bibfnamefont {K.}~\bibnamefont
  {Freese}}\ and\ \bibinfo {author} {\bibfnamefont {M.~W.}\ \bibnamefont
  {Winkler}},\ }\href {\doibase 10.1103/PhysRevD.107.083522} {\bibfield
  {journal} {\bibinfo  {journal} {Phys. Rev. D}\ }\textbf {\bibinfo {volume}
  {107}},\ \bibinfo {pages} {083522} (\bibinfo {year} {2023})},\ \Eprint
  {http://arxiv.org/abs/2302.11579} {arXiv:2302.11579 [astro-ph.CO]}
  \BibitemShut {NoStop}%
\bibitem [{\citenamefont {Elor}\ \emph {et~al.}(2023)\citenamefont {Elor},
  \citenamefont {Jinno}, \citenamefont {Kumar}, \citenamefont {McGehee},\ and\
  \citenamefont {Tsai}}]{Elor:2023xbz}%
  \BibitemOpen
  \bibfield  {author} {\bibinfo {author} {\bibfnamefont {G.}~\bibnamefont
  {Elor}}, \bibinfo {author} {\bibfnamefont {R.}~\bibnamefont {Jinno}},
  \bibinfo {author} {\bibfnamefont {S.}~\bibnamefont {Kumar}}, \bibinfo
  {author} {\bibfnamefont {R.}~\bibnamefont {McGehee}}, \ and\ \bibinfo
  {author} {\bibfnamefont {Y.}~\bibnamefont {Tsai}},\ }\href@noop {} {\
  (\bibinfo {year} {2023})},\ \Eprint {http://arxiv.org/abs/2311.16222}
  {arXiv:2311.16222 [hep-ph]} \BibitemShut {NoStop}%
\bibitem [{\citenamefont {Akrami}\ \emph
  {et~al.}(2020{\natexlab{a}})\citenamefont {Akrami} \emph
  {et~al.}}]{Planck:2018jri}%
  \BibitemOpen
  \bibfield  {author} {\bibinfo {author} {\bibfnamefont {Y.}~\bibnamefont
  {Akrami}} \emph {et~al.} (\bibinfo {collaboration} {Planck}),\ }\href
  {\doibase 10.1051/0004-6361/201833887} {\bibfield  {journal} {\bibinfo
  {journal} {Astron. Astrophys.}\ }\textbf {\bibinfo {volume} {641}},\ \bibinfo
  {pages} {A10} (\bibinfo {year} {2020}{\natexlab{a}})},\ \Eprint
  {http://arxiv.org/abs/1807.06211} {arXiv:1807.06211 [astro-ph.CO]}
  \BibitemShut {NoStop}%
\bibitem [{\citenamefont {Ghosh}\ \emph {et~al.}(2022)\citenamefont {Ghosh},
  \citenamefont {Kumar},\ and\ \citenamefont {Tsai}}]{Ghosh:2021axu}%
  \BibitemOpen
  \bibfield  {author} {\bibinfo {author} {\bibfnamefont {S.}~\bibnamefont
  {Ghosh}}, \bibinfo {author} {\bibfnamefont {S.}~\bibnamefont {Kumar}}, \ and\
  \bibinfo {author} {\bibfnamefont {Y.}~\bibnamefont {Tsai}},\ }\href {\doibase
  10.1088/1475-7516/2022/05/014} {\bibfield  {journal} {\bibinfo  {journal}
  {JCAP}\ }\textbf {\bibinfo {volume} {05}},\ \bibinfo {pages} {014} (\bibinfo
  {year} {2022})},\ \Eprint {http://arxiv.org/abs/2107.09076} {arXiv:2107.09076
  [astro-ph.CO]} \BibitemShut {NoStop}%
\bibitem [{\citenamefont {Adshead}\ \emph {et~al.}(2020)\citenamefont
  {Adshead}, \citenamefont {Holder},\ and\ \citenamefont
  {Ralegankar}}]{Adshead:2020htj}%
  \BibitemOpen
  \bibfield  {author} {\bibinfo {author} {\bibfnamefont {P.}~\bibnamefont
  {Adshead}}, \bibinfo {author} {\bibfnamefont {G.}~\bibnamefont {Holder}}, \
  and\ \bibinfo {author} {\bibfnamefont {P.}~\bibnamefont {Ralegankar}},\
  }\href {\doibase 10.1088/1475-7516/2020/09/016} {\bibfield  {journal}
  {\bibinfo  {journal} {JCAP}\ }\textbf {\bibinfo {volume} {09}},\ \bibinfo
  {pages} {016} (\bibinfo {year} {2020})},\ \Eprint
  {http://arxiv.org/abs/2006.01165} {arXiv:2006.01165 [astro-ph.CO]}
  \BibitemShut {NoStop}%
\bibitem [{\citenamefont {Guth}\ and\ \citenamefont
  {Weinberg}(1983)}]{GUTH1983321}%
  \BibitemOpen
  \bibfield  {author} {\bibinfo {author} {\bibfnamefont {A.~H.}\ \bibnamefont
  {Guth}}\ and\ \bibinfo {author} {\bibfnamefont {E.~J.}\ \bibnamefont
  {Weinberg}},\ }\href {\doibase https://doi.org/10.1016/0550-3213(83)90307-3}
  {\bibfield  {journal} {\bibinfo  {journal} {Nuclear Physics B}\ }\textbf
  {\bibinfo {volume} {212}},\ \bibinfo {pages} {321} (\bibinfo {year}
  {1983})}\BibitemShut {NoStop}%
\bibitem [{\citenamefont {Liddle}\ and\ \citenamefont
  {Wands}(1991)}]{10.1093/mnras/253.4.637}%
  \BibitemOpen
  \bibfield  {author} {\bibinfo {author} {\bibfnamefont {A.~R.}\ \bibnamefont
  {Liddle}}\ and\ \bibinfo {author} {\bibfnamefont {D.}~\bibnamefont {Wands}},\
  }\href@noop {} {\bibfield  {journal} {\bibinfo  {journal} {Monthly Notices of
  the Royal Astronomical Society}\ }\textbf {\bibinfo {volume} {253}},\
  \bibinfo {pages} {637} (\bibinfo {year} {1991})}\BibitemShut {NoStop}%
\bibitem [{\citenamefont {Turner}\ \emph {et~al.}(1992)\citenamefont {Turner},
  \citenamefont {Weinberg},\ and\ \citenamefont {Widrow}}]{Turner:1992tz}%
  \BibitemOpen
  \bibfield  {author} {\bibinfo {author} {\bibfnamefont {M.~S.}\ \bibnamefont
  {Turner}}, \bibinfo {author} {\bibfnamefont {E.~J.}\ \bibnamefont
  {Weinberg}}, \ and\ \bibinfo {author} {\bibfnamefont {L.~M.}\ \bibnamefont
  {Widrow}},\ }\href {\doibase 10.1103/PhysRevD.46.2384} {\bibfield  {journal}
  {\bibinfo  {journal} {Phys. Rev. D}\ }\textbf {\bibinfo {volume} {46}},\
  \bibinfo {pages} {2384} (\bibinfo {year} {1992})}\BibitemShut {NoStop}%
\bibitem [{\citenamefont {Copeland}\ \emph {et~al.}(1994)\citenamefont
  {Copeland}, \citenamefont {Liddle}, \citenamefont {Lyth}, \citenamefont
  {Stewart},\ and\ \citenamefont {Wands}}]{Copeland:1994vg}%
  \BibitemOpen
  \bibfield  {author} {\bibinfo {author} {\bibfnamefont {E.~J.}\ \bibnamefont
  {Copeland}}, \bibinfo {author} {\bibfnamefont {A.~R.}\ \bibnamefont
  {Liddle}}, \bibinfo {author} {\bibfnamefont {D.~H.}\ \bibnamefont {Lyth}},
  \bibinfo {author} {\bibfnamefont {E.~D.}\ \bibnamefont {Stewart}}, \ and\
  \bibinfo {author} {\bibfnamefont {D.}~\bibnamefont {Wands}},\ }\href
  {\doibase 10.1103/PhysRevD.49.6410} {\bibfield  {journal} {\bibinfo
  {journal} {Phys. Rev. D}\ }\textbf {\bibinfo {volume} {49}},\ \bibinfo
  {pages} {6410} (\bibinfo {year} {1994})},\ \Eprint
  {http://arxiv.org/abs/astro-ph/9401011} {arXiv:astro-ph/9401011} \BibitemShut
  {NoStop}%
\bibitem [{\citenamefont {Baccigalupi}\ and\ \citenamefont
  {Perrotta}(2000)}]{Baccigalupi:1999rz}%
  \BibitemOpen
  \bibfield  {author} {\bibinfo {author} {\bibfnamefont {C.}~\bibnamefont
  {Baccigalupi}}\ and\ \bibinfo {author} {\bibfnamefont {F.}~\bibnamefont
  {Perrotta}},\ }\href {\doibase 10.1046/j.1365-8711.2000.03274.x} {\bibfield
  {journal} {\bibinfo  {journal} {Mon. Not. Roy. Astron. Soc.}\ }\textbf
  {\bibinfo {volume} {314}},\ \bibinfo {pages} {1} (\bibinfo {year} {2000})},\
  \Eprint {http://arxiv.org/abs/astro-ph/9911530} {arXiv:astro-ph/9911530}
  \BibitemShut {NoStop}%
\bibitem [{\citenamefont {Barir}\ \emph {et~al.}(2023)\citenamefont {Barir},
  \citenamefont {Geller}, \citenamefont {Sun},\ and\ \citenamefont
  {Volansky}}]{Barir:2022kzo}%
  \BibitemOpen
  \bibfield  {author} {\bibinfo {author} {\bibfnamefont {J.}~\bibnamefont
  {Barir}}, \bibinfo {author} {\bibfnamefont {M.}~\bibnamefont {Geller}},
  \bibinfo {author} {\bibfnamefont {C.}~\bibnamefont {Sun}}, \ and\ \bibinfo
  {author} {\bibfnamefont {T.}~\bibnamefont {Volansky}},\ }\href {\doibase
  10.1103/PhysRevD.108.115016} {\bibfield  {journal} {\bibinfo  {journal}
  {Phys. Rev. D}\ }\textbf {\bibinfo {volume} {108}},\ \bibinfo {pages}
  {115016} (\bibinfo {year} {2023})},\ \Eprint
  {http://arxiv.org/abs/2203.00693} {arXiv:2203.00693 [hep-ph]} \BibitemShut
  {NoStop}%
\bibitem [{\citenamefont {Blas}\ \emph {et~al.}(2011)\citenamefont {Blas},
  \citenamefont {Lesgourgues},\ and\ \citenamefont {Tram}}]{Blas:2011rf}%
  \BibitemOpen
  \bibfield  {author} {\bibinfo {author} {\bibfnamefont {D.}~\bibnamefont
  {Blas}}, \bibinfo {author} {\bibfnamefont {J.}~\bibnamefont {Lesgourgues}}, \
  and\ \bibinfo {author} {\bibfnamefont {T.}~\bibnamefont {Tram}},\ }\href
  {\doibase 10.1088/1475-7516/2011/07/034} {\bibfield  {journal} {\bibinfo
  {journal} {JCAP}\ }\textbf {\bibinfo {volume} {07}},\ \bibinfo {pages} {034}
  (\bibinfo {year} {2011})},\ \Eprint {http://arxiv.org/abs/1104.2933}
  {arXiv:1104.2933 [astro-ph.CO]} \BibitemShut {NoStop}%
\bibitem [{\citenamefont {Lesgourgues}(2011)}]{Lesgourgues:2011re}%
  \BibitemOpen
  \bibfield  {author} {\bibinfo {author} {\bibfnamefont {J.}~\bibnamefont
  {Lesgourgues}},\ }\href@noop {} {\  (\bibinfo {year} {2011})},\ \Eprint
  {http://arxiv.org/abs/1104.2932} {arXiv:1104.2932 [astro-ph.IM]} \BibitemShut
  {NoStop}%
\bibitem [{\citenamefont {Lesgourgues}\ and\ \citenamefont
  {Tram}(2011)}]{Lesgourgues:2011rh}%
  \BibitemOpen
  \bibfield  {author} {\bibinfo {author} {\bibfnamefont {J.}~\bibnamefont
  {Lesgourgues}}\ and\ \bibinfo {author} {\bibfnamefont {T.}~\bibnamefont
  {Tram}},\ }\href {\doibase 10.1088/1475-7516/2011/09/032} {\bibfield
  {journal} {\bibinfo  {journal} {JCAP}\ }\textbf {\bibinfo {volume} {09}},\
  \bibinfo {pages} {032} (\bibinfo {year} {2011})},\ \Eprint
  {http://arxiv.org/abs/1104.2935} {arXiv:1104.2935 [astro-ph.CO]} \BibitemShut
  {NoStop}%
\bibitem [{\citenamefont {Akrami}\ \emph
  {et~al.}(2020{\natexlab{b}})\citenamefont {Akrami} \emph
  {et~al.}}]{Planck:2019kim}%
  \BibitemOpen
  \bibfield  {author} {\bibinfo {author} {\bibfnamefont {Y.}~\bibnamefont
  {Akrami}} \emph {et~al.} (\bibinfo {collaboration} {Planck}),\ }\href
  {\doibase 10.1051/0004-6361/201935891} {\bibfield  {journal} {\bibinfo
  {journal} {Astron. Astrophys.}\ }\textbf {\bibinfo {volume} {641}},\ \bibinfo
  {pages} {A9} (\bibinfo {year} {2020}{\natexlab{b}})},\ \Eprint
  {http://arxiv.org/abs/1905.05697} {arXiv:1905.05697 [astro-ph.CO]}
  \BibitemShut {NoStop}%
\bibitem [{\citenamefont {Montandon}\ \emph {et~al.}(2021)\citenamefont
  {Montandon}, \citenamefont {Patanchon},\ and\ \citenamefont {van
  Tent}}]{Montandon:2020kuk}%
  \BibitemOpen
  \bibfield  {author} {\bibinfo {author} {\bibfnamefont {T.}~\bibnamefont
  {Montandon}}, \bibinfo {author} {\bibfnamefont {G.}~\bibnamefont
  {Patanchon}}, \ and\ \bibinfo {author} {\bibfnamefont {B.}~\bibnamefont {van
  Tent}},\ }\href {\doibase 10.1088/1475-7516/2021/01/004} {\bibfield
  {journal} {\bibinfo  {journal} {JCAP}\ }\textbf {\bibinfo {volume} {01}},\
  \bibinfo {pages} {004} (\bibinfo {year} {2021})},\ \Eprint
  {http://arxiv.org/abs/2007.05457} {arXiv:2007.05457 [astro-ph.CO]}
  \BibitemShut {NoStop}%
\bibitem [{\citenamefont {Linde}(1983)}]{Linde:1981zj}%
  \BibitemOpen
  \bibfield  {author} {\bibinfo {author} {\bibfnamefont {A.~D.}\ \bibnamefont
  {Linde}},\ }\href {\doibase 10.1016/0550-3213(83)90293-6,
  10.1016/0550-3213(83)90072-X} {\bibfield  {journal} {\bibinfo  {journal}
  {Nucl. Phys.}\ }\textbf {\bibinfo {volume} {B216}},\ \bibinfo {pages} {421}
  (\bibinfo {year} {1983})},\ \bibinfo {note} {[Erratum: Nucl.
  Phys.B223,544(1983)]}\BibitemShut {NoStop}%
%%CITATION = NUPHA,B216,421;%%
\bibitem [{\citenamefont {Coleman}(1977)}]{Coleman:1977py}%
  \BibitemOpen
  \bibfield  {author} {\bibinfo {author} {\bibfnamefont {S.~R.}\ \bibnamefont
  {Coleman}},\ }\href {\doibase 10.1103/PhysRevD.15.2929,
  10.1103/PhysRevD.16.1248} {\bibfield  {journal} {\bibinfo  {journal} {Phys.
  Rev.}\ }\textbf {\bibinfo {volume} {D15}},\ \bibinfo {pages} {2929} (\bibinfo
  {year} {1977})},\ \bibinfo {note} {[Erratum: Phys.
  Rev.D16,1248(1977)]}\BibitemShut {NoStop}%
%%CITATION = PHRVA,D15,2929;%%
\bibitem [{\citenamefont {Callan}\ and\ \citenamefont
  {Coleman}(1977)}]{Callan:1977pt}%
  \BibitemOpen
  \bibfield  {author} {\bibinfo {author} {\bibfnamefont {C.~G.}\ \bibnamefont
  {Callan}, \bibfnamefont {Jr.}}\ and\ \bibinfo {author} {\bibfnamefont
  {S.~R.}\ \bibnamefont {Coleman}},\ }\href {\doibase 10.1103/PhysRevD.16.1762}
  {\bibfield  {journal} {\bibinfo  {journal} {Phys. Rev. D}\ }\textbf {\bibinfo
  {volume} {16}},\ \bibinfo {pages} {1762} (\bibinfo {year}
  {1977})}\BibitemShut {NoStop}%
\bibitem [{\citenamefont {Coleman}\ and\ \citenamefont
  {De~Luccia}(1980)}]{PhysRevD.21.3305}%
  \BibitemOpen
  \bibfield  {author} {\bibinfo {author} {\bibfnamefont {S.}~\bibnamefont
  {Coleman}}\ and\ \bibinfo {author} {\bibfnamefont {F.}~\bibnamefont
  {De~Luccia}},\ }\href {\doibase 10.1103/PhysRevD.21.3305} {\bibfield
  {journal} {\bibinfo  {journal} {Phys. Rev. D}\ }\textbf {\bibinfo {volume}
  {21}},\ \bibinfo {pages} {3305} (\bibinfo {year} {1980})}\BibitemShut
  {NoStop}%
\bibitem [{\citenamefont {Ma}\ and\ \citenamefont
  {Bertschinger}(1995)}]{Ma:1995ey}%
  \BibitemOpen
  \bibfield  {author} {\bibinfo {author} {\bibfnamefont {C.-P.}\ \bibnamefont
  {Ma}}\ and\ \bibinfo {author} {\bibfnamefont {E.}~\bibnamefont
  {Bertschinger}},\ }\href {\doibase 10.1086/176550} {\bibfield  {journal}
  {\bibinfo  {journal} {Astrophys. J.}\ }\textbf {\bibinfo {volume} {455}},\
  \bibinfo {pages} {7} (\bibinfo {year} {1995})},\ \Eprint
  {http://arxiv.org/abs/astro-ph/9506072} {arXiv:astro-ph/9506072} \BibitemShut
  {NoStop}%
\bibitem [{\citenamefont {Savelainen}\ \emph {et~al.}(2013)\citenamefont
  {Savelainen}, \citenamefont {Valiviita}, \citenamefont {Walia}, \citenamefont
  {Rusak},\ and\ \citenamefont {Kurki-Suonio}}]{Savelainen:2013iwa}%
  \BibitemOpen
  \bibfield  {author} {\bibinfo {author} {\bibfnamefont {M.}~\bibnamefont
  {Savelainen}}, \bibinfo {author} {\bibfnamefont {J.}~\bibnamefont
  {Valiviita}}, \bibinfo {author} {\bibfnamefont {P.}~\bibnamefont {Walia}},
  \bibinfo {author} {\bibfnamefont {S.}~\bibnamefont {Rusak}}, \ and\ \bibinfo
  {author} {\bibfnamefont {H.}~\bibnamefont {Kurki-Suonio}},\ }\href {\doibase
  10.1103/PhysRevD.88.063010} {\bibfield  {journal} {\bibinfo  {journal} {Phys.
  Rev. D}\ }\textbf {\bibinfo {volume} {88}},\ \bibinfo {pages} {063010}
  (\bibinfo {year} {2013})},\ \Eprint {http://arxiv.org/abs/1307.4398}
  {arXiv:1307.4398 [astro-ph.CO]} \BibitemShut {NoStop}%
\bibitem [{\citenamefont {Audren}\ \emph {et~al.}(2013)\citenamefont {Audren},
  \citenamefont {Lesgourgues}, \citenamefont {Benabed},\ and\ \citenamefont
  {Prunet}}]{Audren:2012wb}%
  \BibitemOpen
  \bibfield  {author} {\bibinfo {author} {\bibfnamefont {B.}~\bibnamefont
  {Audren}}, \bibinfo {author} {\bibfnamefont {J.}~\bibnamefont {Lesgourgues}},
  \bibinfo {author} {\bibfnamefont {K.}~\bibnamefont {Benabed}}, \ and\
  \bibinfo {author} {\bibfnamefont {S.}~\bibnamefont {Prunet}},\ }\href
  {\doibase 10.1088/1475-7516/2013/02/001} {\bibfield  {journal} {\bibinfo
  {journal} {JCAP}\ }\textbf {\bibinfo {volume} {02}},\ \bibinfo {pages} {001}
  (\bibinfo {year} {2013})},\ \Eprint {http://arxiv.org/abs/1210.7183}
  {arXiv:1210.7183 [astro-ph.CO]} \BibitemShut {NoStop}%
\bibitem [{\citenamefont {Brinckmann}\ and\ \citenamefont
  {Lesgourgues}(2019)}]{Brinckmann:2018cvx}%
  \BibitemOpen
  \bibfield  {author} {\bibinfo {author} {\bibfnamefont {T.}~\bibnamefont
  {Brinckmann}}\ and\ \bibinfo {author} {\bibfnamefont {J.}~\bibnamefont
  {Lesgourgues}},\ }\href {\doibase 10.1016/j.dark.2018.100260} {\bibfield
  {journal} {\bibinfo  {journal} {Phys. Dark Univ.}\ }\textbf {\bibinfo
  {volume} {24}},\ \bibinfo {pages} {100260} (\bibinfo {year} {2019})},\
  \Eprint {http://arxiv.org/abs/1804.07261} {arXiv:1804.07261 [astro-ph.CO]}
  \BibitemShut {NoStop}%
\bibitem [{\citenamefont {Beutler}\ \emph {et~al.}(2011)\citenamefont
  {Beutler}, \citenamefont {Blake}, \citenamefont {Colless}, \citenamefont
  {Jones}, \citenamefont {Staveley-Smith}, \citenamefont {Campbell},
  \citenamefont {Parker}, \citenamefont {Saunders},\ and\ \citenamefont
  {Watson}}]{Beutler:2011hx}%
  \BibitemOpen
  \bibfield  {author} {\bibinfo {author} {\bibfnamefont {F.}~\bibnamefont
  {Beutler}}, \bibinfo {author} {\bibfnamefont {C.}~\bibnamefont {Blake}},
  \bibinfo {author} {\bibfnamefont {M.}~\bibnamefont {Colless}}, \bibinfo
  {author} {\bibfnamefont {D.~H.}\ \bibnamefont {Jones}}, \bibinfo {author}
  {\bibfnamefont {L.}~\bibnamefont {Staveley-Smith}}, \bibinfo {author}
  {\bibfnamefont {L.}~\bibnamefont {Campbell}}, \bibinfo {author}
  {\bibfnamefont {Q.}~\bibnamefont {Parker}}, \bibinfo {author} {\bibfnamefont
  {W.}~\bibnamefont {Saunders}}, \ and\ \bibinfo {author} {\bibfnamefont
  {F.}~\bibnamefont {Watson}},\ }\href {\doibase
  10.1111/j.1365-2966.2011.19250.x} {\bibfield  {journal} {\bibinfo  {journal}
  {Mon. Not. Roy. Astron. Soc.}\ }\textbf {\bibinfo {volume} {416}},\ \bibinfo
  {pages} {3017} (\bibinfo {year} {2011})},\ \Eprint
  {http://arxiv.org/abs/1106.3366} {arXiv:1106.3366 [astro-ph.CO]} \BibitemShut
  {NoStop}%
\bibitem [{\citenamefont {Ross}\ \emph {et~al.}(2015)\citenamefont {Ross},
  \citenamefont {Samushia}, \citenamefont {Howlett}, \citenamefont {Percival},
  \citenamefont {Burden},\ and\ \citenamefont {Manera}}]{Ross:2014qpa}%
  \BibitemOpen
  \bibfield  {author} {\bibinfo {author} {\bibfnamefont {A.~J.}\ \bibnamefont
  {Ross}}, \bibinfo {author} {\bibfnamefont {L.}~\bibnamefont {Samushia}},
  \bibinfo {author} {\bibfnamefont {C.}~\bibnamefont {Howlett}}, \bibinfo
  {author} {\bibfnamefont {W.~J.}\ \bibnamefont {Percival}}, \bibinfo {author}
  {\bibfnamefont {A.}~\bibnamefont {Burden}}, \ and\ \bibinfo {author}
  {\bibfnamefont {M.}~\bibnamefont {Manera}},\ }\href {\doibase
  10.1093/mnras/stv154} {\bibfield  {journal} {\bibinfo  {journal} {Mon. Not.
  Roy. Astron. Soc.}\ }\textbf {\bibinfo {volume} {449}},\ \bibinfo {pages}
  {835} (\bibinfo {year} {2015})},\ \Eprint {http://arxiv.org/abs/1409.3242}
  {arXiv:1409.3242 [astro-ph.CO]} \BibitemShut {NoStop}%
\bibitem [{\citenamefont {Alam}\ \emph {et~al.}(2017)\citenamefont {Alam} \emph
  {et~al.}}]{BOSS:2016wmc}%
  \BibitemOpen
  \bibfield  {author} {\bibinfo {author} {\bibfnamefont {S.}~\bibnamefont
  {Alam}} \emph {et~al.} (\bibinfo {collaboration} {BOSS}),\ }\href {\doibase
  10.1093/mnras/stx721} {\bibfield  {journal} {\bibinfo  {journal} {Mon. Not.
  Roy. Astron. Soc.}\ }\textbf {\bibinfo {volume} {470}},\ \bibinfo {pages}
  {2617} (\bibinfo {year} {2017})},\ \Eprint {http://arxiv.org/abs/1607.03155}
  {arXiv:1607.03155 [astro-ph.CO]} \BibitemShut {NoStop}%
\bibitem [{\citenamefont {Gelman}\ and\ \citenamefont
  {Rubin}(1992)}]{Gelman:1992zz}%
  \BibitemOpen
  \bibfield  {author} {\bibinfo {author} {\bibfnamefont {A.}~\bibnamefont
  {Gelman}}\ and\ \bibinfo {author} {\bibfnamefont {D.~B.}\ \bibnamefont
  {Rubin}},\ }\href {\doibase 10.1214/ss/1177011136} {\bibfield  {journal}
  {\bibinfo  {journal} {Statist. Sci.}\ }\textbf {\bibinfo {volume} {7}},\
  \bibinfo {pages} {457} (\bibinfo {year} {1992})}\BibitemShut {NoStop}%
\bibitem [{\citenamefont {Lewis}(2019)}]{Lewis:2019xzd}%
  \BibitemOpen
  \bibfield  {author} {\bibinfo {author} {\bibfnamefont {A.}~\bibnamefont
  {Lewis}},\ }\href@noop {} {\  (\bibinfo {year} {2019})},\ \Eprint
  {http://arxiv.org/abs/1910.13970} {arXiv:1910.13970 [astro-ph.IM]}
  \BibitemShut {NoStop}%
\bibitem [{\citenamefont {Sch\"oneberg}\ \emph {et~al.}(2022)\citenamefont
  {Sch\"oneberg}, \citenamefont {Franco~Abell\'an}, \citenamefont
  {P\'erez~S\'anchez}, \citenamefont {Witte}, \citenamefont {Poulin},\ and\
  \citenamefont {Lesgourgues}}]{Schoneberg:2021qvd}%
  \BibitemOpen
  \bibfield  {author} {\bibinfo {author} {\bibfnamefont {N.}~\bibnamefont
  {Sch\"oneberg}}, \bibinfo {author} {\bibfnamefont {G.}~\bibnamefont
  {Franco~Abell\'an}}, \bibinfo {author} {\bibfnamefont {A.}~\bibnamefont
  {P\'erez~S\'anchez}}, \bibinfo {author} {\bibfnamefont {S.~J.}\ \bibnamefont
  {Witte}}, \bibinfo {author} {\bibfnamefont {V.}~\bibnamefont {Poulin}}, \
  and\ \bibinfo {author} {\bibfnamefont {J.}~\bibnamefont {Lesgourgues}},\
  }\href {\doibase 10.1016/j.physrep.2022.07.001} {\bibfield  {journal}
  {\bibinfo  {journal} {Phys. Rept.}\ }\textbf {\bibinfo {volume} {984}},\
  \bibinfo {pages} {1} (\bibinfo {year} {2022})},\ \Eprint
  {http://arxiv.org/abs/2107.10291} {arXiv:2107.10291 [astro-ph.CO]}
  \BibitemShut {NoStop}%
\bibitem [{\citenamefont {M\'egevand}\ and\ \citenamefont
  {Membiela}(2023)}]{Megevand:2023nin}%
  \BibitemOpen
  \bibfield  {author} {\bibinfo {author} {\bibfnamefont {A.}~\bibnamefont
  {M\'egevand}}\ and\ \bibinfo {author} {\bibfnamefont {F.~A.}\ \bibnamefont
  {Membiela}},\ }\href {\doibase 10.1088/1475-7516/2023/06/007} {\bibfield
  {journal} {\bibinfo  {journal} {JCAP}\ }\textbf {\bibinfo {volume} {06}},\
  \bibinfo {pages} {007} (\bibinfo {year} {2023})},\ \Eprint
  {http://arxiv.org/abs/2302.13349} {arXiv:2302.13349 [gr-qc]} \BibitemShut
  {NoStop}%
\bibitem [{\citenamefont {Malik}\ and\ \citenamefont
  {Wands}(2009)}]{Malik:2008im}%
  \BibitemOpen
  \bibfield  {author} {\bibinfo {author} {\bibfnamefont {K.~A.}\ \bibnamefont
  {Malik}}\ and\ \bibinfo {author} {\bibfnamefont {D.}~\bibnamefont {Wands}},\
  }\href {\doibase 10.1016/j.physrep.2009.03.001} {\bibfield  {journal}
  {\bibinfo  {journal} {Phys. Rept.}\ }\textbf {\bibinfo {volume} {475}},\
  \bibinfo {pages} {1} (\bibinfo {year} {2009})},\ \Eprint
  {http://arxiv.org/abs/0809.4944} {arXiv:0809.4944 [astro-ph]} \BibitemShut
  {NoStop}%
\bibitem [{\citenamefont {Liu}\ \emph {et~al.}(2023)\citenamefont {Liu},
  \citenamefont {Bian}, \citenamefont {Cai}, \citenamefont {Guo},\ and\
  \citenamefont {Wang}}]{Liu:2022lvz}%
  \BibitemOpen
  \bibfield  {author} {\bibinfo {author} {\bibfnamefont {J.}~\bibnamefont
  {Liu}}, \bibinfo {author} {\bibfnamefont {L.}~\bibnamefont {Bian}}, \bibinfo
  {author} {\bibfnamefont {R.-G.}\ \bibnamefont {Cai}}, \bibinfo {author}
  {\bibfnamefont {Z.-K.}\ \bibnamefont {Guo}}, \ and\ \bibinfo {author}
  {\bibfnamefont {S.-J.}\ \bibnamefont {Wang}},\ }\href {\doibase
  10.1103/PhysRevLett.130.051001} {\bibfield  {journal} {\bibinfo  {journal}
  {Phys. Rev. Lett.}\ }\textbf {\bibinfo {volume} {130}},\ \bibinfo {pages}
  {051001} (\bibinfo {year} {2023})},\ \Eprint
  {http://arxiv.org/abs/2208.14086} {arXiv:2208.14086 [astro-ph.CO]}
  \BibitemShut {NoStop}%
\end{thebibliography}%

\end{document}